\documentclass{emulateapj}
\usepackage{graphicx}

\def\ltsima{$\; \buildrel < \over \sim \;$}
\def\simlt{\lower.5ex\hbox{\ltsima}}
\def\gtsima{$\; \buildrel > \over \sim \;$}
\def\simgt{\lower.5ex\hbox{\gtsima}}
\def\kms{{\rm\,km\,s^{-1}}}

\def\kpc{{\rm\,kpc}}
\def\mpc{{\rm\,Mpc}}
\def\msun{{\rm\,M_\odot}}

\def\lsun{{\rm\,L_\odot}}

\newcommand{\fmmm}[1]{\mbox{$#1$}}
\newcommand{\scnd}{\mbox{\fmmm{''}\hskip-0.3em .}}
\newcommand{\scnp}{\mbox{\fmmm{''}}}
\newcommand{\mcnd}{\mbox{\fmmm{'}\hskip-0.3em .}}

\def\deg{^\circ}
\def\degg{\hbox{$\null^\circ$\hskip-3pt .}}

\def\s{\ifmmode \widetilde \else \~\fi}
\def\={\overline}

\def\spose#1{\hbox to 0pt{#1\hss}}

\def\lta{\mathrel{\spose{\lower 3pt\hbox{$\mathchar"218$}}
     \raise 2.0pt\hbox{$\mathchar"13C$}}}
\def\gta{\mathrel{\spose{\lower 3pt\hbox{$\mathchar"218$}}
     \raise 2.0pt\hbox{$\mathchar"13E$}}}
\def\Dt{\spose{\raise 1.5ex\hbox{\hskip3pt$\mathchar"201$}}}    
\def\dt{\spose{\raise 1.0ex\hbox{\hskip2pt$\mathchar"201$}}}    

\def\dotsfill{\leaders\hbox to 1em{\hss.\hss}\hfill}

\def\Gyr{{\rm\,Gyr}}

\begin{document} 

\title{The Haunted Halos of Andromeda and Triangulum:\\
A panorama of galaxy formation in action}

\author{R. Ibata\altaffilmark{1}, 
N. F. Martin\altaffilmark{2},
M. Irwin\altaffilmark{3}, 
S. Chapman\altaffilmark{3}, 
A. M. N. Ferguson\altaffilmark{4}, 
G. F. Lewis\altaffilmark{5}, 
A. W. McConnachie\altaffilmark{6}}
\altaffiltext{1}{Observatoire Astronomique, UniversitŽ de Strasbourg, CNRS, 11 rue de l'universitŽ, 67000 Strasbourg, France}
\altaffiltext{2}{Max-Planck-Institut f\"ur Astronomie, K\"onigstuhl 17, D-69117 Heidelberg, Germany}
\altaffiltext{3}{Institute of Astronomy, Madingley Road, Cambridge, CB3 0HA, U.K.}
\altaffiltext{4}{Institute for Astronomy, University of Edinburgh,
Royal Observatory, Blackford Hill, Edinburgh, UK EH9 3HJ}
\altaffiltext{5}{Institute of Astronomy, School of Physics, A29, University of Sydney, NSW
2006, Australia}
\altaffiltext{6}{Department of Physics and Astronomy, University of Victoria, Victoria, B.C., V8W
3P6, Canada}

\begin{abstract}
We present a deep photometric survey of the Andromeda galaxy, conducted with the
wide-field cameras of the CFHT and INT telescopes.  The surveyed area covers
the inner $50\kpc$ of the galaxy and the Southern quadrant
out to a projected distance of $\sim 150\kpc$. A survey extension to M33 
at $>200\kpc$ probes the interface between the halos of these two galaxies.
This survey is the first systematic panoramic study of this very outermost region
of galaxies. We detect a multitude of large-scale 
structures of low surface brightness, including several streams. 
Significant variations in stellar populations due to intervening stream-like structures
are detected in the inner halo along the minor
axis. This, together with the fact that the light profile 
between $0\degg5 < R < 1\degg3$ follows the exponential ``extended disk'',
is particularly important in shedding light on the mixed and
sometimes conflicting results reported in previous studies.
Two new relatively luminous (${\rm M_V \sim -9}$)
dwarf galaxies And~XV and XVI are found in the study; And~XVI is a particularly
interesting specimen being located $270\kpc$ in front of M31, towards the Milky Way.
Underlying the many substructures that we have uncovered
lies a faint, smooth and extremely extended halo component, 
reaching out to 150~kpc, whose stellar populations
are predominantly metal-poor. This is consistent with recent claims
based on spectroscopy of a small sample of stars.
We find that the smooth halo component in M31
has a radially-decreasing profile that can be fit
with a Hernquist model of immense scale radius $\sim 55\kpc$,
almost a factor of 4 larger than theoretical predictions.
Alternatively a power-law with $\Sigma_V \propto R^{-1.91 \pm 0.11}$
can be fit to the projected profile, similar to the density
profile in the Milky Way. If it is symmetric, the total luminosity of this structure
is $\sim 10^9 \lsun$, again similar to the stellar halo of the Milky Way.
This vast, smooth, underlying halo is reminiscent of
a classical ``monolithic'' model and completely unexpected 
from modern galaxy formation models where stars form in 
the most massive subhalos and are preferentially delivered 
into the inner regions of the galaxy.
Furthermore, over the region surveyed,
the smooth stellar halo follows closely the profile of the dark matter distribution
predicted from earlier kinematic analyses.
M33 is also found to have an extended metal-poor halo component, which can be fit with 
a Hernquist model also of scale radius $\sim 55\kpc$. These extended slowly-decreasing halos
will provide a challenge and strong constraints for further modeling. 
\end{abstract}


\section{Introduction}

\renewcommand{\thefootnote}{\fnsymbol{footnote}} \footnotetext[1]{Based on
observations obtained with MegaPrime/MegaCam, a joint project of CFHT and
CEA/DAPNIA, at the Canada-France-Hawaii Telescope (CFHT) which is operated
by the National Research Council (NRC) of Canada, the Institute National des
Sciences de l'Univers of the Centre National de la Recherche Scientifique of
France, and the University of Hawaii.}

The outskirts of galaxies hold fundamental clues about their formation
history. It is into these regions that new material continues to arrive
as part of their on-going assembly, and it was also into these regions
that material was deposited during the violent interactions in the galaxy's 
distant past. 
Moreover, the long dynamical timescales for structures beyond the 
disk ensure that the debris of accreted material takes a very long time
to be erased by the process of phase mixing, which in turn means that we can hope
to detect many of these signatures of formation as coherent spatial structures
\citep{johnston96}.

Much theoretical effort has been devoted in recent years to understanding
the fine-scale structure of galaxies \citep{abadi03, bullock05, abadi06}, 
as researchers realized that 
cosmological models could be tested not only with the classical large-scale
probes such as galaxy clusters, filaments and voids, but also
with observations on galactic and sub-galactic scales \citep{freeman02}.
Indeed, it is precisely
in these latter regions that the best constraints on cosmology
are expected to be put \citep{springel06} in the coming decades.
$\Lambda$-CDM cosmologies, in particular, are now sufficiently well
developed theoretically (e.g., \citealt{bullock01, bullock05}) that the Local Group provides a means of
directly testing and constraining these theories, by observing
the profiles of density, age, and metallicity of the structure and
substructure predicted to be found in the outer parts of galaxy disks
and in galaxy halos.

\subsection{The Andromeda galaxy}

Andromeda, like the Milky Way, is a canonical galaxy, and a laboratory
for examining in close detail many of the astrophysical processes that
are investigated in the more distant field. 
Studying Andromeda and Triangulum in the Local Group has the advantage 
that it affords us a view free from the problems that
plague Galactic studies due to our position within the Milky Way, yet their location
within the Local Group allows us to resolve and study individual stars
and deduce population properties in incomparably greater detail than is possible
in distant systems.

Andromeda is the closest giant spiral galaxy to
our own, and the only other giant galaxy in the Local Group. 
In many ways Andromeda is the ``sister'' to the Milky Way,
having very similar total
masses (including the dark matter, \citealt{evans00b, ibata04}),
having shared a common origin, and probably sharing the
same ultimate fate when they finally merge in the distant future.
However,  there
are significant differences between these ``twins''.  M31 is slightly more luminous
than the Milky Way, it has a higher rotation speed, and a bulge with higher
velocity dispersion. M31 possesses a globular cluster system with $\sim 500$
members, approximately three times more numerous than that of the Milky Way.  
The disk of Andromeda is also
much more extensive, with a scale-length of $5.9 \pm 0.3 \kpc$ (R-band value
corrected for a distance of $785\kpc$, \citealt{walterbos88}) compared to
$2.3\pm0.1$ for the Milky Way \citep{ruphy}; but which is currently forming
stars at a lower rate than the Galaxy \citep{avila-reese, walterbos94}.
There are indications that the Milky Way has
undergone an exceptionally low amount of merging and has unusually 
low specific angular momentum, whereas M31 appears to be
a much more normal galaxy in these respects \citep{hammer07}.
Though possibly the consequence of low-number statistics, it is tempting to
attribute significance to the fact that Andromeda has a compact elliptical (M32)
and three dwarf elliptical
galaxies (NGC~205, NGC~147, NGC~185) among its entourage of satellites,
and no star-forming dwarf irregulars (dIrrs) within $200\kpc$, whereas the Milky Way has no
ellipticals but two dIrrs. 
However, it is perhaps in their purported halo
populations that the differences between the two galaxies are most curious
and most interesting.

\subsection{Comparing the halos of Andromeda and the Milky Way}

A large number of studies of the Milky Way halo (e.g., \citealt{ryan91, chiba00}, 
and references therein), have revealed that this structure is very metal-poor,
with a median ${\rm \langle [Fe/H] \rangle = -1.6}$. It has a
high velocity dispersion, with $(U,V,W)$ values in the solar
neighborhood of $(141\kms:106\kms:94\kms)$, and a small prograde 
rotation of 30~--~$50\kms$. There is broad agreement
that the stellar halo is flattened with $b/a \sim 0.6$ 
(e.g., \citealt{morrison00, yanny00, chen01, siegel02}), though there are indications
that the distribution becomes spherical beyond 15~--~$20\kpc$ \citep{chiba00}.

The volume density profile and extent of this structure have been harder to pin down.
This is perhaps not surprising given the patchy sky coverage of most studies,
since current expectations are that the stellar
halo is significantly lumpy \citep{bullock05}. 
The stellar volume density is generally modeled as  $\rho(r) \propto r^{-\alpha}$,
and recent studies \citep{wetterer96, morrison00, yanny00, ivezic00, siegel02,
vivas06} have found values of the exponent ranging from 
$\alpha = 3.55\pm0.13$ \citep{chiba00} to $\alpha = 2.5\pm0.3$ \citep{chen01},
with a general consensus of $\rho(r) \propto r^{-3}$.
Note that in external systems, where we
observe the projected density, $\rho(r) \propto r^{-3}$ would correspond to
$\Sigma(R) \propto R^{-2}$.

Recent wide-field studies have gone a long way in improving
our knowledge of the 
radial extent of the Milky Way halo. Using the SDSS database,
\citet{yanny00} were able to follow A-colored stars in the halo
to $\sim 25\kpc$, and blue-straggler candidates out to $\sim 50\kpc$.
From the same survey, 
\citet{ivezic00} followed the profile of RRLyrae candidates, and found
a sharp drop in the star-counts between 50~--~$60\kpc$, though this
discontinuity in density has since been found to be due to the intervening stream of the Sgr dwarf 
galaxy \citep{ibata01c}. From VLT spectroscopy of 34 faint A-stars selected
from the SDSS, \citet{clewey05} were able to show that the
stellar halo extends out to at least $100\kpc$, although again
a sub-sample of their stars appears to be associated to the 
stream of Carbon stars emanating from the Sgr dwarf \citep{ibata01a}.
Several other studies have found 
evidence for further lumpy structures in the halo 
(e.g., \citealt{vivas06, martin07a, belokurov07}, and references therein).

It has been believed for many years that M31
possesses a stellar halo that is fundamentally different to
that deduced from the above and earlier observations in
the Milky Way.  The first deep CCD studies by 
\citet{mould86} in a field in the inner halo of M31 found a surprisingly
high mean metallicity of ${\rm \langle [M/H] \rangle = -0.6}$. 
While the surface brightness profile measured along the minor axis from integrated light
\citep{pritchet94} is consistent with a de Vaucouleurs 
$R^{1/4}$-law out to $R=20\kpc$, quite unlike the power-law behavior 
deduced for the halo of the Milky Way.
Both the de Vaucouleurs profile and the high metallicity
are suggestive of an active merger history at the time of halo (or bulge)
formation.

The existence of the metal-rich halo population 
was confirmed by several subsequent studies; notably among these
the wide-field ($0.16$~deg$^2$) photometric study by \citet{durrell01} 
in a location $20\kpc$ out along the minor axis. In addition to
the main ${\rm \langle [M/H] \rangle = -0.5}$ component, 
\citet{durrell01} also discovered
that 30-40\% of of the stars at this location belong to a metal-poor population.
The surface density of the metal-poor sub-sample falls off rapidly as 
$\Sigma(R) \propto R^{-5.25\pm 0.63}$, but slower than the 
$\Sigma(R) \propto R^{-6.54\pm 0.59}$ relation for the metal-rich sub-sample.
These results were later complemented by the same authors with a minor axis field at $R=30\kpc$
\citep{durrell04} , which showed essentially identical 
abundance properties to their $20\kpc$ field, leading them to conclude that the outer halo 
shows little or no radial metallicity gradient.

As an alternative to the above ``wide-field'' approach,
\citet{bellazzini03} analyzed a set of 16 HST/WFPC2 
fields with much deeper photometry, 
mostly in and around the M31 disk, but with some fields
extending out to a distance of $35\kpc$.
Throughout this area they 
detect the previously-discussed dominant metal-rich component
with ${\rm [Fe/H] \sim -0.6}$, but also an additional high metallicity component
with ${\rm [Fe/H] \sim -0.2}$. Interestingly, they found that the 
fraction of metal-poor stars is constant from field to field, though
metal-rich stars are enhanced in regions containing substructure, especially 
along the extended path of the Giant Stream \citep{ibata01b}.

The inclusion of kinematic information has been extremely useful,
but has also added another dimension of complexity to the 
puzzle.
\citet{reitzel02} analyzed a sample of 29 stars in a field at 
$R=19\kpc$ on the minor axis, and found the mean metallicity to be in the
range ${\rm \langle [M/H] \rangle = -1.9}$ to $-1.1$, dependent on
calibration and sample selection issues, but significantly lower than the 
results deduced from the above photometric analyses. 

A wider-field view was obtained by \citet{chapman06}, who sampled
the halo at 54 locations between 10~--~$70\kpc$, isolating 827 
out of a sample of $\sim 10^4$ stars as having kinematics 
consistent with being halo members. The population
was found to have ${\rm \langle [Fe/H] \rangle \sim -1.4}$ with a dispersion of
0.2~dex, indicating that kinematic selection reveals a
halo similar to that of the Milky Way
underneath the ``halo'' substructures, which in many cases
are metal-rich, and in general 
cannot have halo-like kinematics. The (central) velocity dispersion
of $152\kms$ deduced from the sample, is also comparable
to that of the Milky Way.

In an impressive effort of finding needles in a haystack,
\citet{kalirai06b} and \citet{gilbert06} extended the 
kinematic coverage out to $165\kpc$, and claim a detection of the
halo at $R>100\kpc$ based on a sample of 3 stars. 
To minimize contamination they 
implemented a complex non-linear algorithm to assign likelihoods to the observed
stars, and as the algorithm was trained on the inner region of M31,
the biasses for the outer halo population are not well known.

\subsection{The Triangulum galaxy}

If Andromeda is the twin of the Milky Way, the Triangulum galaxy (M33)
with a mass $\sim 10$ times lower than either of these two giants, 
is their little sister. M33 is the third brightest galaxy in the Local Group ($M_V = -18.9$),
and probably a satellite of M31. The relatively undisturbed optical appearance
of M33 places strong constraints on the past interaction of these two galaxies \citep{loeb05},
though it should be noted that the gaseous component is extremely warped \citep{rogstad76}.

The early CCD study of the halo of M33 by \citet{mould86} claimed
an inner halo component with a more ``normal'' metallicity (${\rm \langle [M/H] \rangle = -2.2}$)
than deduced for M31. In reality however, this field lies within the disk of M33
and does not probe the ``halo'', as we show below in \S9.
Further progress in understanding 
the elusive halo component 
of this galaxy was only achieved recently. In their kinematic study of star clusters in M33,
\citet{chandar02} find evidence for two sub-populations, with old clusters showing
evidence for a large velocity dispersion, which they interpret as the sign of a halo population.
Further signs of this halo component were detected in the spectroscopic study of
\citet{mcconnachie06b} with Keck/DEIMOS, who distinguished halo field stars
from stars in the disk via their kinematics, and deduce a mean metallicity for the
halo component of ${\rm \langle [Fe/H] \rangle = -1.5}$, with a narrower
spread of abundance than the disk stars.

\subsection{Halos of more distant disk galaxies}

Due to their extremely faint nature the halos of spiral galaxies
beyond the Local Group have been extremely challenging to
observe. A major advance in detecting extra-planar light in distant
galaxies was made by stacking 1047 edge-on
spiral galaxies observed in the SDSS \citep{zibetti04a}. The resulting 
stack showed a flattened ($c/a \sim 0.6$) distribution
with a power-law density profile $\rho(r) \propto r^{-3}$,
similar to the properties of the halo of the Milky Way deduced from the studies
reviewed above.
This structure could be detected out to approximately 10 exponential
scalelengths of the disk (i.e., approximately $25\kpc$ for the case of the Milky Way).
An analogous structure was also detected directly from the surface
brightness around a single isolated
galaxy in an ultra-deep HST survey \citep{zibetti04b}.

Extra-planar populations have also been detected via star-counts
of resolved RGB populations
in nearby ($<10\mpc$) galaxies from deep HST imaging. 
Notable among these is the survey of \citet{mouhcine05a, mouhcine05b},
who employed WFPC2 to survey 8 nearby spirals. Their fields probed the minor axis
halo out to $R=13\kpc$. Interestingly, they find a correlation between galaxy 
luminosity and the metallicity of the extra-planar population, with low luminosity 
galaxies containing metal-poor stars with a narrow abundance spread, while 
luminous galaxies contain metal-rich stars and a wide abundance spread. 
Their results for galaxies of similar luminosity to M31 are
in good agreement with the metallicity distribution of minor axis fields in Andromeda at 
10~--~$20\kpc$. 

However, as we will show below, the minor axis fields in M31, from which most of the
information on the ``halo'' or ``spheroid'' is derived, do not
directly probe that component. Furthermore, as we have reviewed above,
kinematically-selected halo stars in M31 display a similar metallicity to genuine
halo stars in the Milky Way \citep{chapman06}. These considerations suggest that the
Mouhcine relation is caused by small structures accreted into
the inner regions of the halo, and which are largely supported by rotation, rather than
random motions. The correlation of the metallicity of the extraplanar stars 
with galaxy luminosity found by Mouhcine et al. may then simply reflect that
more massive host galaxies are able to accrete larger dwarf galaxies which themselves
have a higher metallicity. 

Nevertheless, we stress that all of these observations beyond the Local Group 
are derived from regions close to the centre of the galaxy, and there is concern
that contamination from other components, such as streams or a warped disk could be
affecting the observations. Extending further out in radius, as we will do
in this contribution, will allow us to eliminate this uncertainty. But most importantly
it will allow us to examine a different region of the halo, one that is less
dominated by the remnants of massive accretions.

\subsection{Theoretical motivation}

Several theoretical studies have been undertaken in recent years to
attempt to understand and reproduce the above observations and to make
useful predictions for the next generation of surveys. 

\citet{bullock05} implemented a hybrid N-body plus semi-analytic approach. 
Their simulations provide very high spatial resolution compared to the
other studies discussed below, which they achieve by concentrating on 
each merger event in turn, with the rest of the galaxy modeled 
with analytic (but time varying) potentials. The drawback of this method
is that the dynamical evolution of the system is not fully self-consistent, and
star-formation is implemented with empirical recipes.
They find that the present-day density profile of 
stars within $10\kpc$ of a Milky Way or M31-like galaxy should be shallow, $\rho(r) \propto r^{-1}$, 
steepening to $\rho(r) \propto r^{-4}$ beyond $50\kpc$, resembling a Hernquist profile with scale
radius of $\sim 15\kpc$.
They also find that the bulk of the stars that constitute a stellar halo were formed more than $8\Gyr$ ago,
with most of these stars originating from massive accretions ($M_{vir} > 2 \times 10^{10} \msun$).
Beyond $30\kpc$, substructure begins to predominate in their simulations,
and they find that most of the stars beyond this radius arrived after the last major merger.

The problem of stellar halo formation was also tackled by \citet{renda05}, who
used a chemodynamical code to treat self-consistently
gravity, gas dynamics, radiative cooling, star formation and chemical enrichment.
The drawback of this approach was a very much lower spatial resolution 
compared to \citet{bullock05}.
\citet{renda05} find a large ($\sim 1$~dex) spread in the
mean metallicity of halos of galaxies of a given (final) luminosity,
where the large variations in the metallicity distribution between their galaxy models
is related to the diversity in the galactic mass assembly history. 
This is somewhat at odds with the finding that M31 and the Milky Way have
underlying halos of similar metallicity \citep{chapman06}.
They also find that a more extended
assembly history gives more massive stellar halos, and a higher halo surface brightness.

Yet another approach was adopted by \citet{abadi06}, who
undertook SPH simulations that follow the gas evolution in a small sample of galaxy models forming
in a $\Lambda$CDM cosmology. Overcooling early on leads to large spheroid component in their simulations,
though they claim that the insensitivity of the halo parameters to the final stellar halo mass implies
that their simulations are also applicable to Milky Way-like systems.
In their models stars formed in situ in the galaxy are all confined to the inner luminous region, while
accreted stars dominate beyond $20\kpc$, 
and are the main population contributing to the spheroid.
The stellar surface density profile is very similar in all their simulations,
and has $\Sigma(R) \propto R^{-2.3}$ at $r \sim 20\kpc$, steepening to
$\Sigma(R) \propto R^{-2.9}$ at $r \sim 100\kpc$, and steepening
further to at $\Sigma(R) \propto R^{-3.5}$ near the virial radius.
Furthermore, they find that the stellar halo is a mildly triaxial 
structure ($\langle c/b \rangle = 0.90$, $\langle c/a \rangle = 0.84$,
with no obvious alignment of the triaxial halo with the angular momentum vector of the galaxy.
Old stars disrupted in the early history of the galaxy are ejected into highly 
eccentric and energetic orbits during close perigalactic passages, and it is 
these stars that primarily populate the outer halo.

Complementary studies using pure N-body simulations were undertaken 
by \citet{diemand05} and \citet{gauthier06}.
\citet{diemand05} focus on the evolution of high density peaks in cosmological simulations
that form Milky Way-like galaxies.
Their result of relevance to the present study is the asymptotic density profile of these peaks
in the galaxy simulation: they find that the outer profile behaves as $\rho(r) \propto r^{-3.26}$ 
for $1\sigma$ peaks, steepening to
$\rho(r) \propto r^{-4.13}$ for $2.5\sigma$ and $\rho(r) \propto r^{-5.39}$ for $4\sigma$ peaks.
In contrast, \citet{gauthier06} simulate the evolution of satellites around a fully-formed
M31-like galaxy, with the satellites modeled as a collection of NFW density profiles \citep{navarro97}.
They do not consider star-formation in the satellites, but instead identify
the 10\% most bound particles as tracers of the stars in the satellite. They predict
that disrupted satellites give rise to a halo luminosity profile that falls as
$\rho(r) \propto r^{-3.5}$ at large radii. Since massive satellites correspond to rare
overdense peaks in cosmological simulations, the difference in the profile slope compared to
\citet{diemand05} suggests that taking the full cosmological evolution 
of the host galaxy into account is important.

\subsection{Purpose of the present study}

In this contribution we are building upon an earlier wide-field survey of Andromeda
with the Wide Field Camera camera at the Isaac Newton Telescope \citep{ibata01b, ferguson02, irwin05}. 
This panoramic survey covered the entirety
of the disk and inner halo of the galaxy out to $\sim 55\kpc$ (see Fig.~1), which
combined with follow-up kinematics from Keck/DEIMOS \citep{ibata04, ibata05, chapman06} and deep 
HST/ACS photometry in selected fields \citep{ferguson05, faria07}
opened up a new violent vision of an apparently normal
disk galaxy. We found that M31 possesses
of order half a dozen substructures, probably debris fragments from merging
galaxies that have not yet lost all spatial coherence \citep{ferguson02, ferguson05}; that it
is surrounded by a vast 
rotating disk-like structure, extending out to $\sim 40\kpc$ \citep{ibata05}; that
it contains a giant stellar stream of width greater than the diameter of the disk of the Milky Way
and $> 100\kpc$ long \citep{ibata01b, mcconnachie03}; and that underlying all of this substructure 
there is a kinematically hot, metal-poor halo \citep{chapman06}.

Thus the inner halo region covered by the INT survey is 
completely contaminated by these various structures.
Indeed it was a surprising result of that survey that it is necessary 
to observe at much larger radius to obtain
a  clear measurement of the accretion rate, the incidence of
sub-structures, the stellar mass of the accreted objects, and the
global properties of the halo.
We therefore embarked on the deep imaging campaign
of the outer halo presented in this contribution, undertaken
with MegaCam, a state-of-the-art wide-field camera at the CFHT.

One of the main aims of the present survey was to investigate the
prediction of CDM cosmology that upward of 500 satellites
reside in the halo of a galaxy like M31 \citep{klypin99, moore99}. The possibility remains that
many dwarf galaxies are being missed in current surveys. However
we defer all discussion of this issue to a companion
paper (Martin et al. 2007b). 

The layout of this paper is as follows. In \S2\ we first present the 
photometric data and data processing. The color-magnitude distribution of
detected sources is discussed in \S3, and their spatial distribution
in \S4. The resulting maps of the stellar populations of interest
are presented in \S5, continuing in \S6 with the detected streams
and other spatial substructures, and in \S7 with the properties
of the outer halo. The radial profiles of the stellar populations in M31 are
analyzed in \S8. A short discussion of the properties of the halo
of M33 are presented in \S9. Finally in \S10 we discuss the implications
of our findings and compare to previous studies, and draw conclusions
in \S11.

Throughout this
work, we  assume a distance of  $785\kpc$ to M31  \citep{mcconnachie05}.
We also adopt the convention of using $R$ to denote projected radius, 
$s$ an elliptical projected radius, and $r$ a three-dimensional distance or radius.

\begin{figure}
\begin{center}
\includegraphics[angle=-90, width=\hsize]{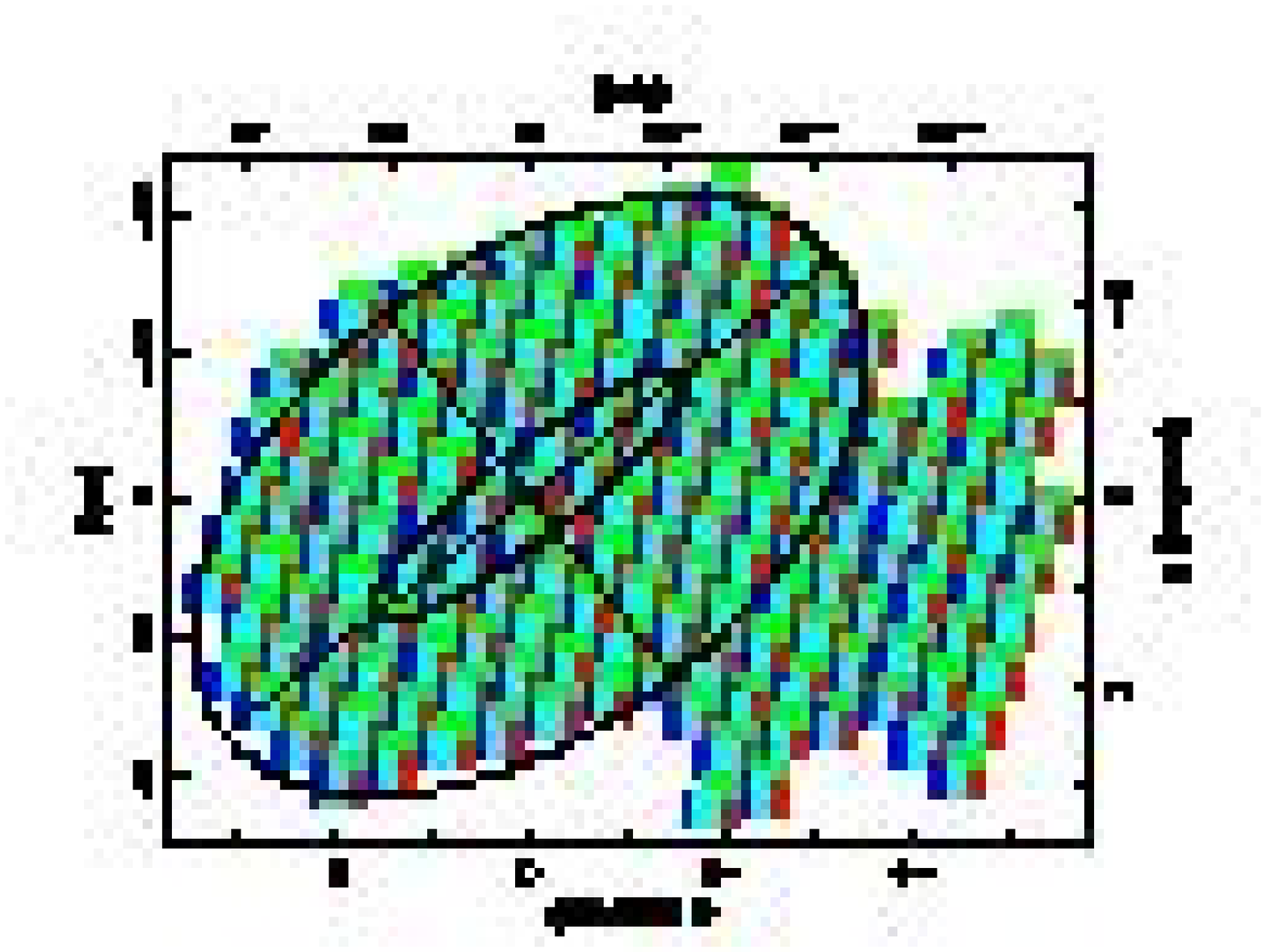}
\end{center}
\caption{The coverage of our large panoramic survey of M31 with the
INT camera, in standard coordinates $(\xi,\eta)$. The  inner
ellipse represents a disk of inclination $77\deg$ 
and radius $2\deg$ ($27\kpc$), the approximate end of the 
regular HI disk. The
outer ellipse shows a $55\kpc$ radius ellipse flattened to $c/a=0.6$, 
and the major
and minor axis are indicated with straight lines out to this ellipse. This
map is constructed from a total of 164 INT/WFC individual pointings.}
\end{figure}

\section{Observations}

\subsection{INT observations}

The Wide Field Camera (WFC) of the Isaac Newton Telescope (INT) was used in
four observing runs between 1998 and 2003 to map the Andromeda galaxy over
the area displayed in Fig.~1. The observations were taken with the V and i
filters, with exposures of $1200$~sec and $900$~sec, respectively, in each
of these two bandpasses. The data were obtained in dark skies, with typical
seeing of $1\scnp$. A total of 164 individual fields were observed, each
covering an ``L''-shaped region of 0.33~deg$^2$. A small $\sim 5$\% overlap
between adjacent fields was adopted to ensure a homogenous photometric
survey.

The images were processed by the Cambridge Astronomical Survey Unit (CASU)
pipeline \citep{irwin01}, in an identical manner to that described
in \citet{segall06}.  This includes corrections for
bias, flat-fielding, and for the fringing pattern. The software then
proceeds to detect sources, and measures their photometry, the image profile
and shape. Based upon the information contained in the curve of growth, 
the algorithm classifies the
objects into noise detections, galaxies, and probable stars.
(For comparison to previous studies using this 
classification algorithm, throughout this
paper we adopt as stars
those objects that have classifications of either -1 or -2 in both colors; this
corresponds to stars up to $2\sigma$ from the stellar locus).

\begin{figure}
\begin{center}
\includegraphics[angle=-90, width=\hsize]{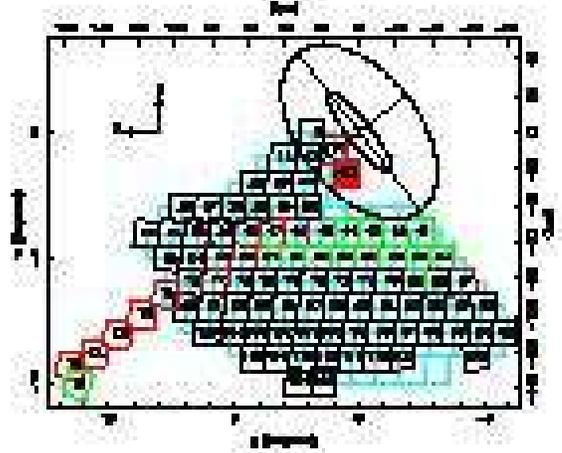}
\end{center}
\caption{The main area surveyed with the CFHT MegaCam instrument.
As we describe below, the image stability over the field of view of the camera
varied slightly from one year to another.
We therefore show the year that the field was observed in by a color code: red, green and black
mark fields obtained in 2003, 2004 and 2005--2006, respectively. The field T6, centered on M33,
was observed in primarily in 2004, with some data in 2003. The offset fields colored turquoise
mark the positions of the short exposure fields.
In the case of field H13, we also 
display the layout of the 36 CCDs. The meaning of the ellipses
centered on M31 is described in Fig.~1.}
\end{figure}

\subsection{CFHT observations}

\begin{figure*}
\begin{center}
\includegraphics[angle=-90, width=\hsize]{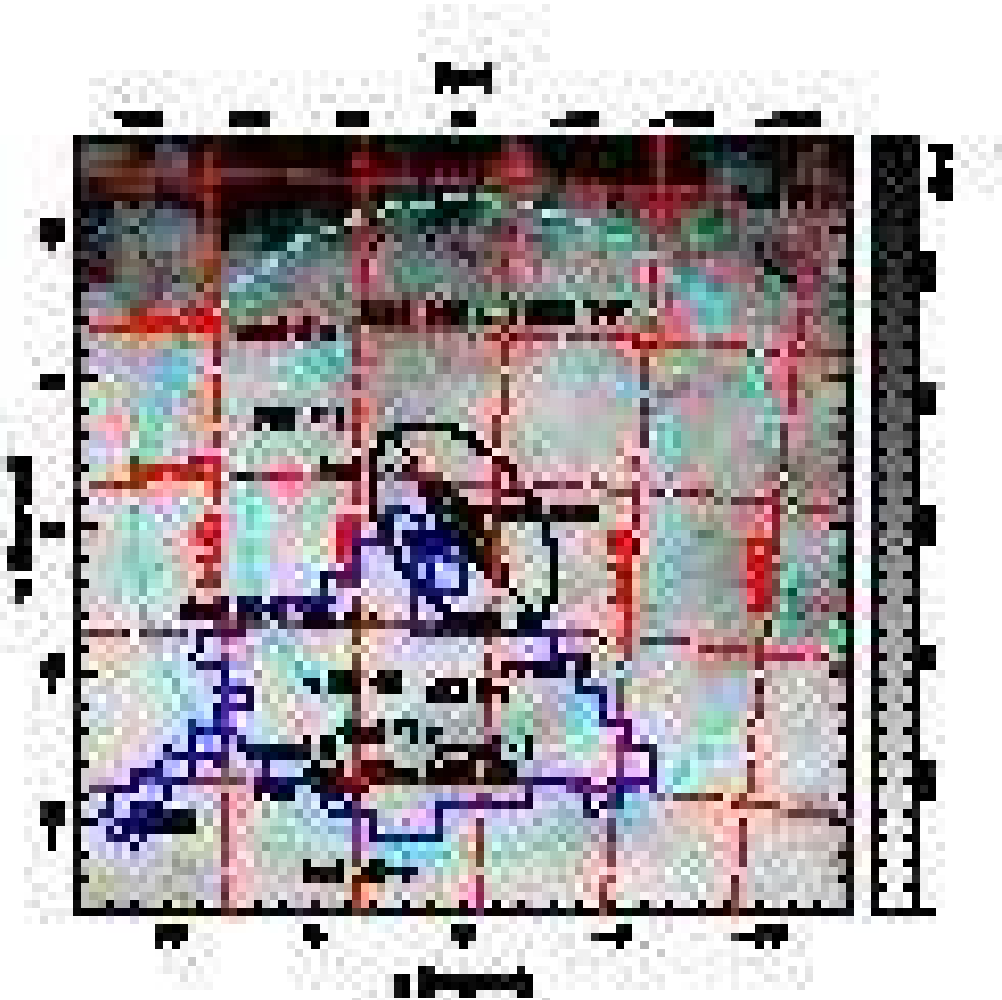}
\end{center}
\caption{The survey region (irregular blue polygon) is overlaid on a schematic 
diagram of M31 and surrounding Local Group structure. Note that the 
survey extension along the M31 minor axis reaches M33 and therefore probes
the halos of both these disk galaxies. In addition to the ellipses reproduced from 
Fig.~1, the two concentric (dashed-line) circles show projected radii of $100\kpc$ and
$150\kpc$. A grid in Galactic longitude and latitude has been marked.
The extinction over the surveyed region, interpolated from the maps of
\citet{schlegel} is also shown.}
\end{figure*}

The survey of the inner halo of M31 with the INT was complemented with a deeper
survey with the CFHT MegaCam wide-field camera to probe the outer reaches of
the halo of this galaxy.  MegaCam consists of a mosaic of 36 $2048\times4612$ pixel
CCDs, covering a $0.96^\circ \times 0.94^\circ$ field, with a pixel scale of
$0.187 {\rm \, arcsec/pixel}$. The greater photometric depth and field-of-view
achievable with this instrument makes it particularly powerful in such regions
of extremely low surface density of stars. The g and i-band filters were
used, totalling $5 \times 290$~sec of exposure per field in each passband. Figure~2
displays the survey fields, while Fig.~3 shows this area in relation to the environment
around M31. The survey comprises 89 deep fields, observed
in service mode over the 2003 to 2006 seasons. We chose a tiling
pattern with no 
overlap between the deep fields, using instead short ($45$~sec) exposures in
g and i to establish a consistent photometric level over the survey. These
short exposure images were taken offset by half a field size in the
right-ascension and declination directions.  The fields were observed in
photometric conditions in good seeing conditions (typically better than
$0\scnd8$). 
In addition, the two inner halo fields marked H11 and H13 were retrieved from the CFHT archive.
These g and i-band images are somewhat deeper that the main survey
fields with exposures of $5\times 289$~sec in each passband.
A further field centered on M33 (marked field T6 in Fig.~2) was 
obtained from the archive. After elimination of frames with poorer seeing ($>1\scnp$)
or CCD controller problems, 37 g-band frames and 32 i-band frames were
combined, for a total of 18306~sec in the g-band and 19165~sec in the i-band.

The solid angle covered by the INT survey corresponds to a
projected area of $\sim 9500 \kpc^2$ at the distance of M31
($\sim 7400 \kpc^2$ not overlapping with the MegaCam survey),
while the MegaCam survey area subtends $1.6\times10^4 \kpc^2$.
This vast area encompasses several previously known structures,
as we show in Fig.~3. These are the dwarf galaxies M32, NGC~205,
And~I, And~II (though we miss its center), And~III, And~IX, as well as the new discoveries from this work: And~XI, And~XII, And~XIII, all discussed in \citep{martin06}, and 
And~XV, and And~XVI presented below. We also mark the positions of the known
globular clusters in the MegaCam region: GC~5, GC~6, EC~4
\citep{mackey06, mackey07}, and
GC-M06 \citep{martin06}.

In addition to the INT fields and the 92 contiguous MegaCam fields, we consider below
two additional fields, which will be used as background references: 
a comparison field taken for a study of the Draco
dwarf spheroidal (dSph) galaxy (field D7 of \citealt{segall06}, located at $\ell=81\degg5$, $b=34\degg9$), 
and the field D3 of the Legacy Survey of the CFHT (CFHTLS). 
The observations on the Draco dSph comparison field 
had slightly different exposure times to those taken for the M31 survey (950~s in g and 1700~s in i), 
though similar image quality. From the public release data of the CFHTLS field D3 
(located at $\ell=96\degg3$, $b=59\degg7$), 
we selected a subset of the best seeing frames, totaling 2702~s in the g-band and 4520~s in the i-band.

The MegaCam data were pre-processed by CFHT staff using the ``Elixir''
pipeline; which accomplishes the usual bias, flat and fringe corrections,
and also determines the photometric zero-point of the observations. These
images were then processed by the Cambridge Astronomical Survey Unit 
photometry pipeline in an identical
manner to that described above for the INT data. Using the multiple overlaps between deep and
shallow fields we correct the photometric solution provided by the
``Elixir'' algorithm (by up to $\sim 0.5$~mag), finding a global solution over all 92 deep fields that
has an RMS scatter of $0.02$~mags. 

Using observations of the Draco dwarf spheroidal galaxy for which we had
both INT-WFC and CFHT-MegaCam data in the (V,i) and (g,r,i) bandpasses, respectively,
we determined colour transformations to put the INT (Vega-calibrated)
photometry onto the MegaCam AB photometric system. The advantage of using
the Draco field is that the region has also been covered by the Sloan
Digital Sky Survey (SDSS), providing an external check to the
photometry. Note that the MegaCam (g,~i) bands are not identical to the SDSS
(g$^\prime$,~i$^\prime$), though the conversions between these two systems have
been determined by the CFHT staff. We refer the interested reader to
\citet{segall06} for further details. The conversion between INT (V,i) and
MegaCam (g,i) were found to be:
\begin{displaymath}
i_{MC} = i_{INT} - 0.105 \, ,
\end{displaymath}
\begin{displaymath}
g_{MC} = 
\left\{
\begin{array}{rl}
0.030 + 1.400 \times (V - i)_{INT} + i_{MC}  & \\
                                                                     \textrm{for} \, \, \, \, (V - i)_{INT} & < 1.3 \, , \\
                                                                     & \\
0.491 + 1.046 \times (V - i)_{INT} + i_{MC}  & \\
                                                                     \textrm{for} \, \, \, \,  (V - i)_{INT} & > 1.3 \, .
\end{array} \right.
\end{displaymath}

In order to enable the construction of maps over the combined area of the
INT and CFHT surveys, we converted the INT photometry to (g,i) using these
relations. The conversion appears to be adequately accurate, judging from
the photometry of bright stars (with magnitudes in the range $18 < g < 20$
and $18 < i < 20$) in the large overlap region between the two surveys: the
RMS scatter around zero offset was found to be $< 0.02$~mags in both bands.

Given the huge area of the survey it is necessary to be aware of variations
in the interstellar extinction which will affect the depth of the
photometry. In Fig.~3 the surveyed area is superimposed on a map of the
extinction derived from \citet{schlegel}; the maximum i-band extinction
over the halo region observed with MegaCam is $A_i = 0.27$~mags, with a mean of
$A_i=0.1$~mags. Thus the extinction is neither very high nor very variable,
though we nevertheless correct for it using the \citet{schlegel} maps. In
all the discussion below, $g_0$ and $i_0$ will refer to extinction-corrected
magnitudes.

\begin{figure}
\begin{center}
\includegraphics[angle=0, width=\hsize]{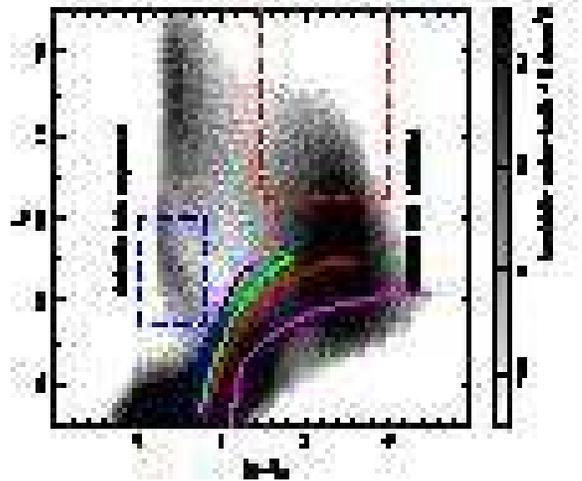}
\end{center}
\caption{The combined CMD of the MegaCam survey fields of M31, 
except fields T5 and T6 which are excluded because they are dominated by stars 
from M33 (including young stars in the disk), and fields 6, H11, and H13 which are close
to the M31 disk. 
The fiducial RGBs correspond to, from left to right, NGC~6397, NGC~1851, 47~Tuc, NGC~6553, 
which have metallicity of ${\rm [Fe/H] = -1.91}$, $-1.29$, $-0.71$, and $-0.2$, respectively.
The sequences have been shifted to a distance modulus of ${\rm (m-M)_0=24.47}$.
The dashed-line rectangles show the regions selected to probe the
foreground Galactic halo (blue) and Galactic disk (red).}
\end{figure}

\section{Color-magnitude distribution of sources}

As well as encompassing a large fraction of the halo of M31, the survey
also intersects a substantial volume of the foreground Milky Way. This is
clearly seen in Fig.~4, where
we show the combined color-magnitude distribution of all stars 
in the deep MegaCam
fields of the main survey, except for fields T5 and T6, close to M33, and fields
6, H11 and H13 close to M31.
Prominent at ${\rm (g-i)_0 > 1.5}$ and ${\rm i_0 < 23}$ is the sequence of
Galactic disk dwarfs; the vertical sequence is the result of low-mass
stars accumulating in a narrow color range, yet being seen over a 
large range in distance along the line of sight.
In addition, on the blue side of this diagram,
at ${\rm (g-i)_0 < 0.8}$ and ${\rm i_0 < 23}$, resides the Galactic halo sequence.
Usually, this is seen as a smooth vertical structure, due to stars at or 
close to the main-sequence turnoff at increasing distance through the Galactic halo.
Curiously, however in these fields towards M31 the sequence bifurcates --- indicating that
the Galactic ``halo'' is not spatially smooth along this line of sight. This issue is explored in 
detail in a companion article \citep{martin07a}.

The stellar populations of immediate interest to this study are revealed by the
red giant branch (RGB) stars that span the globular cluster fiducial sequences
that have been overlaid on the CMD. The bluemost and redmost sequence correspond
to clusters of metallicity ${\rm [Fe/H] = -1.91}$ and ${\rm [Fe/H] = -0.2}$, respectively,
so the survey is sensitive to stars of a wide range of abundance.
At the limiting magnitude of ${\rm i_0 \sim 24.5}$, the survey can in principle detect 
horizontal branch stars (see \citealt{martin06}), though of course the contamination
at these magnitudes, mostly from unresolved background galaxies and noise artifacts, is very large.  Nevertheless down to ${\rm i_0 \sim 24.0}$ the photometric quality
remains excellent, as we show in Fig.~5, with $\delta i < 0.1$~mag.

\begin{figure}
\begin{center}
\includegraphics[angle=-90, width=\hsize]{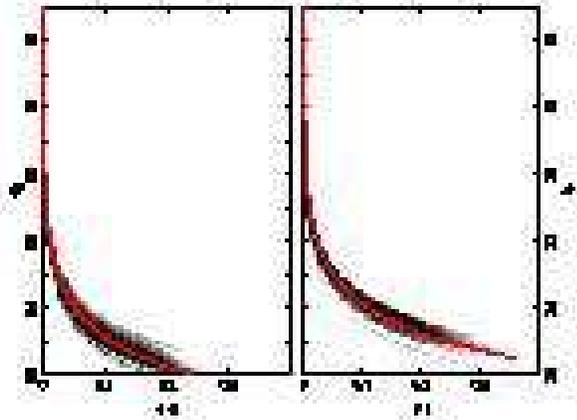}
\end{center}
\caption{The left and right panels show the distributions of photometric uncertainty
in $g_0$ and $i_0$, respectively, together with simple exponential 
fits (red lines). Some fields have slightly better photometry
than others, giving rise to the inhomogenous aspect at faint magnitudes.}
\end{figure}

\begin{figure}
\begin{center}
\includegraphics[angle=0, width=\hsize]{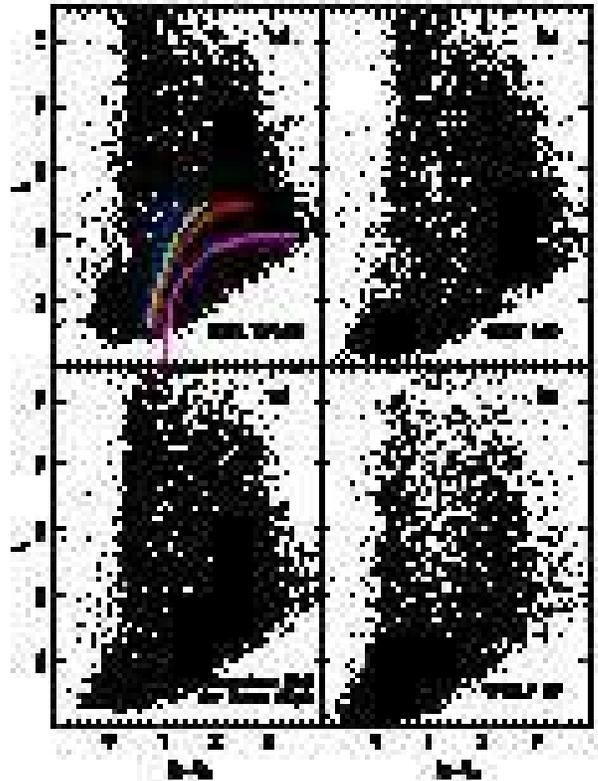}
\end{center}
\caption{The upper panels show sample CMDs of point-sources in the MegaCam survey. 
The panel `a' is for field 46, in a dense region within the giant stream, 
while panel `b' is for field 106, in the outer halo. The lower panels
correspond to the comparison fields: `c' lies near the Draco dSph, while `d'
is contructed from the CFHTLS field D3. As in Fig.~4, the lines in panel `a'
are the RGB ridge-lines of globular clusters of metallicity (from left to right) 
${\rm [Fe/H] = -1.91}$, $-1.29$, $-0.71$, and $-0.2$. The dense 
grouping of objects with ${\rm -0.5 < (g-i)_0 < 1.5}$ are mostly
due to misclassified compact galaxies.}
\end{figure}

\begin{figure}
\begin{center}
\includegraphics[angle=0, width=\hsize]{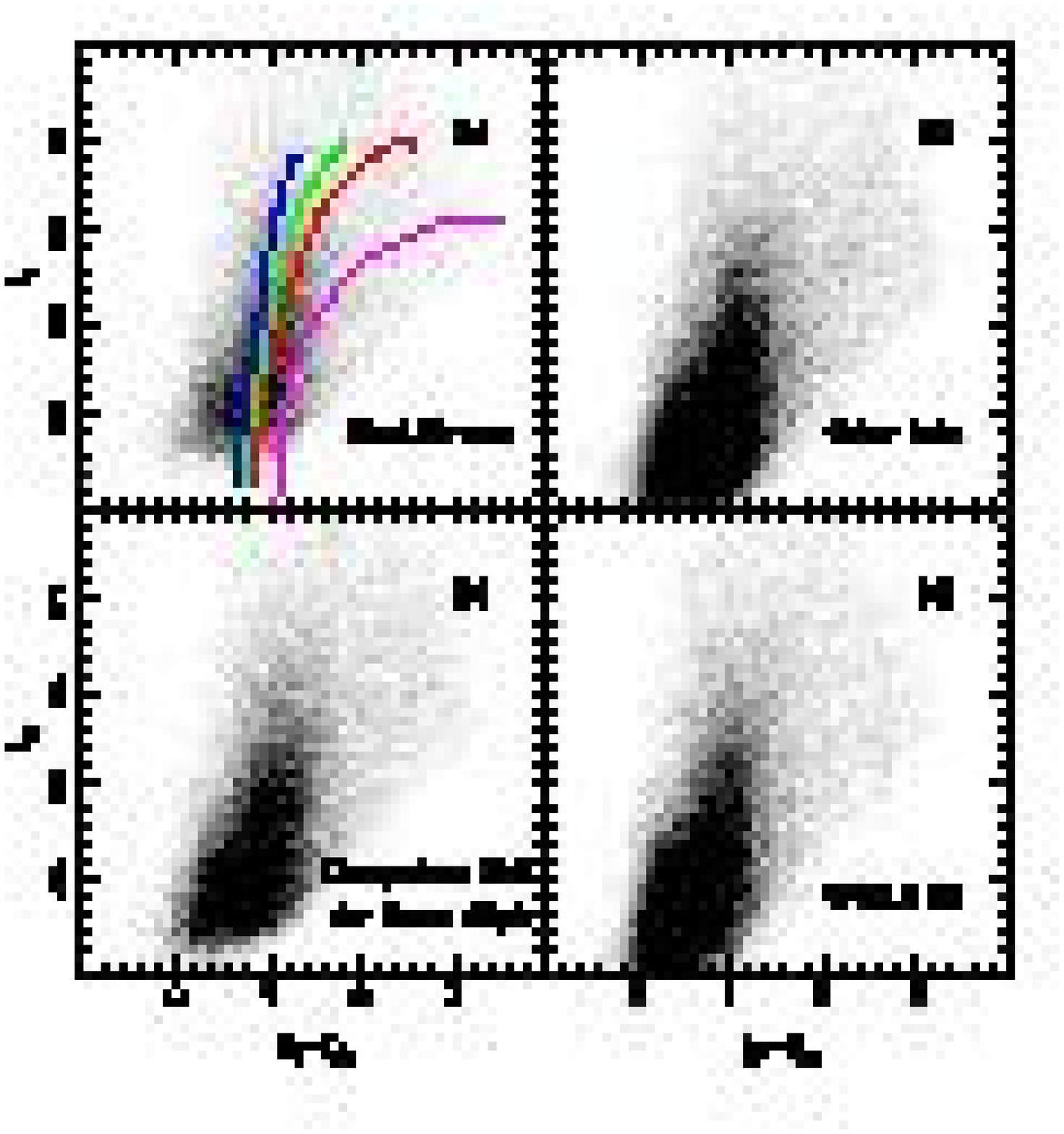}
\end{center}
\caption{As Fig.~6, but for sources classified as galaxies by the image analysis algorithm.}
\end{figure}

There are substantial variations of stellar populations between fields, as we demonstrate
in Fig.~6. Here, panel `a' displays the CMD of field 46, which lies in a dense
area of the so-called ``Giant Stream'' \citep{ibata01b}, and clearly contains a numerous population
of RGB sources with a wide spread of metallicity. Panel `b' shows the photometry of
field 106 in the far outer halo; no obvious RGB is discernible visually in this diagram,
though as we shall see later in \S7, the combination of this with several other outer fields does
allow a detection of the stellar halo of M31. 
For comparison, we also display the CMDs of the reference fields
near the Draco dSph (panel `c') and the CFHTLS field D3 (panel `d'). The photometric depth
of the survey clearly varies slightly from field to field (note that the images
from which the CMDs in panels `a' and `b' were constructed had identical exposure times).
The data taken in the 2005 and 2006 runs (of which panel `b' is an example) were very homogenous
in depth, whereas the earlier 2003 and 2004 runs were more patchy. It is likely that the improvement
in the 2005 and 2006 seasons was a result of the correction of the detector plane tilt
\footnote{See {\tt http://www.cfht.hawaii.edu/News/Projects/MPIQ/}}, allowing a
uniform focus to be achieved over the $0\degg96\times0\degg94$ field of view.
(For comparison to Fig.~6, in Fig.~7 we show the color-magitude distribution
of sources classified as galaxies).

Though the globular cluster RGB ridge-lines shown in Figs.~4 and 6 are
useful to show the behavior of known stellar populations, the set of 4 templates
is too sparse to allow accurate comparisons to be made with the distant
M31 population. Instead we chose to adopt the Padova isochrones 
\citep{girardi}, which conveniently have been calculated in the Sloan passbands.
Figure~8 shows the isochrones we used, converted into the MegaCam photometric system,
which were chosen for a population
age of $10\Gyr$. For each star in the survey, a photometric metallicity was calculated by interpolating
between the RGB curves. The assumption that the stellar populations
have an age of $10\Gyr$ over the entirety of the survey is clearly incorrect
\citep{brown06b}, but this is probably a reasonable estimate for the majority
of the stars at large radius.

\begin{figure}
\begin{center}
\includegraphics[angle=-90, width=\hsize]{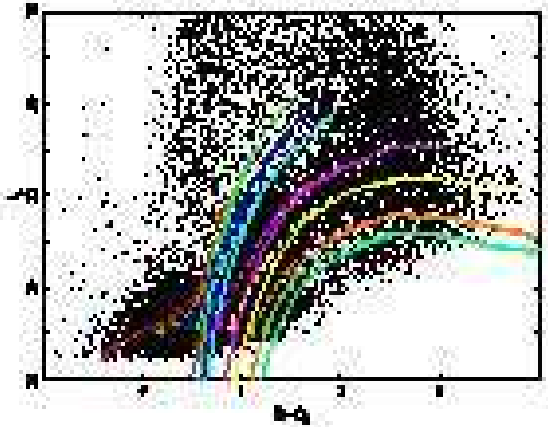}
\end{center}
\caption{The Padova isochrones superimposed on the CMD of field 47. The isochrone models
are all  for $10\Gyr$, and ${\rm [Fe/H]}$ metallicity (from left to right) of $-3$ (actually $Z=0$), $-2.3$, $-1.7$, $-1.3$, $-0.7$, $-0.4$, $0.0$ and $+0.2$. The continuous line part of each of these curves corresponds to the RGB, 
while the horizontal branch and asymptotic giant branch are indicated with dashed lines.}
\end{figure}

As we have shown in Fig.~3, the region surveyed with MegaCam includes
several known sources. For comparison to the populations we will
encounter below, we display their CMD structure in Fig.~9.

\begin{figure}
\begin{center}
\includegraphics[angle=-90, width=\hsize]{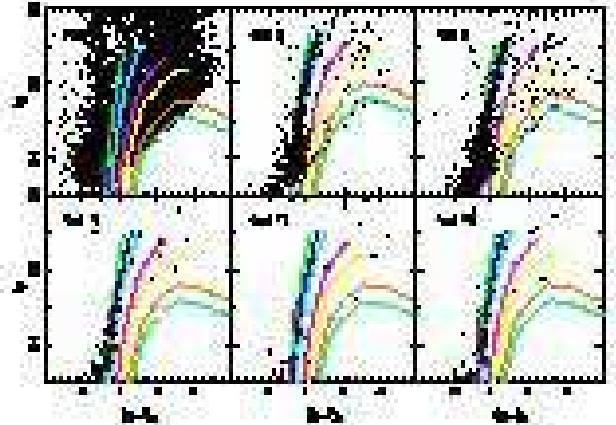}
\end{center}
\caption{CMDs of known satellite galaxies in the MegaCam survey region.
The Padova isochrones from Fig.~8 are reproduced here.
For M33 we show the sources within an annulus between $1\deg$ and $2\deg$
while for And~II, And~III and the remaining dwarfs, we show the sources
within a circular region of $12\arcmin$, $6\arcmin$, and $12\arcmin$, respectively.
For the purposes of overlaying the isochrones, we adopt the following
distance moduli. M33: $24.54 \pm 0.06$; And~II: $24.07 \pm 0.06$ (both from \citealt{mcconnachie04a};
And~III: $24.37 \pm 0.07$ \citep{mcconnachie05}; while
for And XI, XII and XIII \citep{martin06} we assume the distance modulus of M31:
$24.47 \pm 0.07$ \citep{mcconnachie05}.}
\end{figure}

\section{Spatial distribution of sources}

Although the MegaCam camera covers a large area, there are large inter-CCD
gaps in the mosaic, that were not filled by our chosen dithering pattern
with 5 sub-exposures. These gaps are partially filled by the short exposures,
but of course reach to a much shallower limiting depth. These inter-CCD gaps
are seen in Fig.~10, which shows the stellar density in one of the MegaCam
fields.  Another problem that is not limited to the MegaCam data are the
``halos'' of bright stars that effectively render useless certain regions of
the detector mosaic. The effect of these halos is also illustrated in
Fig.~10. Both the gaps and bright star holes could easily be accounted for in
the analysis of the surface density, by simply correcting
for the missing area. However, we found this approach to be somewhat unsatisfactory
when making maps of spatial resolution smaller than the area of the bright star ``halos''.
Instead we chose to replace the affected areas with nearby counts: the inter-CCD gaps
were filled with the detections of the CCD immediately to the South, while the bright star
halos were filled with detections either to the East or West of the hole (depending on the 
location of the field edge or other nearby bright stars). Figure~10 shows an example of the
procedure adopted. A further problem was that in several fields observed in 2003
the data for CCD~4 of the MegaCam mosaic was absent due to a CCD controller
malfunction. For these fields, which comprise fields 
48, 63, 77, 92, H11, H13, T2, T3, T4 and T5, we copied over the sources from CCD~3,
adjacent on the mosaic.
All the sources that were added artificially in these various ways were flagged.

\begin{figure}
\begin{center}
\includegraphics[angle=-90, width=\hsize]{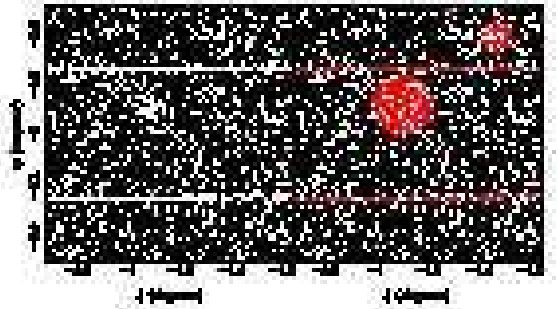}
\end{center}
\caption{As an example of our correction technique for the effect of bright stars, we show
in the left-hand panel the distribution of stellar sources in field 70, a field
containing several unusually bright stars. 
The two horizontal gaps
are due to a physical gap between the first two and the last two
rows of detectors on the mosaic camera. The lower source density at $\xi \sim -1\degg1$,
$\eta \sim -5\degg9$ is due to a bright star ``halo''. In the right-hand panel, we show
the corrected counts in this region, where the stars in the affected region
have been deleted, and replaced with 
artificial sources (red points) that were copied from adjacent areas of the sky.}
\end{figure}

The final catalog contains a total of 19 million sources.
However, many of these sources are foreground and background
contaminants, so we must assess their numbers and distribution
before being able to analyze the distribution of genuine M31 stars.
In Fig.~11 we show the spatial
distribution of Galactic disk dwarf stars with $1.5 < (g-i)_0 < 3.0$ and $15.0 < i_0 < 19.5$; 
from an inspection of Fig.~4 it can
be seen that these stars are located at brighter magnitudes than the tip of
the M31 red giant branch (RGB) and should therefore be an almost pure Galactic
sample. Figure~11 shows that this is not entirely correct, as a strong enhancement
of sources is seen in the inner regions of M31 and M33, 
due to the presence of blue loop stars and asymptotic giant branch (AGB) stars
in the disks of those galaxies.
Ignoring these disk regions, we detect a smooth gradient 
towards the Galactic plane in the North, with no obvious structures.

\begin{figure}
\begin{center}
\includegraphics[angle=-90, width=\hsize]{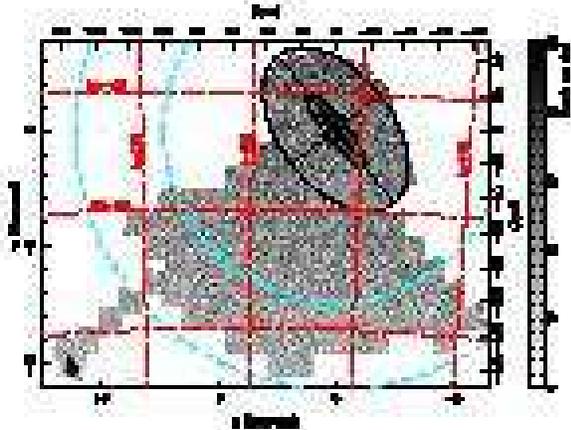}
\end{center}
\caption{The distribution of 
stars within the color-magnitude selection box 
$1.5 < (g-i)_0 < 3.0$ and $15.0 < i_0 < 19.5$, which outside of the
inner regions of M31 and M33, which contain blue loop and AGB stars, gives a clean sample
of Milky Way disk dwarf stars. The map is a linear representation of the 
star counts, with pixels of size $0\degg1\times0\degg1$.}
\end{figure}

In addition to the Galactic disk dwarfs, there is some contamination from distant 
bright main-squence halo stars, as we showed in Fig.~4. We select a representative
sample of this population by choosing stars within the box 
$0.0 < (g-i)_0 < 0.8$ and $20.0 < i_0 < 22.5$. The resulting spatial distribution is
presented in Fig.~12. The contamination to this sample from the disks of M31 and M33
is not at all surprising, as young blue supergiant stars in these galaxies will fall into this
color-magnitude selection box. However, excluding a $2\deg$ and $1\deg$ circle
around M31 and M33, respectively, shows the remaining Galactic population
to have a very uniform density over the survey region.

\begin{figure}
\begin{center}
\includegraphics[angle=-90, width=\hsize]{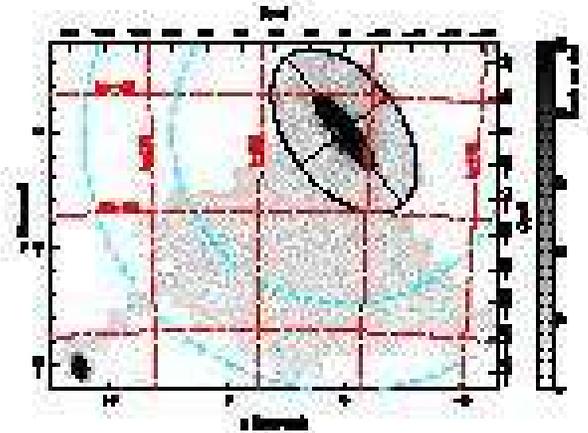}
\end{center}
\caption{As Fig.~11, but showing the distribution of Milky Way halo stars over the survey region,
selected within the color-magnitude box $0.0 < (g-i)_0 < 0.8$ and $20.0 < i < 22.5$. 
The concentration of sources inside the disks of M31 and M33 is due to young
blue main sequence stars in those galaxies.}
\end{figure}

A further source of contaminants are background galaxies. Most of these are
readily identifiable from their image parameters, though there will be
some distant compact galaxies that are unresolved with the
typical depth and seeing achieved in this survey.  The map of the sources classified as galaxies by the
algorithm is displayed in Fig.~13. Apart from the usual filamentary
signature of large-scale structure there is no apparent correlation with
either the Milky Way, Andromeda or M33, beyond the disks of these latter two
galaxies (where some sources are classified as being extended due to image crowding). 
The colour-magnitude distribution of
these contaminants is displayed in Fig.~7 for four selected fields. 
These resolved galaxies are approximately as numerous as the point-sources
in the dense Giant Stream fields, but become up to 6 times more numerous than point-sources
in the outer halo fields.
Clearly a small error in image classification towards fainter magnitudes could have a 
significant repercussion in the measured density of point-sources. 
We return to this issue below.

\begin{figure}
\begin{center}
\includegraphics[angle=-90, width=\hsize]{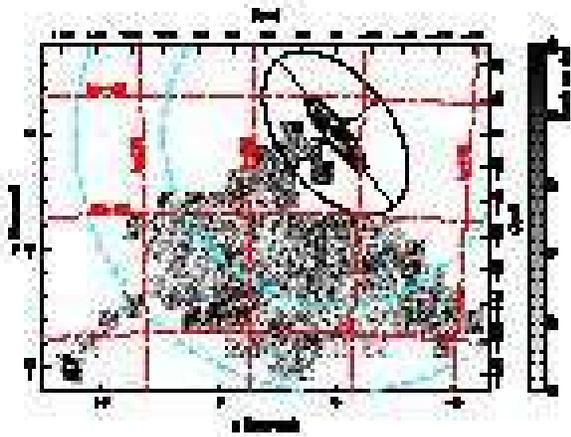}
\end{center}
\caption{As Fig.~11, but showing the distribution of objects classified 
as extended sources over the survey region. Due to the high
source density in the disks of M31 and M33, some point sources
are blended and are classified as galaxies by the photometry software.
A pixel size of $0\degg05\times0\degg05$
has been used.}
\end{figure}

\begin{figure}
\begin{center}
\includegraphics[angle=-90, width=\hsize]{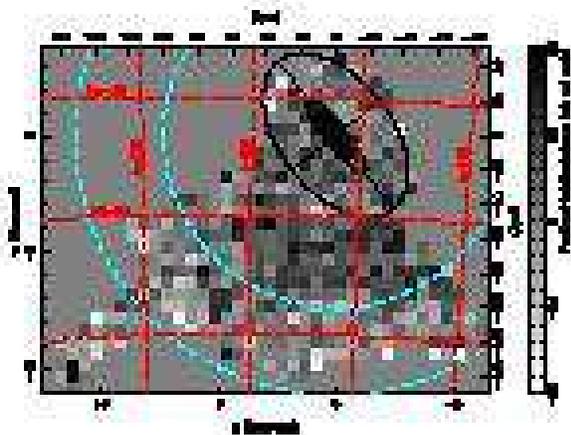}
\end{center}
\caption{The map of the fractional residuals 
between the Galactic disk selection previously presented in
Fig.~11, and the Besan{\c c}on model predictions 
(calculated as ${\rm (Data-Model)/Model}$ for each 
$0\degg5 \times 0\degg5$ bin). Ignoring a $2\deg$ circle around M31 
and a $1\deg$ circle around M33, the average difference is less than 2\%.}
\end{figure}

\subsection{Foreground subtraction}

We had envisaged using the MegaCam comparison fields presented in Fig.~6
to subtract off the background counts, however since the Galactic contamination
varies substantially from these fields to our M31 fields of interest, and even 
varies significantly over the main area of this vast survey, we decided to investigate
whether Galactic models could be used instead to predict the contamination more
reliably. To this 
end we tessellated the survey area with $0\degg5 \times 0\degg5$ bins, and 
generated simulated catalogues using the Besan{\c c}on Galactic 
populations model. All stellar populations in the model with i-band 
magnitudes between $15 < i_0 < 26$ were accepted. To reduce shot noise
in the randomly generated catalogs, at each spatial bin we simulated 
a 10 times larger solid angle, and later corrected the density maps for this factor.
Finally, the artificial photometry was convolved with the observed magnitude-dependent 
uncertainty function (from Fig.~5).

\begin{figure}
\begin{center}
\includegraphics[angle=0, width=\hsize]{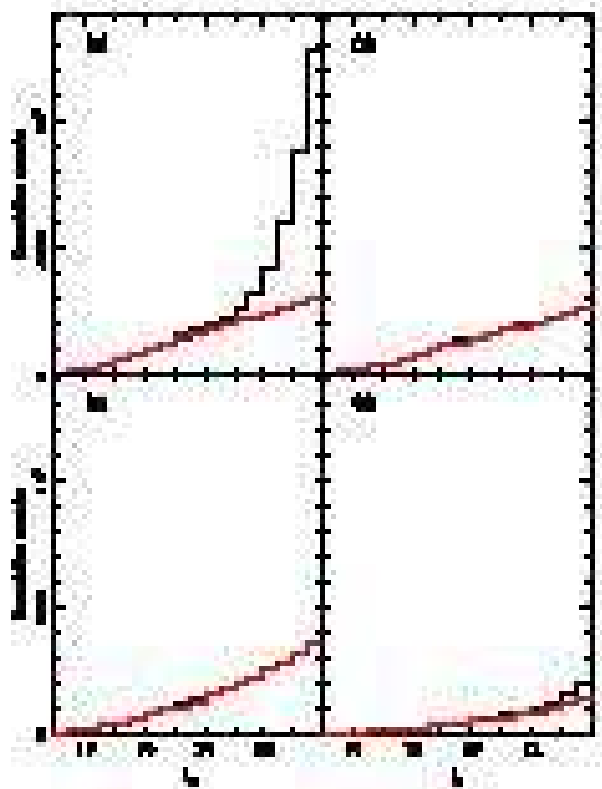}
\end{center}
\caption{The luminosity function of point sources in the color range $0.8 < (g-i)_0 < 1.8$
for the sample fields shown previously in Figs.~6 and 7: field 46 (a), field 106 (b), the Draco dSph 
comparison field (c) and the CFHTLS field D3 (d). The observed luminosity functions are shown
in black, while the red lines show the Besan{\c c}on model predictions. In panel `a' the stellar
populations of the Giant Stream cause the large increase in counts beyond $i_0=21$.
The correspondence between observations and model in panels `b' and `c' is excellent,
though there is a significant departure in panel `d'. A limiting g-band magnitude
of $g_0 < 25.5$ was imposed to data and models alike.}
\end{figure}

We were impressed to discover the accuracy to which the Besan{\c c}on model
predicts the starcounts towards our fields. For the Galactic
disk sample selected with $1.5 < (g-i)_0 < 3.0$ and $15.0 < i_0 < 19.5$ (red
dashed-line box in Fig.~4), whose observed spatial distribution was presented
previously in Fig.~11, the Besan{\c c}on model correctly predicts the observed counts over the 
survey area to better than 2\%. The fractional residuals between the observations and the model are
shown in Fig.~14.

Evidently the Besan{\c c}on model has the correct ingredients to reproduce 
very accurately the Galactic disk
starcounts towards these fields around M31. However, we need to investigate the
model further before we can use it with confidence. The color-magnitude
region that is of particular interest to us, is the region where the RGB of M31 has
its greatest contrast over the contaminants. We will return to this in more quantitative
detail later, when we discuss the matched filter method, yet a visual inspection of
Fig.~4 shows that the color interval will be approximately in the range
$0.8 < (g-i)_0 < 1.8$, where we avoid the bulk of the Galactic disk contamination,
and also the faint blue contaminants, which are most likely unresolved background galaxies. 
In Fig.~15 we display the observed luminosity function in this color interval (in black), as well
as the corresponding Besan{\c c}on model predictions (in red) for the two representative fields
and the two reference fields that we presented previously
in Figs.~6 and 7. The correspondence is excellent from $i_0 = 15$ down to $i_0 = 20.0$, with
Kolmogorov-Smirnof (KS) test probability that the observations are drawn from the model
of greater than 10\% for all four fields. In panel `a' the observations depart strongly 
from the model for $i > 21$, this is however completely expected, as the field contains the
RGB of Andromeda at these magnitudes. Panel `b' is for field 106 in the outer halo, 
and panel `c' is the Draco dSph comparison field; in both cases
the model predictions are extremely close to what is observed: the KS test over the range
$15 < i_0 < 24$ gives 27\% and 9\% probability, respectively, that the 
observed and modeled distributions are identical, and the total counts 
agree to within better than $2\sigma$. However, for the CFHTLS field D3, shown in panel `d', 
the Besan{\c c}on model predictions over the full range 
$15 < i_0 < 24$, do not
accurately match the observations (KS-test probability $<0.01$\%). This failure towards the direction
($\ell=96\degg3$, $b=59\degg7$), 
is likely due to a slightly inaccurate model of the Galactic halo component, or due to local deviations from a globally correct halo model.
Despite this shortcoming, we consider these comparisons to have been very 
encouraging. The Besan{\c c}on model
predicts reasonably well the details of the star counts towards our two comparison fields, and
it predicts perfectly well the star counts in the outer halo field (panel `b'). 
Very similar results were found upon widening the color range to $0.5 < (g-i)_0 < 1.8$, 
to include the bluest RGB stars of interest.
Given the variations in the
luminosity function that are clearly visible in Fig.~15, it is evidently better to use the model
to subtract off the expected contamination rather than use a comparison field located
at a different Galactic latitude and longitude. This is true even for relatively nearby fields: 
the difference in the predicted luminosity function of foreground stars in 
panels `a' and `b' is substantial.

\begin{figure}
\begin{center}
\includegraphics[angle=-90, width=\hsize]{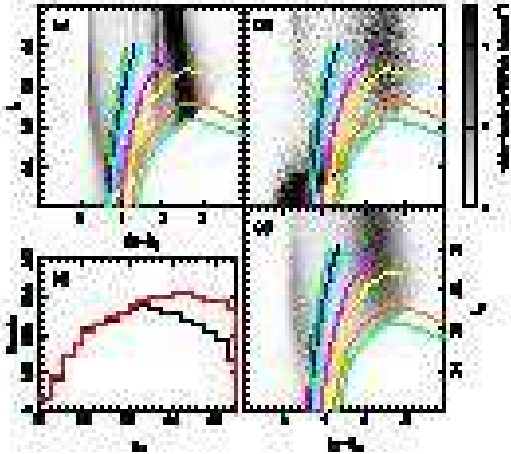}
\end{center}
\caption{The color-magnitude distribution of sources from the Besan{\c c}on model
for the MegaCam comparison fields is shown in panel (a), where the model
predictions have been smoothed with the observational errors in
Fig.~5. The corresponding observed distribution is given in panel (b). 
Clearly, in reality the stellar populations have a much wider color spread
than the model predicts. To alleviate this problem we have introduced
an additional smoothing to the model, as detailed in the text. In panel `c'
the ratio of the luminosity function in the color range 
$2.0 < (g-i)_0 < 3.0$ of the model (red) and the data (black) is used to 
compute an empirical completeness correction, which applied to the 
color-magnitude data, gives the distribution shown in panel `d'.
(A g-band limit of $g_0 = 25.5$ has been imposed throughout).}
\end{figure}

The excellent agreement between the observations and the model predictions in
panels `b' and `c' of Fig.~15 is somewhat surprising given the fact that 
we did not apply any incompleteness corrections to the model, and
have not corrected for contaminating background unresolved galaxies.
We chose not to perform artificial star completeness tests for this survey 
as it would have been a prohibitively expensive undertaking, and refer
instead to a previously computed comparison between MegaCam and Hubble Space Telescope
photometry from the center of the Draco dSph. As we show in Fig.~2 of \citet{segall06},
the completeness of MegaCam down to $i=24$ from data of similar
exposure time is greater than 80\%. Note however,
that this completeness was calculated in a relatively crowded central field of the Draco dSph 
(not the Draco comparison field shown in panel `c' of Figs.~6, 7 and 15), and is therefore 
likely to be substantially worse than what we face in the almost empty fields
in the outer halo of M31.

Despite these successes of the Besan{\c c}on model, it unfortunately 
fails to predict the correct color-magnitude distribution. 
The reason for this is apparent from a visual inspection of panel `a' of
Fig.~16,
in which we present the predicted color-magnitude distribution over the MegaCam
fields 93, 105, 106, 115, 120 and 121, which are all located at the outer edge
of the survey near a projected radius of $150\kpc$ (In the
analysis below we shall refer to these fields as ``background'' fields).
Comparing this distribution to its observed 
counterpart in panel `b', we see that the model has features that are too sharp,
despite the convolution with the photometric uncertainties. This is likely due
to the model not containing a realistic spread of stellar populations types,
in particular the color-magnitude sequences are evidently not as varied 
in the model as in reality. 

To alleviate this problem we have introduced
an additional smoothing to the model. From a Gaussian fit to the 
color distribution of Galactic ``halo''
and Galactic disk populations in the magnitude range $20 < i_0 < 21$
(where the sequences are almost vertical in the CMD),
we measured the intrinsic FWHM of the observed distributions. By introducing a
color-dependent additional Gaussian spread to the model 
of $\sigma = 0.05 + 0.075 (g-i)_0$, we find a similar color spread in the halo
and disk populations to the observations.

In panel (c) we compare the luminosity function in the color range 
$2.0 < (g-i)_0 < 3.0$ in the resulting smoothed model (red) with
that of the data. We see an excellent match down to $g_0 = 23.25$, after 
which the model begins to diverge, due to the effects of incompleteness.
We use the ratio of these distributions beyond $g_0 = 23.25$
to correct the model for incompletness; the resulting final model
for the background region is displayed in panel `d'.

\begin{figure}
\begin{center}
\vbox{
\includegraphics[angle=-90, width=\hsize]{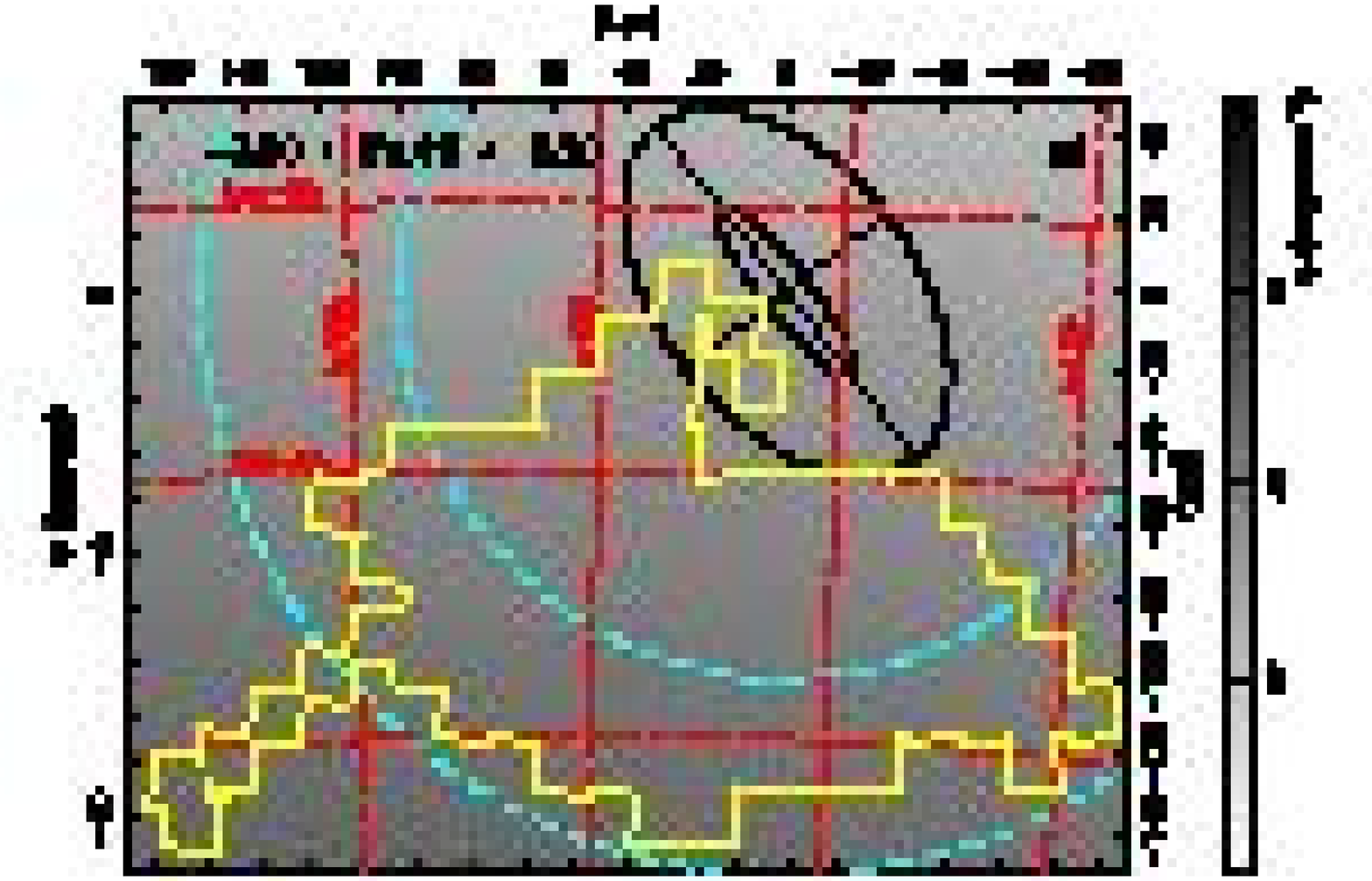}
\includegraphics[angle=-90, width=\hsize]{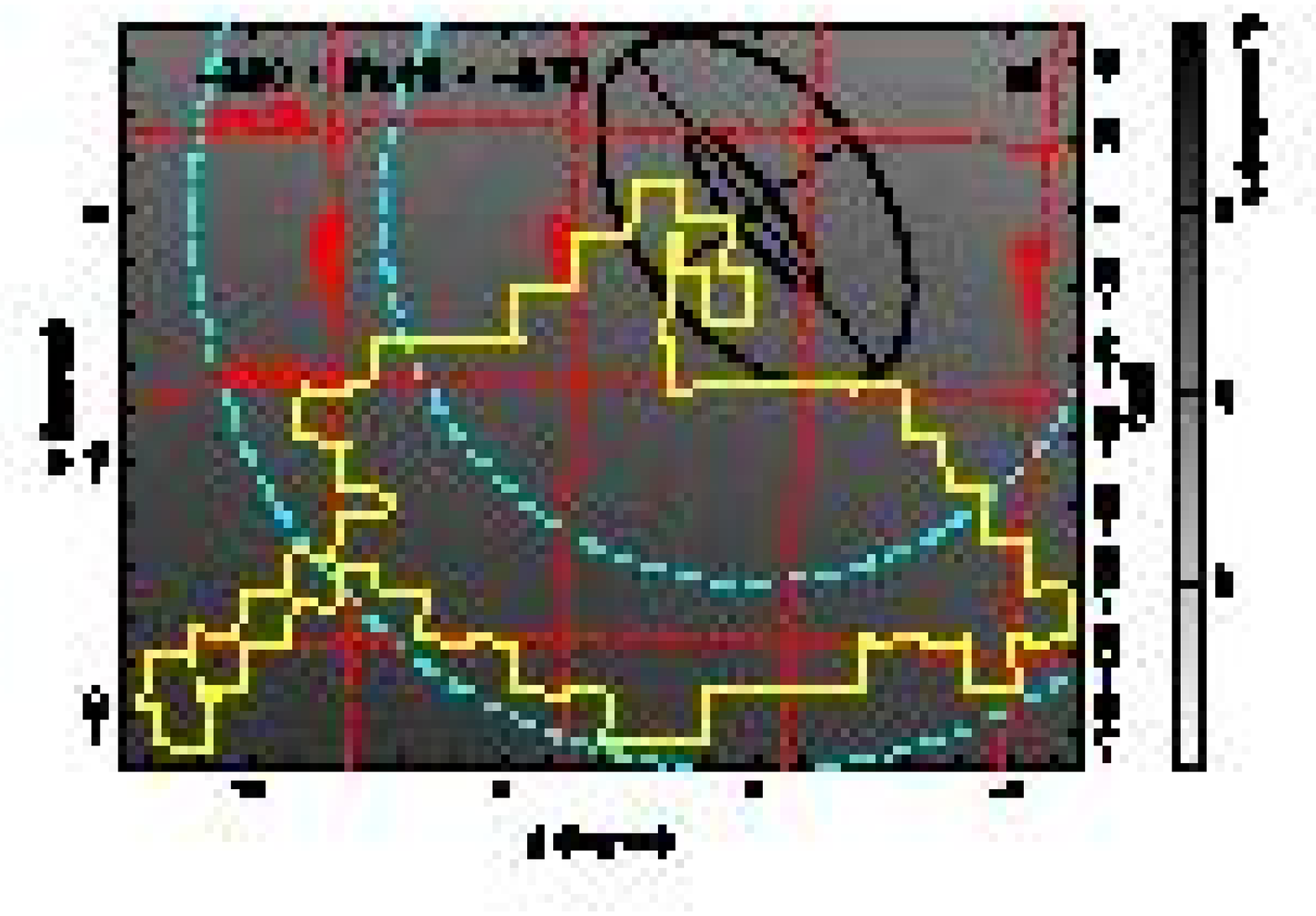}
}
\end{center}
\caption{The spatial distribution of the Besan{\c c}on model 
(calculated for each $0\degg5 \times 0\degg5$ bin) over the survey 
region for two different color-magnitude selections. Panel `a'
is for Galactic stars that have color and magnitude in the
region occupied by stars in M31 of metallicity in the range 
${\rm -3.0 < [Fe/H] < +0.2}$ according to the $10\Gyr$ Padova models.
Panel `b' is for the more restricted range ${\rm -3.0 < [Fe/H] < -0.7}$.}
\end{figure}

The excellent agreement of the Besan{\c c}on model 
with our observations to $g_0=23.25$, indicates that the 
number of background galaxies masquerading as point-sources 
cannot be a substantial fraction of the total counts
down to these photometric limits. Beyond this limit, 
some background galaxy contamination may offset 
the incompleteness, in which case it will be hidden in the empirical
completeness correction adopted for the background fields.

The Besan{\c c}on model, smoothed and corrected for incompleteness,
as discussed previously, can now be used to predict the expected 
foreground contamination, for stars of color and magnitude
that will masquerade as M31 halo stars.
In Fig.~17 we show two
such predictions over the area of the study. The top panel shows
the equivalent surface brightness of the star-count model for stars
with metallicities ${\rm -3 < [Fe/H] < +0.2}$ interpolated from the 
Padova models shifted to the distance of M31. The bottom panel
shows a similar map for ${\rm -3 < [Fe/H] < -0.7}$, which is 
substantially fainter than that of panel `a' because this metallicity 
interval excludes most red stars from the Galactic disk sequence 
(as can be seen in Fig.~16).

To construct Fig.~17 we have converted the predicted Galactic star-counts to an ``equivalent
surface brightness'' $\Sigma_V$ in the V-band, as if these contaminants were RGB stars
in M31. The motivation for converting the measured star-counts into surface brightness is 
of course to be able to compare our observations to previous studies and also to
theoretical predictions. However the procedure requires some further
explanation. Both for the model and for the survey data, we convert
the MegaCam g and i-band photometry into the V-band using the color
equation above. The resulting V-band luminosities are summed for 
the stars in a spatial and/or color-magnitude bin, but we must still
correct for the fact that we are only observing RGB stars which represent
only a fraction of the total luminosity. By comparing the RGB star-counts
of And~III down to a limiting magnitude of $i_0=23.5$ with the 
integrated magnitude of ${\rm m_V = 14.4\pm0.3}$ of this dwarf galaxy 
\citep{mcconnachie06a}, we measure an offset of $2.45$~mag.
This is consistent, and similar, to the value of $2.3$~mag estimated
in the same manner by \citet{martin06} for a limiting magnitude of $i_0=24$.
Furthermore, as we shall see below in \S8,
with this offset we obtain a good correspondence between the profile
of metal-poor stars and the 
V-band surface brightness profile derived from integrated light \citep{irwin05}.
Clearly the uncertainties in this simple correction are large: we are implicitly assuming
that the underlying population has the same luminosity function as And~III
for all metallicities. The equivalent surface brightness measurements we shall present below
must therefore be interpreted with caution, as they are likely to contain 
substantial systematic errors. However, the interested
reader who may wish to convert these
surface brightness profiles back to the reliable measure of luminosity-weighted star-counts 
(to a limiting magnitude of $i_0=23.5$) can do so by simply subtracting $2.45$~mag.

The predicted distributions such as those shown in Fig.~17
are the best means we have to subtract foreground contamination 
from the spatial maps. However, we found that we could improve upon
the foreground subtraction in color-magnitude (Hess) diagrams
by using the observed color-magnitude distribution in the 6 background fields 
(93, 105, 106, 115, 120 and 121) appropriately scaled according to the model
to account for the predicted density variations over the survey.
A different scaling correction is adopted for each metallicity 
interval; we show in Fig.~18 an example of the scaling factor applied to the 
stars with colors consistent with being M31 stars with metallicity in the
interval ${\rm -0.7 < [Fe/H] < -0.4}$, according to the Padova models.
The density of contaminants subtracted from the higher latitude fields
is more than a factor of two larger than from the lower latitude fields.

\begin{figure}
\begin{center}
\includegraphics[angle=-90, width=\hsize]{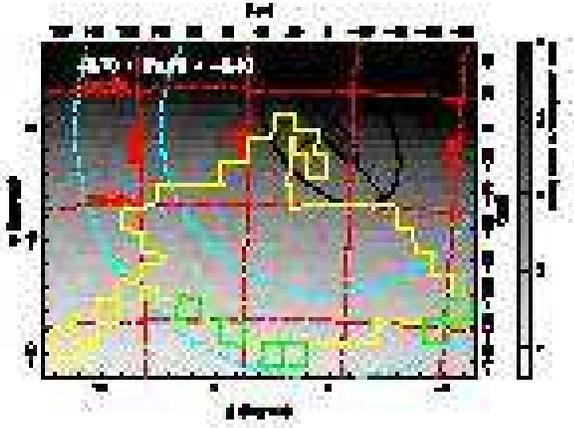}
\end{center}
\caption{An example of a map of the density scaling factor applied to the 
background fields (marked in green) to compensate for the expected variations in foreground
stellar populations over the survey. In this case, we have chosen stars
with colors between the Padova isochrones of metallicity ${\rm -0.7 < [Fe/H] < -0.4}$.}
\end{figure}

Panel `a' of Fig.~19 shows the color-magnitude distribution of the MegaCam fields shown previously
in Fig.~4, with the contamination removed statistically. The subtracted CMD displays
a clear RGB-like population, with a broad range of metallicity, 
although the detection of the more metal-rich populations is clearly hampered by the 
observational g-band limit. In order to investigate the luminosity function along this RGB,
we select stars with interpolated metallicities in the range ${\rm -2.3 < [Fe/H] < -0.7}$
(i.e., between the green and pink isochrones).
The result is shown on panel `b', together with a simple fit.
A linear fit in ${\rm \log(Counts)}$, is precisely what is expected for an RGB population
\citep{bergbusch01}. If this statistical foreground subtraction is reliable, over 
$10^5$ halo RGB stars belonging to M31 are detected over these MegaCam fields.

\begin{figure}
\begin{center}
\includegraphics[angle=-90, width=\hsize]{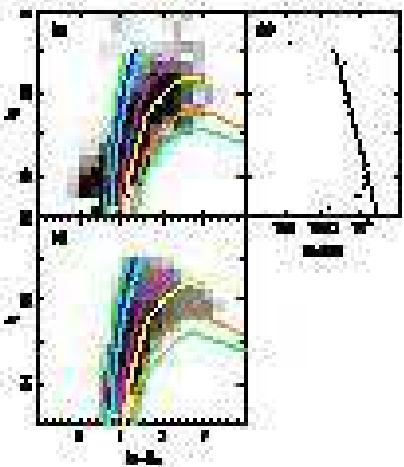}
\end{center}
\caption{Panel `a' shows the Hess diagram of the MegaCam fields previously 
shown in Fig.~4, with foreground
and background contamination subtracted by comparison to
six background fields as detailed in the text. The 
Padova isochrone models from Fig.~8 are reproduced to help guide the eye.
Panel `b': the luminosity function of stars with ${\rm -2.3 > [Fe/H] > -0.7}$.
Panel `c': the matched filter weight map, trimmed to the
color-magnitude region encompassing stars of metallicity ${\rm -3.0 < [Fe/H] < +0.2}$.
(Both gray-scale maps are shown on a linear scale, with the 
photometry limited to $g_0 < 25.5$).}
\end{figure}

\section{Stellar population maps}

Having shown that there is a relatively clean signal of the expected RGB of M31 in the
combined data, we now proceed to mapping out these stellar populations.
A very powerful technique for revealing a signal buried under heavy contamination is
the so-called ``Matched Filter'' method, which is an optimal search strategy 
(in a least-squares sense) if one has a precise idea of the properties of the
signal and the contamination. The properties could be, for instance,
the spatial properties of the
population of interest (a characteristic size or shape)
as well as those of the contamination. Alternatively (or in addition), one may use
the color-magnitude distribution, or whatever other physical properties of these
populations that have been measured.

\begin{figure}
\begin{center}
\includegraphics[angle=-90, width=\hsize]{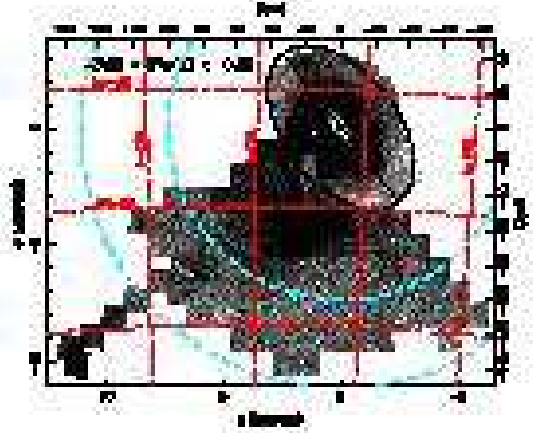}
\end{center}
\caption{Matched-filter map to $i_0=24.5$ ($i_0=22.8$ over the INT survey region). 
The artifacts of the MegaCam fields observed in the 2003 and 2004 seasons are clearly seen. 
A logarithmic scale is used for the representation.}
\end{figure}

\begin{figure}
\begin{center}
\includegraphics[angle=-90, width=\hsize]{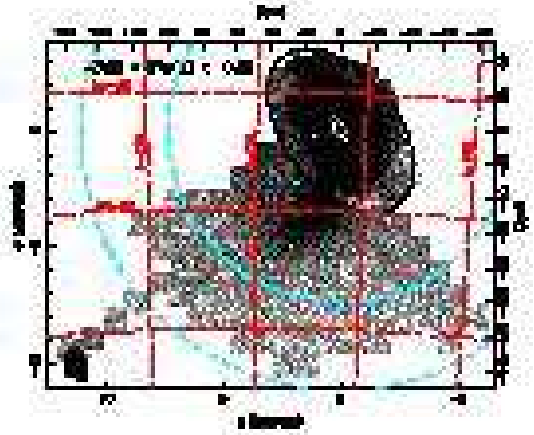}
\end{center}
\caption{As Fig.~20, but to the limiting depth of the INT survey ($i_0=22.8$ for S/N$\sim 10$).
The map is virtually free of obvious artifacts over the entire region observed with MegaCam.}
\end{figure}

To apply the matched filter method one simply weights each datum by the 
ratio of signal to contamination expected for that datum given its parameters.
The resulting ensemble of weighted data can then be analyzed in the usual
way. However, the advantage this effort has afforded us is that the
distribution of weighted data will optimally suppress the contamination,
revealing best whatever signal is present.
In the particular situation confronting us here, we know the color-magnitude distribution
of the signal of interest, as we have just presented in panel `a' of Fig.~19, and as discussed
above the MegaCam ``background'' fields (93, 105, 106, 115, 120 and 121) give us a
reasonable model for the color-magnitude behavior of the contamination
in the absence (or near absence) of that signal. The ratio of these two CMD
distributions gives the weight matrix, which we show in panel `c' of Fig.~19. Here we have trimmed 
the weight matrix down to the maximum possible physical interval (${\rm -3.0 < [Fe/H] < +0.2}$).
Note that, as expected, the greatest weight arises at faint magnitudes in the color range
$0.75 < (g-i)_0 < 1.5$, so of course stars with this photometric property 
will contribute most strongly in the following
matched filter maps.

\begin{figure*}
\begin{center}
\hbox{
\includegraphics[angle=-90, width=8.8cm]{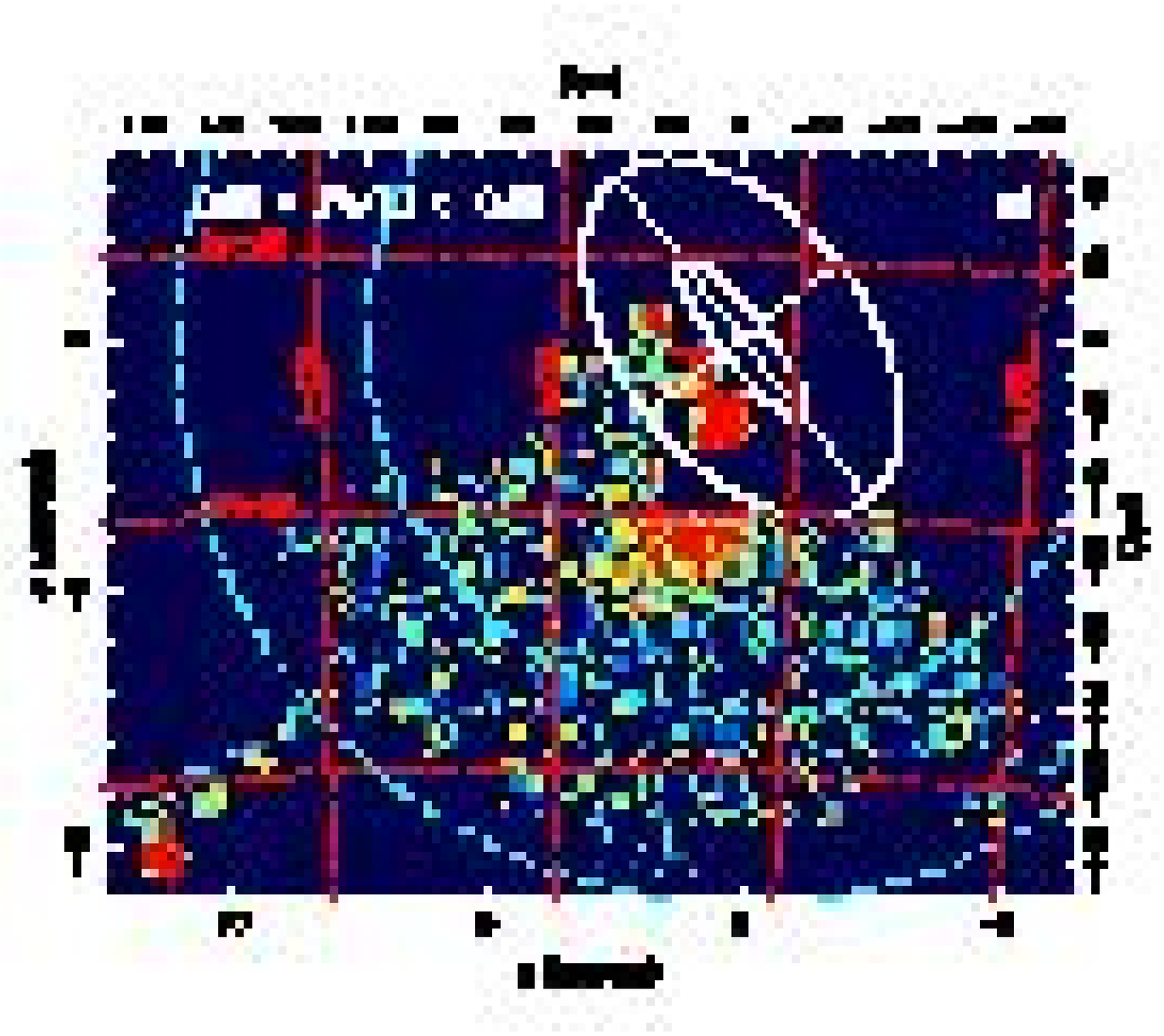}
\includegraphics[angle=-90, width=8.8cm]{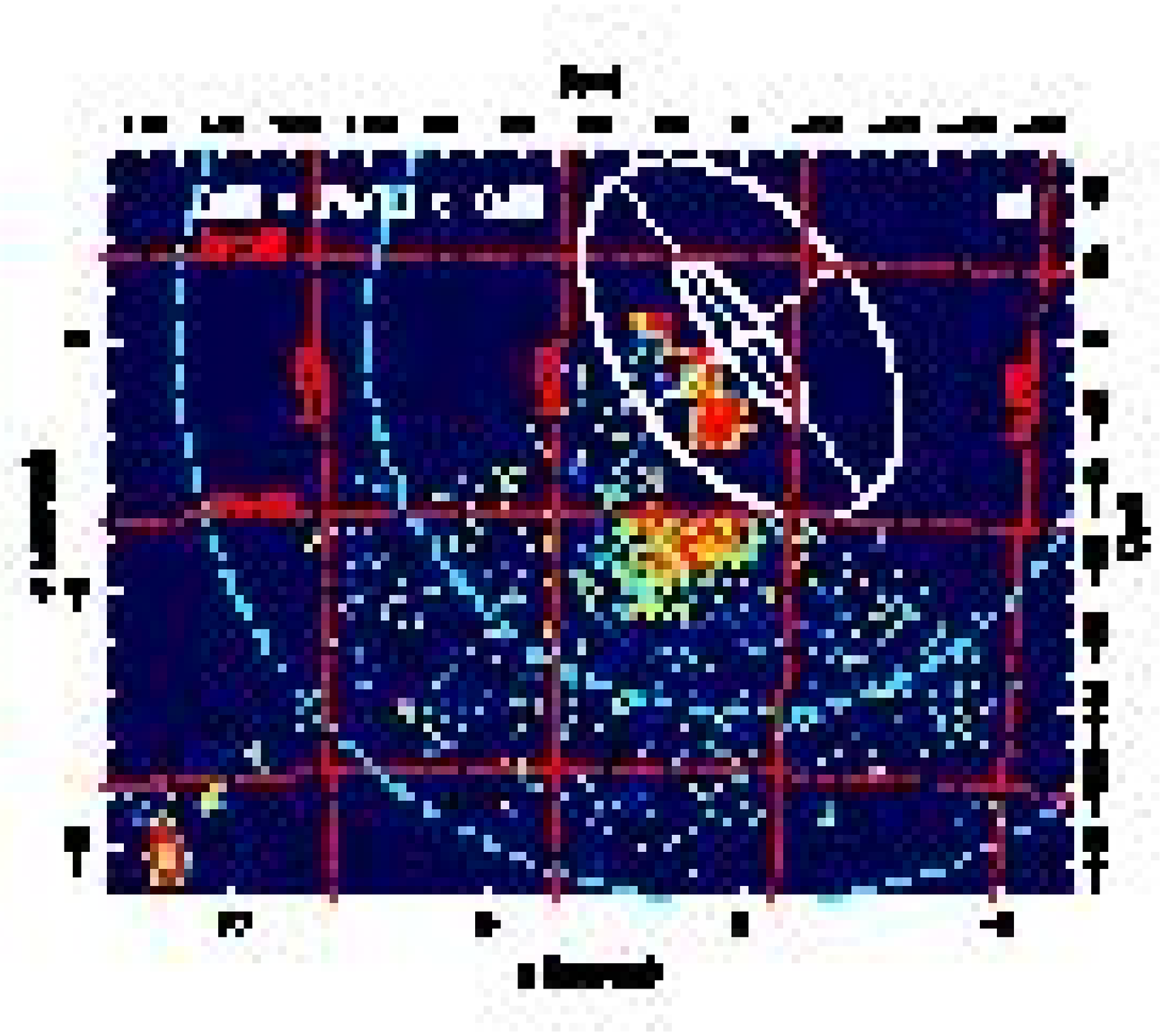}
}
\hbox{
\includegraphics[angle=-90, width=8.8cm]{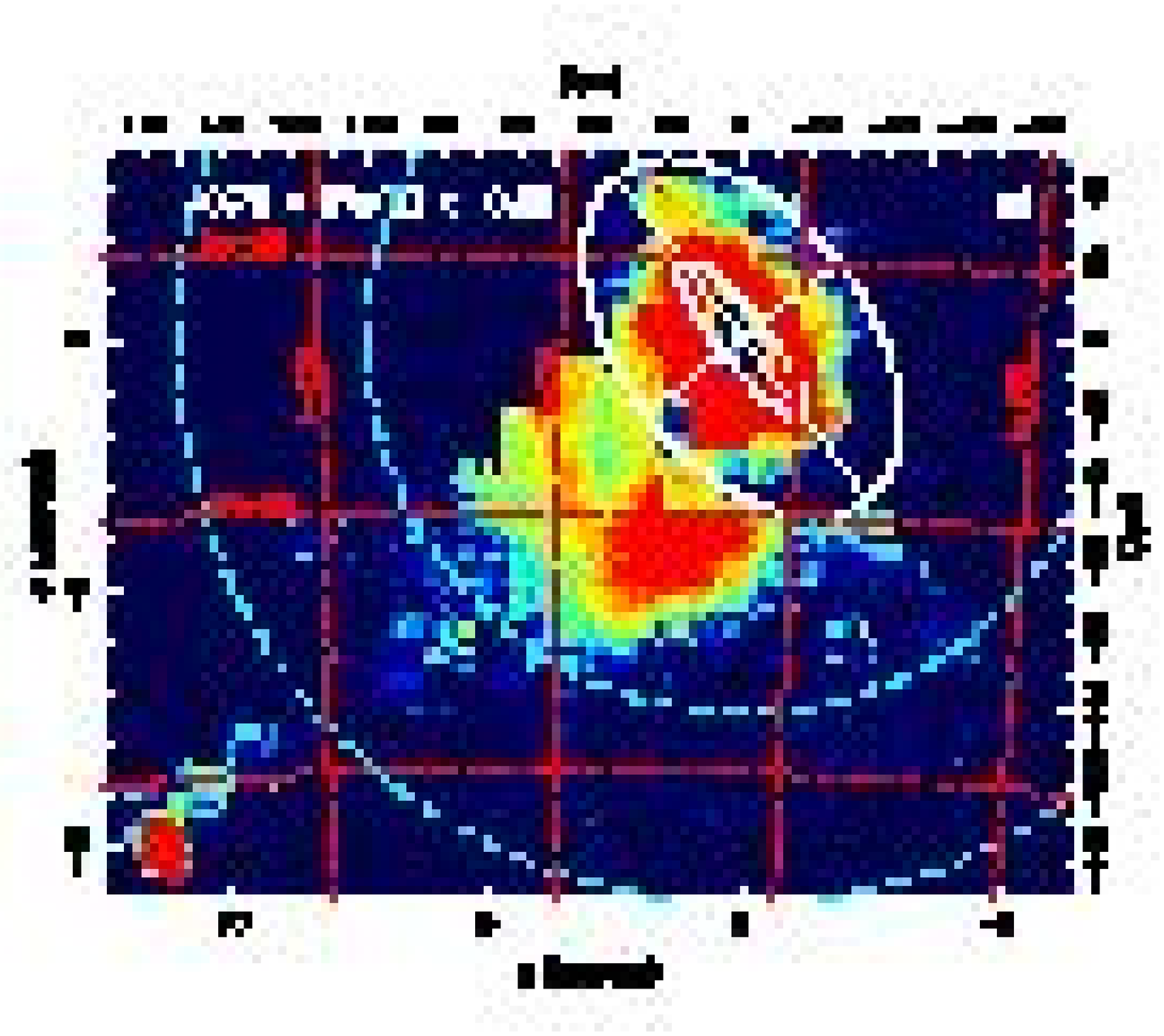}
\includegraphics[angle=-90, width=8.8cm]{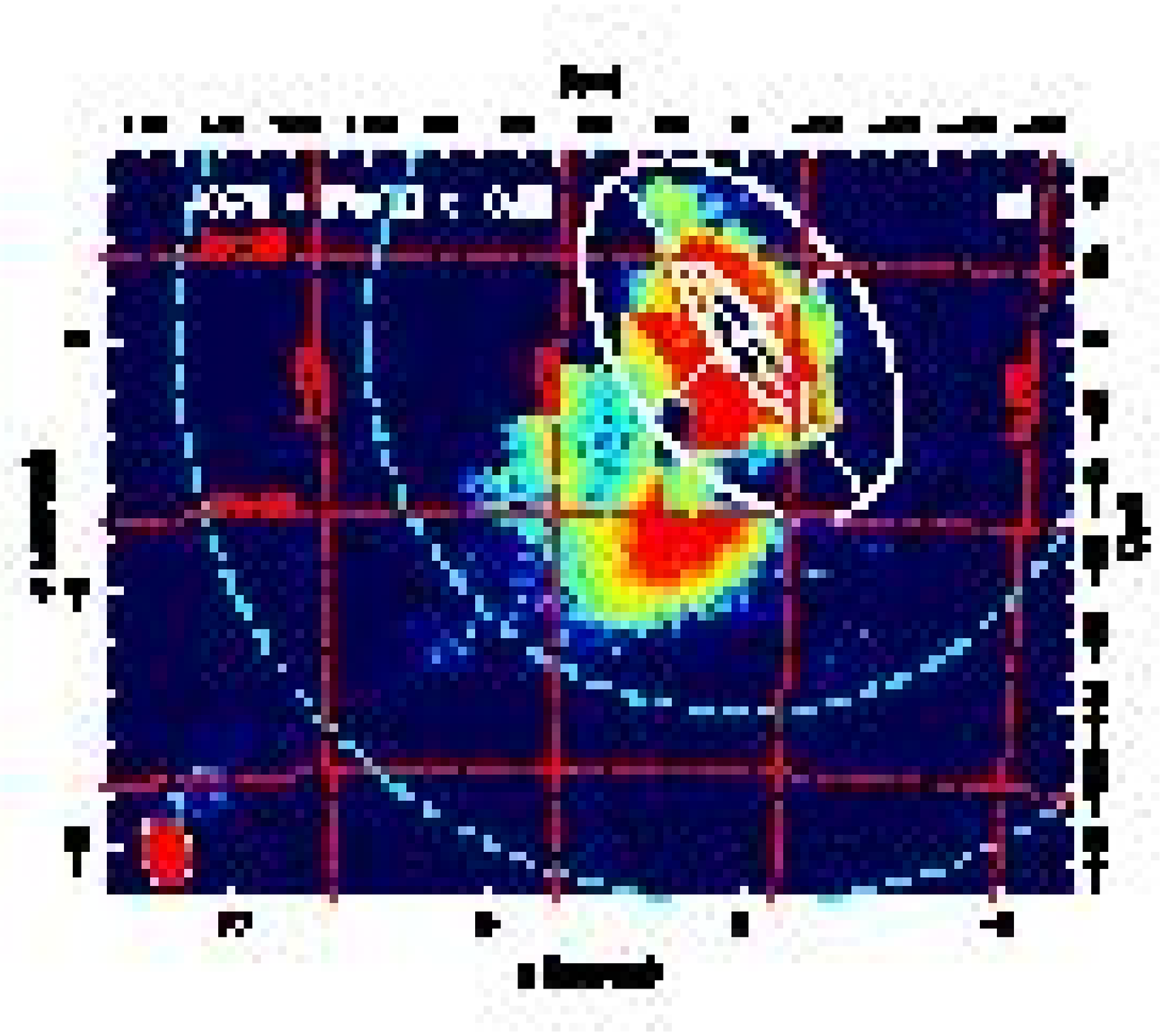}
}
\hbox{
\includegraphics[angle=-90, width=8.8cm]{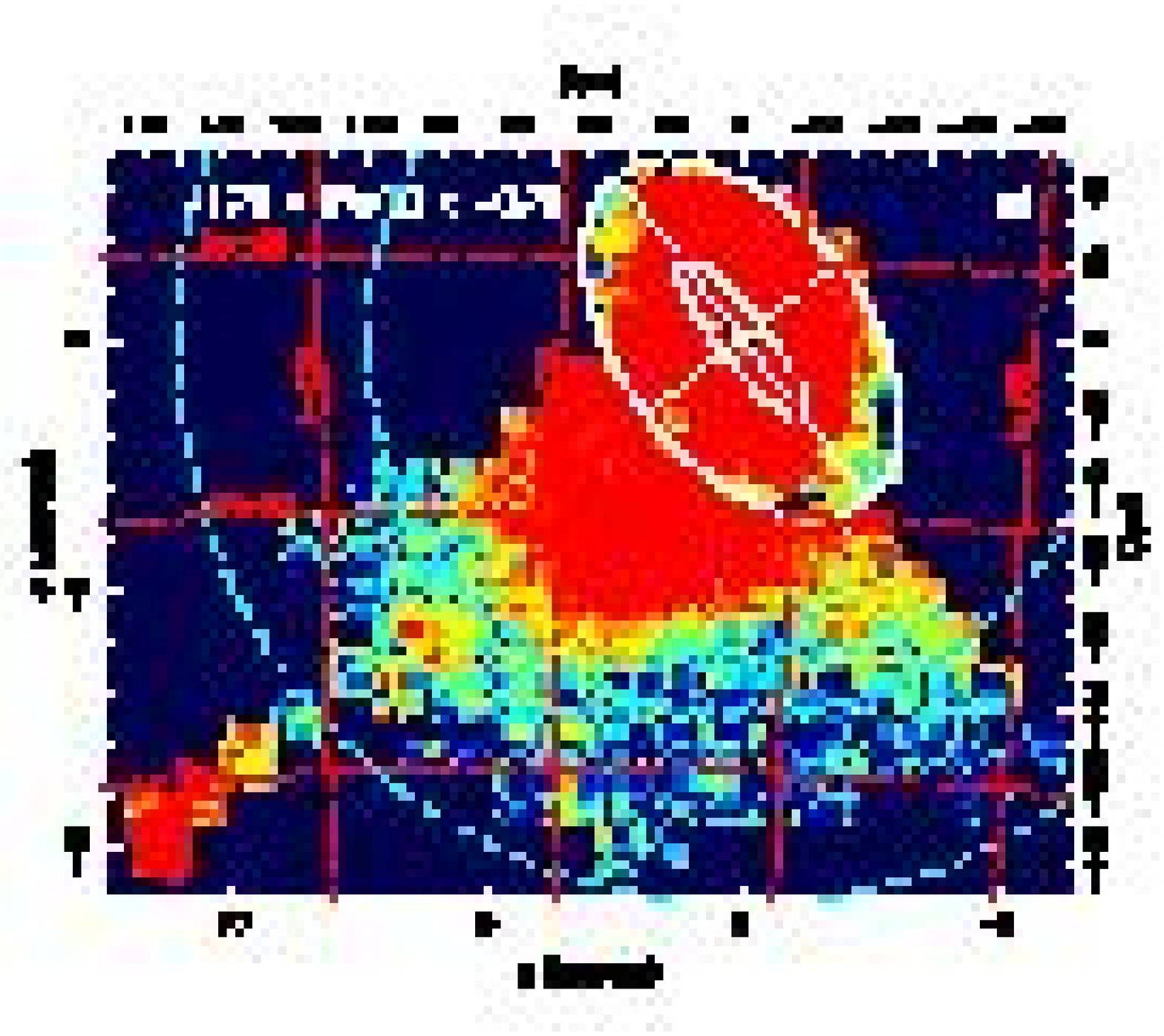}
\includegraphics[angle=-90, width=8.8cm]{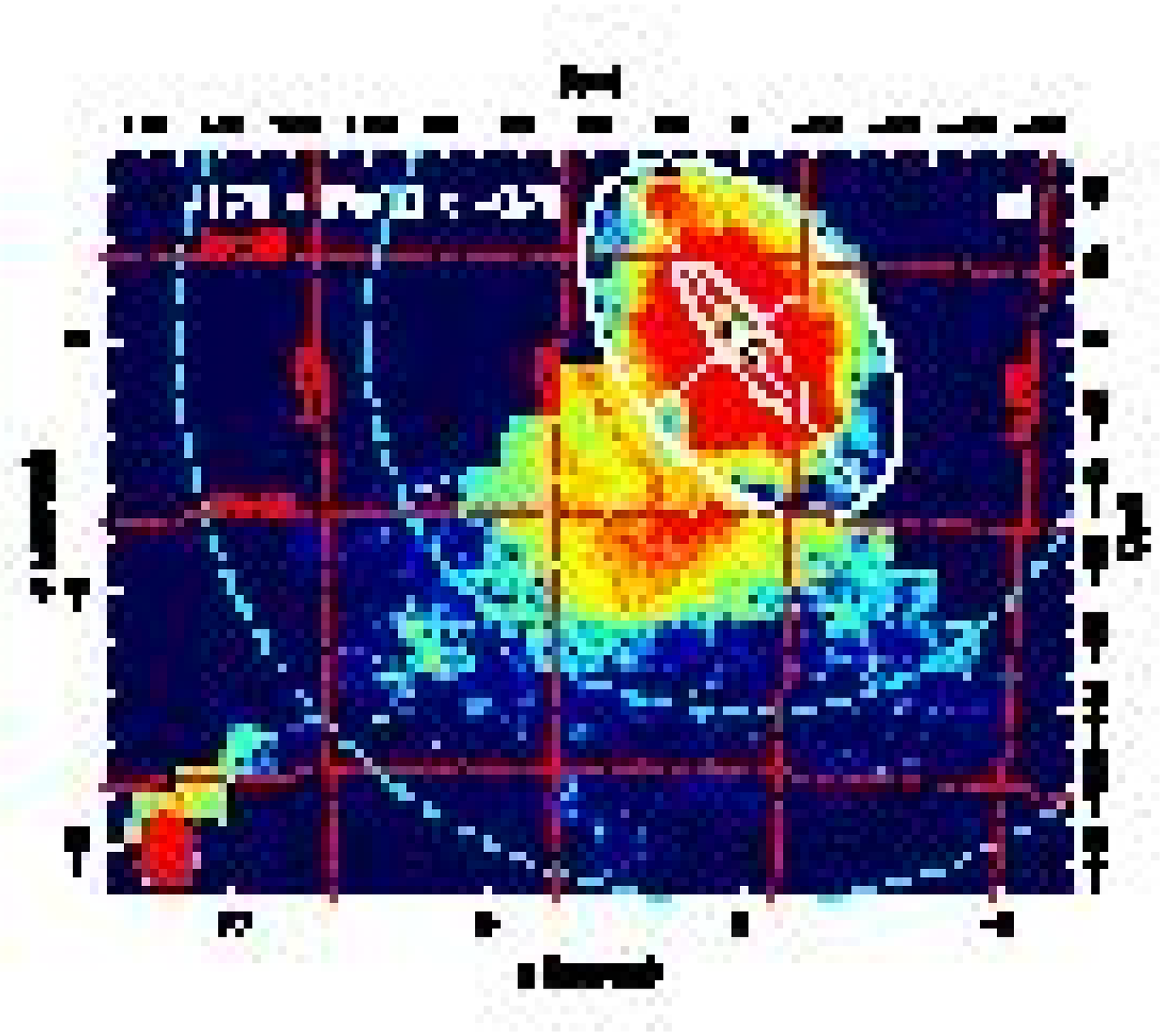}
}
\end{center}
\caption{Logarithmic scale matched-filter maps to a limiting magnitude of $i_0=23.5$, $g_0=25.5$.
Low resolution images ($0\degg2 \times 0\degg2$ pixels) are shown on the left, 
high resolution versions ($0\degg05 \times 0\degg05$ pixels) on the right-hand column.}
\end{figure*}

\begin{figure*}
\figurenum{22 --- continued}
\begin{center}
\hbox{
\includegraphics[angle=-90, width=8.8cm]{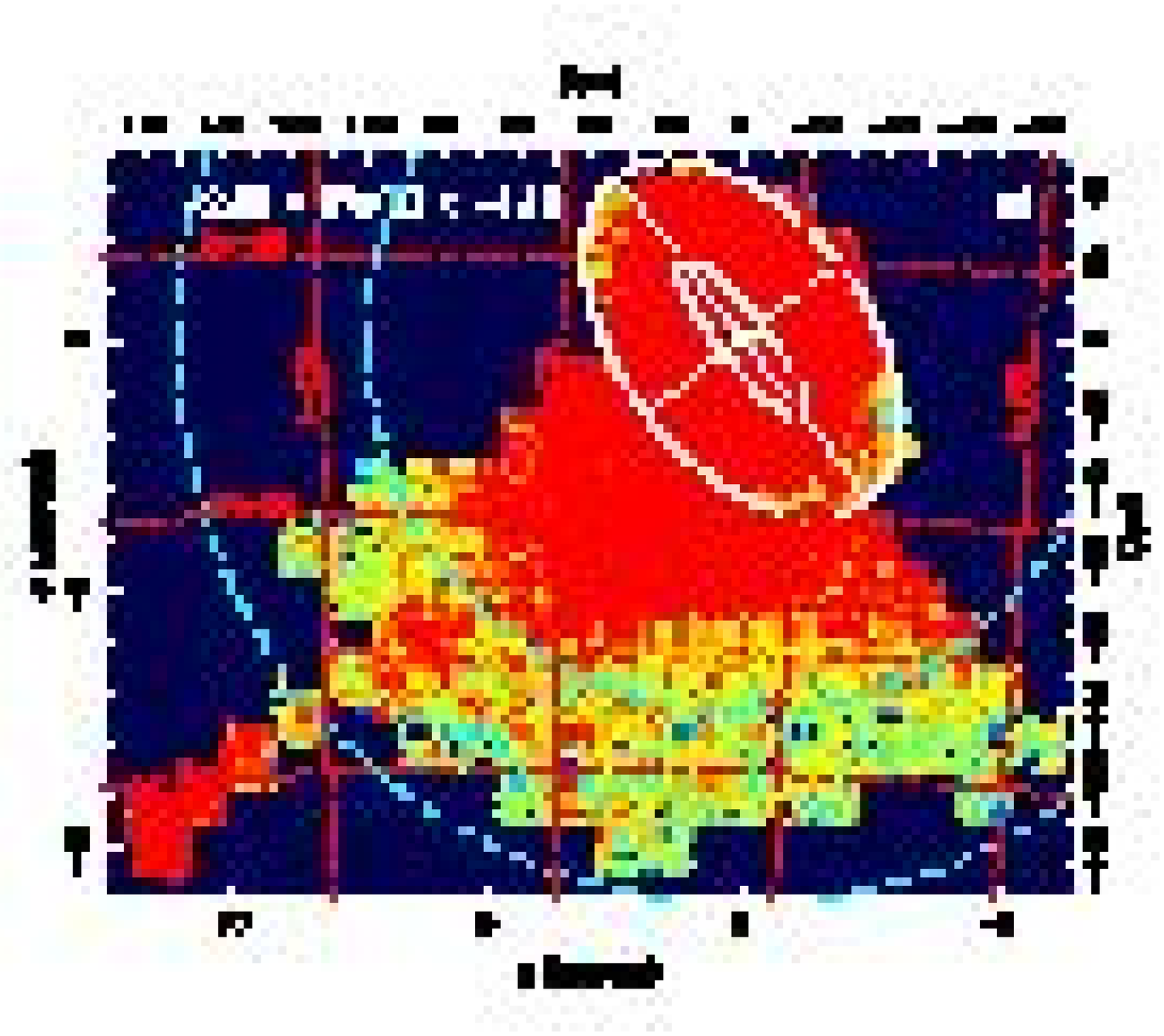}
\includegraphics[angle=-90, width=8.8cm]{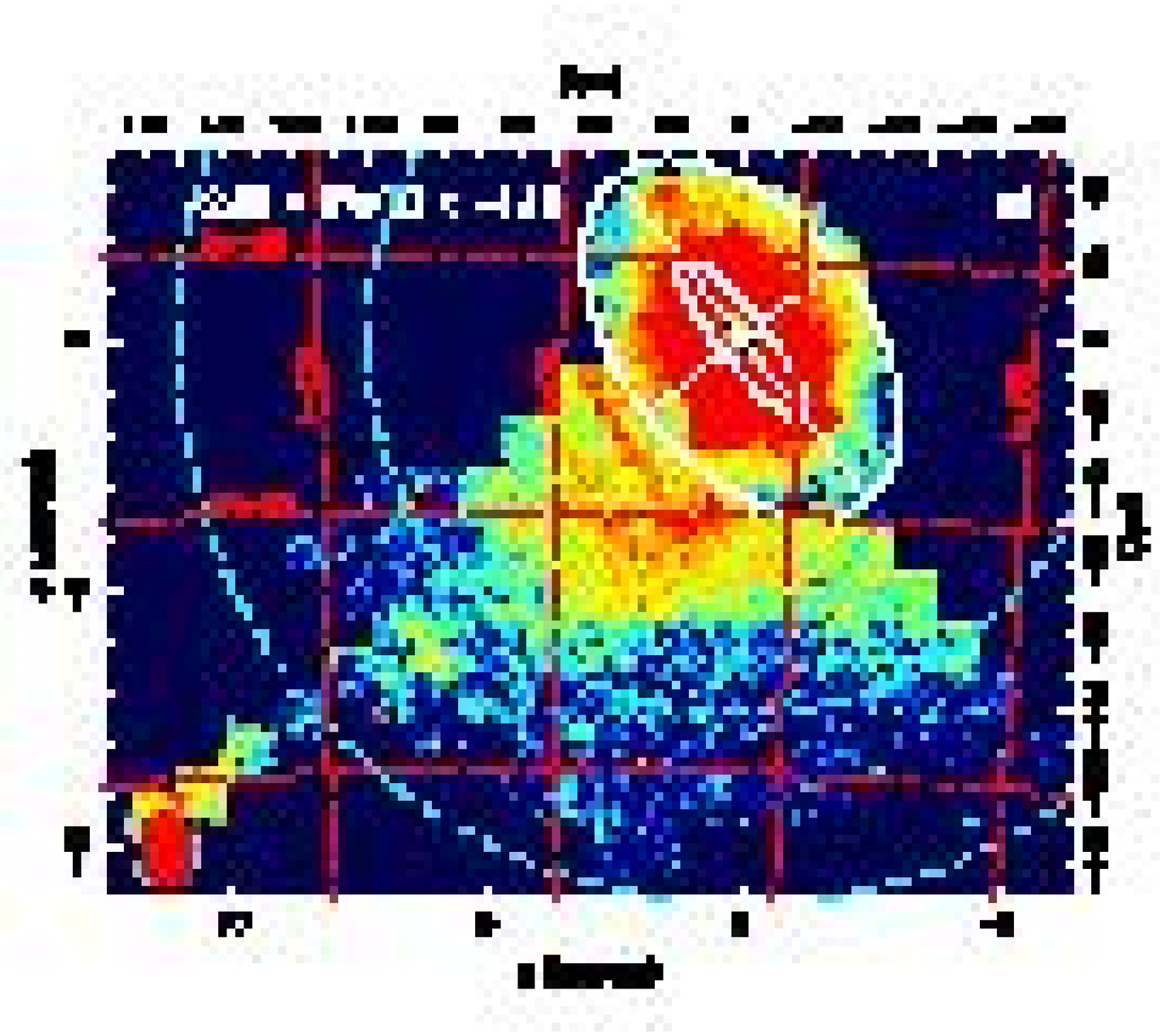}
}
\hbox{
\includegraphics[angle=-90, width=8.8cm]{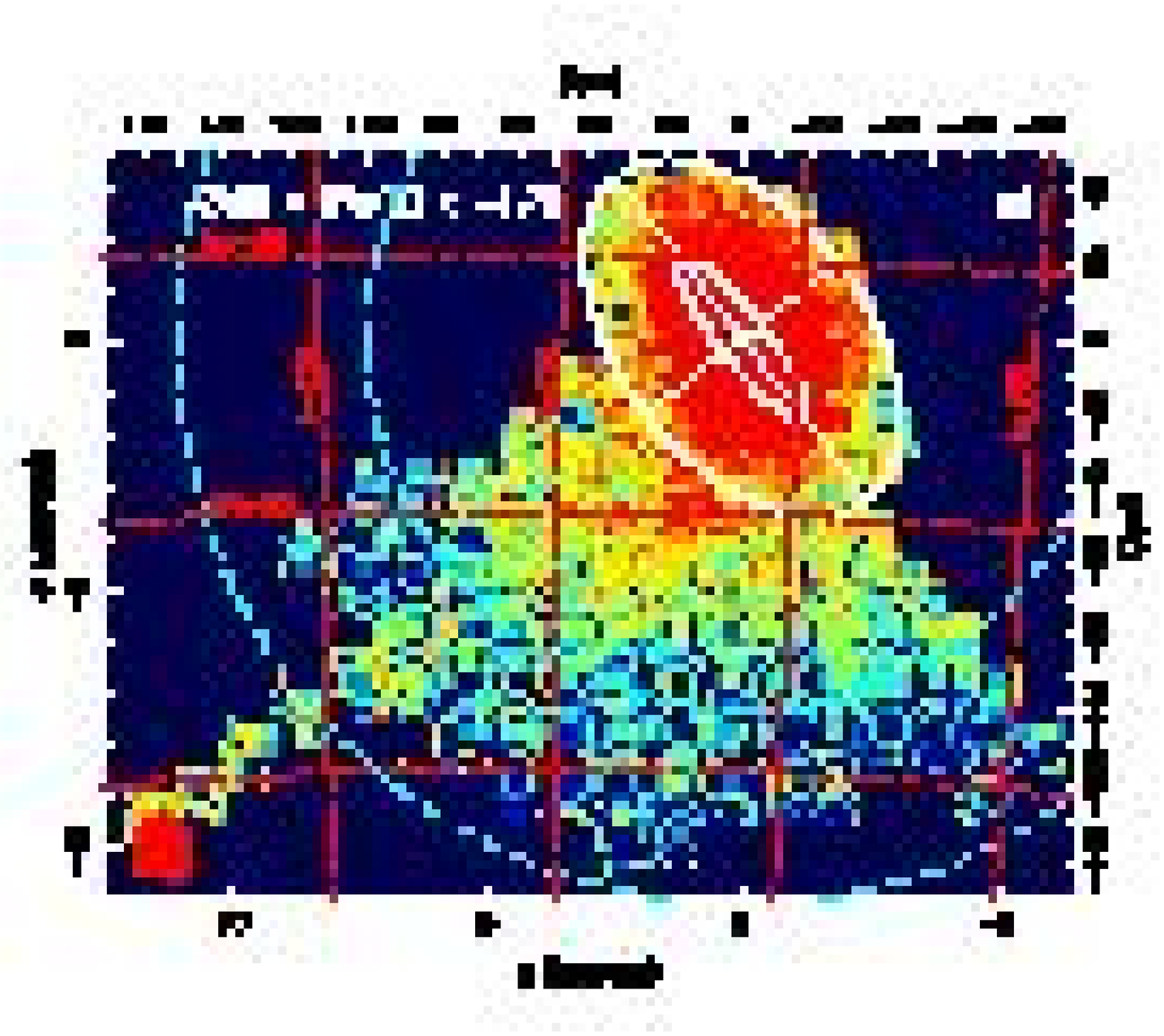}
\includegraphics[angle=-90, width=8.8cm]{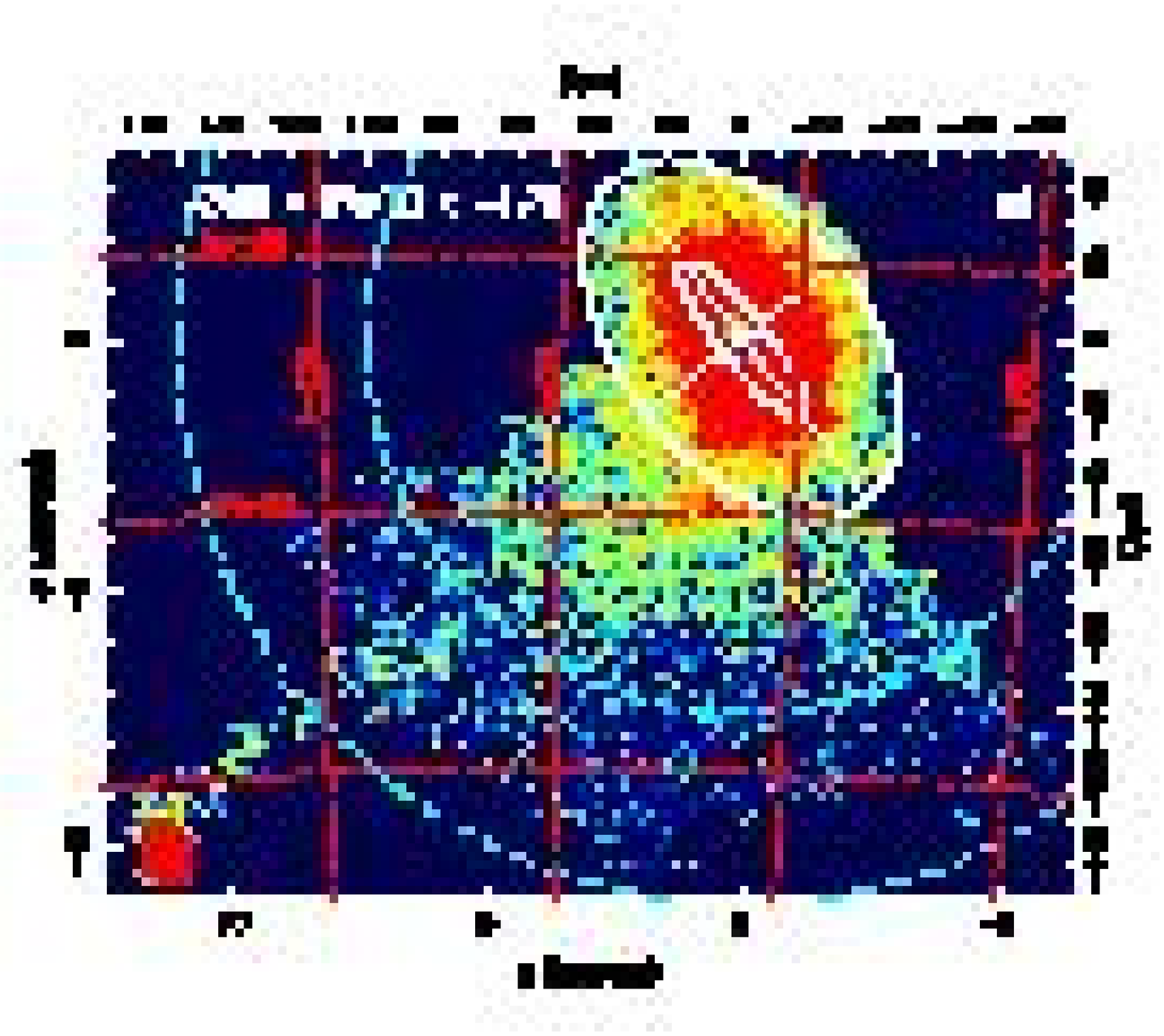}
}
\hbox{
\includegraphics[angle=-90, width=8.8cm]{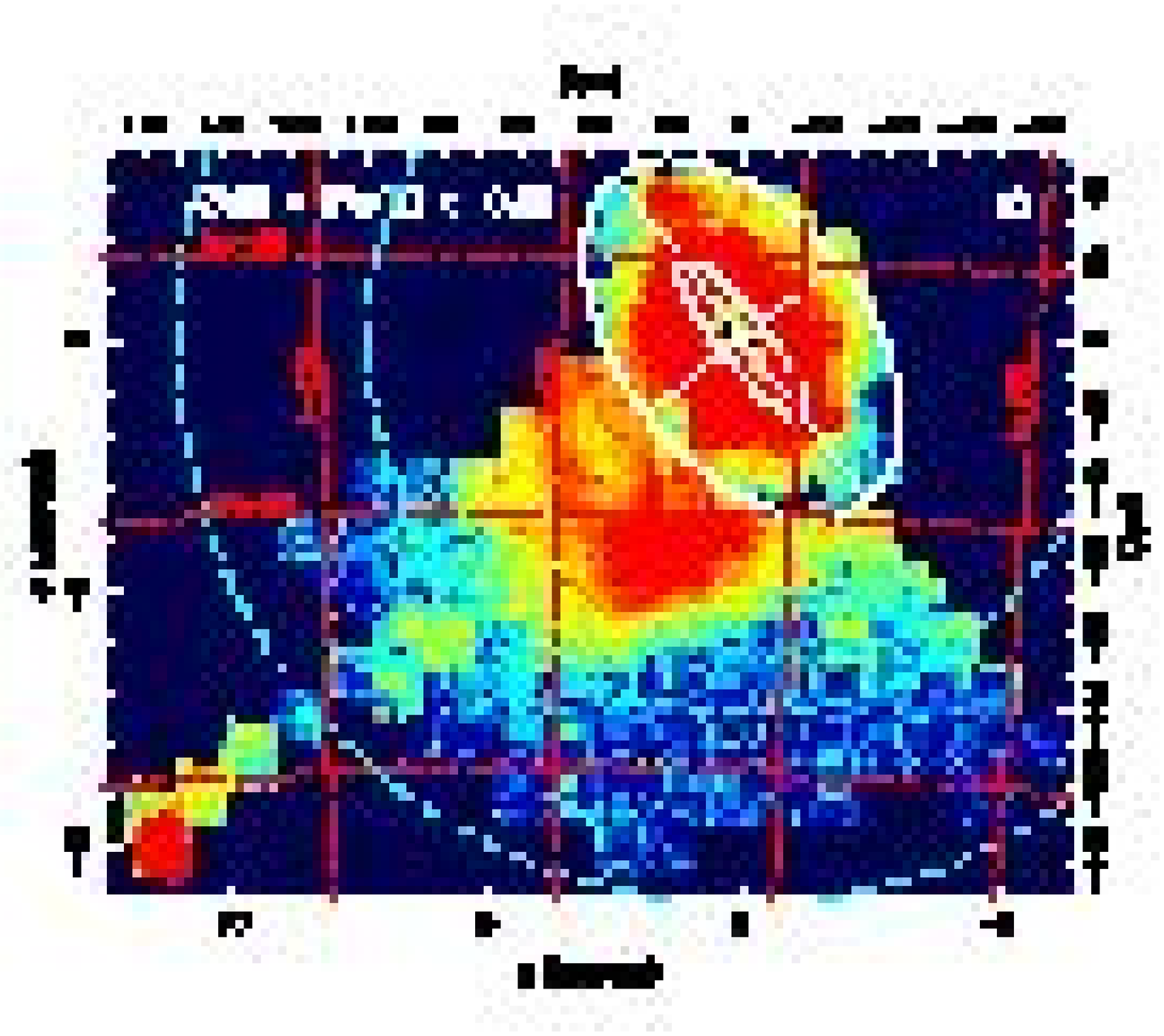}
\includegraphics[angle=-90, width=8.8cm]{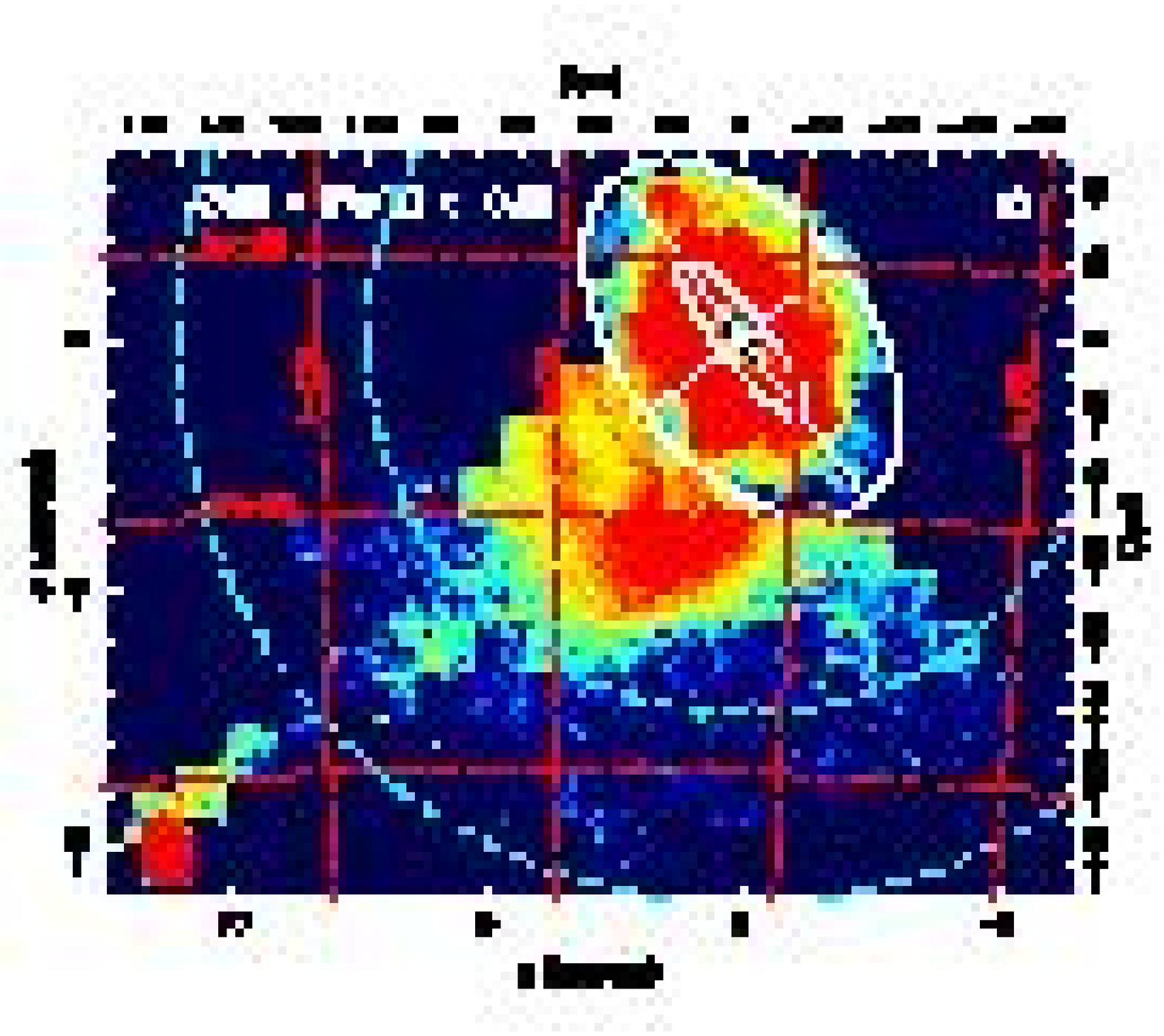}
}
\end{center}
\caption{Logarithmic scale matched-filter maps to a limiting magnitude of $i_0=23.5$, $g_0=25.5$.
Low resolution images ($0\degg2 \times 0\degg2$ pixels) are shown on the left, 
high resolution versions ($0\degg05 \times 0\degg05$ pixels) on the right-hand column.}
\end{figure*}

Figure~20 displays a matched filter map over the entire survey region, where we have chosen
a limiting magnitude ($i_0 = 24.5$), a metallicity range (${\rm -3.0 < [Fe/H] < 0}$) 
and a gray-scale representation to highlight 
the survey defects. The sky region surveyed by the INT is clearly not as deep as the
outer MegaCam region, causing the sharp edge along the MegaCam survey boundary.
However, the most important defect visible here are the long horizontal stripes, which are
present on the top and bottom row of CCDs in the 2003 and 2004 data, 
but not after the camera refurbishment in 2005. The effect is due to a deterioration of
the point spread function (PSF) in those areas, causing stars to appear elliptical and similar to
barely-resolved galaxies. We spent a considerable amount of 
effort adapting our processing software to correct for this effect, but though
substantial improvement was obtained compared to the starcounts derived
assuming a constant PSF,
the problem
could not be removed entirely, since some galaxies intrinsically have ellipticity and
major axis position angle similar to the deformed PSFs. We also attempted to correct
the maps by calculating the equivalent of a flat-field for star-counts 
from the median of many fields. However this was not implemented for the maps 
presented here, 
as the defects were found to be insufficiently
stable, so that the computed corrections introduced other artifacts of almost the same 
amplitude as those they corrected for. Instead, the problem is largely removed by
choosing a brighter limiting magnitude, and virtually disappears if we adopt
$i_0=22.8$ as in Fig.~21, the limit of the INT photometry \citep{ibata01b}. Of the
remaining artifacts, the most obvious remaining are the handful of shallow INT fields
mainly clustered around ($\xi=0\deg$,$\eta=-3\deg$) which were observed
in conditions of poorer seeing than average, and of course the hole
in the star-counts at the center of M31, where the photometry of 
individual stars broke down due to very high crowding.

In Fig.~22 we present the matched filter maps for six different ranges in metallicity.
The limiting magnitude over the MegaCam region was chosen to be $i_0=23.5$,
and we kept a limit of $i_0=22.8$ ($S/N \sim 10$) for the INT survey, 
which gives rise to the obvious discontinuity
around $\eta \sim -3\deg$.
These maps possess a bewildering amount of 
information on a large range of spatial scales and surface densities, so it is impossible
to display all the information at a given pixel scale or with a given color representation.
The diagrams in Fig.~22 have been constructed to show the large-scale distribution
of stellar populations in the MegaCam region of the survey, while retaining some sensitivity
to small structures such as dwarf galaxies which have scales of a few arcmin; in each
row the right-hand panel shows a higher resolution version of the selection in the left-hand panel;
the lower-resolution maps are useful for appreciating the large-scale behavior of the diffuse
components.
We start our discussion with panel `b', which displays the metal-rich selection (${\rm -0.7 < [Fe/H] < 0.0}$). Though noisy,
we can discern many features:
\begin{itemize}
\item The elliptical but irregular distribution of stars with axis ratio $\sim 0.5$ and major axis
diameter $\sim 5\deg$ ($\sim 70\kpc$), containing several previously reported
substructures \citep{ferguson02}. As we have argued elsewhere \citep{ibata05}, 
this is a giant rotating component which is dominant beyond the end of the classical disk, and 
possibly the residue of a significant merger that occurred many Gyr ago \citep{penarrubia06}.
\item The large ($\sim 1\deg$ diameter) overdensity to the north-east 
($\xi \sim 1\degg5$, $\eta \sim 3\deg$), almost certainly unbound debris \citep{zucker04, ibata05}.
\item The ``G1'' clump at ($\xi \sim -1\deg$, $\eta \sim -1\degg5$), a structure surrounding
but unrelated to the luminous globular cluster ``G1'' \citep{ferguson02, rich04, reitzel04, faria07}.
\item The stream-like ``Eastern shelf'' \citep{ferguson02}, at ($\xi \sim 2\deg$, $\eta \sim 0\degg5$).
\item A fainter stream on the western side of the galaxy, the ``Western shelf''
at ($\xi \sim -1\deg$, $\eta \sim 0\degg5$), and seen in the map of \citet{irwin05}.
\item The ``Giant Stream'' \citep{ibata01b, ibata04}, which in the INT data appears to be a linear structure stretching from
very close to the center of M31 to ($\xi \sim 1\degg5$, $\eta \sim -3\deg$), but which shows up as
a substantially wider structure in the MegaCam survey extending to 
($\xi \sim 3\deg$, $\eta \sim -6\deg$).
\item A previously unknown stream is seen extending between ($\xi \sim 4\deg$, $\eta \sim -1\degg5$) and
($\xi \sim 3\deg$, $\eta \sim -4\deg$); we will refer to this as ``Stream C'' in the discussion below.
\item Vast expanses apparently devoid of stars over most of the Southern half of the survey MegaCam.
\item A faint diffuse component is detected approximately $4\deg$ from M33.
\end{itemize}

\begin{figure}
\begin{center}
\includegraphics[angle=-90, width=\hsize]{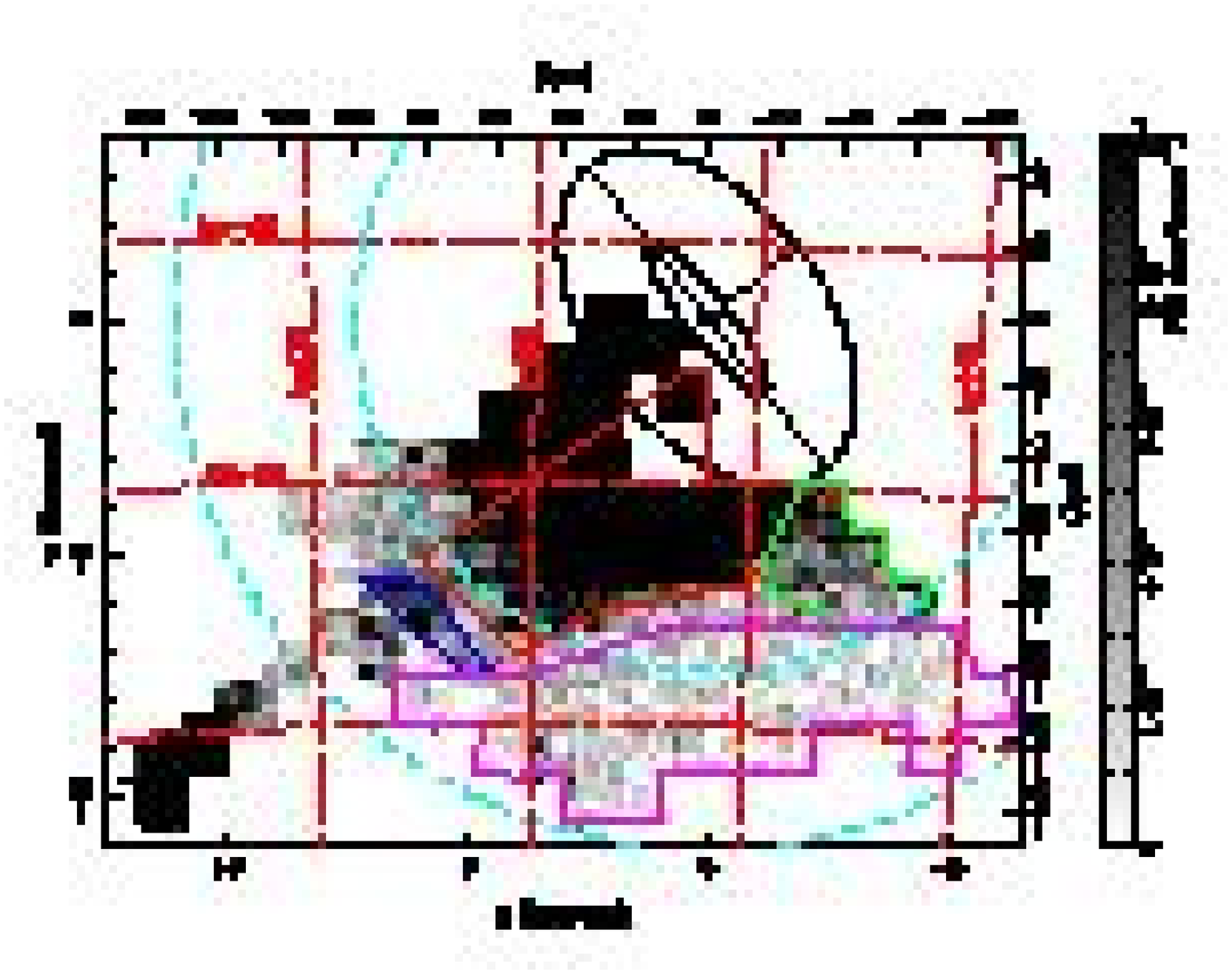}
\end{center}
\caption{Star-count map of the MegaCam region, with the foreground contamination
subtracted using the Besan{\c c}on model. A limiting magnitude of 
$i_0=23.5$ has been adopted. The red, green, blue and pink polygons delineate the
regions chosen to sample, respectively, the Giant Stream, the major axis structure, the
minor axis stream and the empty outer halo region.}
\end{figure}

In panel `c' we show an intermediate metallicity selection (${\rm -1.7 < [Fe/H] < -0.70}$), somewhat
``overexposed'' to bring out better the fainter structures. In addition to the previously-discussed
features, we now notice:
\begin{itemize}
\item The inner ellipse, attributed to the giant rotating component, has become larger and even more irregular.
The more irregular aspect is of course consistent with the expected longer mixing times of debris at larger radius.
An interesting point is that the distribution appears now to be less flattened, suggesting that this extreme
color stretch may be revealing another rounder structure previously hidden beneath the flattened
rotating component.
\item The dwarf galaxies And~II and And~III (cf. Fig.~3) become apparent.
\item Two strong localized structures, at ($\xi \sim 6\degg23$, $\eta \sim -2\degg89$)
and ($\xi \sim 6\degg23$, $\eta \sim -8\degg89$), which as we will discuss below, are two new
dwarf satellite galaxies.
\item A faint low surface brightness fuzz is detected on the extension of the major axis of M31, 
out to ($\xi \sim -5\deg$, $\eta \sim -7\deg$), we will refer to this as the ``Major axis diffuse structure''.
\item A strong stream-like structure is detected between ($\xi \sim 3\deg$, $\eta \sim -1\degg5$) and
($\xi \sim 2\deg$, $\eta \sim -2\degg5$), which we call ``Stream D'' below.
\item A further faint low surface stream-like structure is detected towards ($\xi \sim 6\deg$, $\eta \sim -6\deg$),
which we will refer to as ``Stream A''.
\item The extended structure near M33 is stronger.
\item The region ($\xi < 4\deg$, $\eta < -9\deg$) remains devoid of stars.
\end{itemize}

The more metal-poor selection in panel `d' (${\rm -2.3 < [Fe/H] < -1.1}$) displays 
essentially the same properties as in panel `c', except that
a considerable amount of localized density spikes are detected, covering one to a few contiguous pixels.
Among these are the newly-discovered dwarf galaxies And XI, XII, and XIII \citep{martin06}.
Panel `e' shows the most metal-poor sample (${\rm -3 < [Fe/H] < -1.70}$). Now the Giant Stream
has almost disappeared, and only And II and III are still clearly visible
as substructures, yet one also discerns a radial gradient from M31 over the MegaCam survey region.
For completeness, in panel `a' we show the most metal-rich selection 
considered here (${\rm 0.0 < [Fe/H] < +0.2}$) in which only the inner disk of M33 
and a small portion of the Giant Stream are discernible, while panel `f' shows the map
over the full metallicity range.
The increased sensitivity with the full metallicity range reveals a further 
feature on the minor axis with a stream-like 
structure between ($\xi \sim 5\deg$, $\eta \sim -2\degg5$) and ($\xi \sim 3\deg$, $\eta \sim -5\deg$),
which we will refer to as ``Stream B''.

\begin{figure}
\begin{center}
\includegraphics[bb=50 1 705 600,clip,angle=0, width=\hsize]{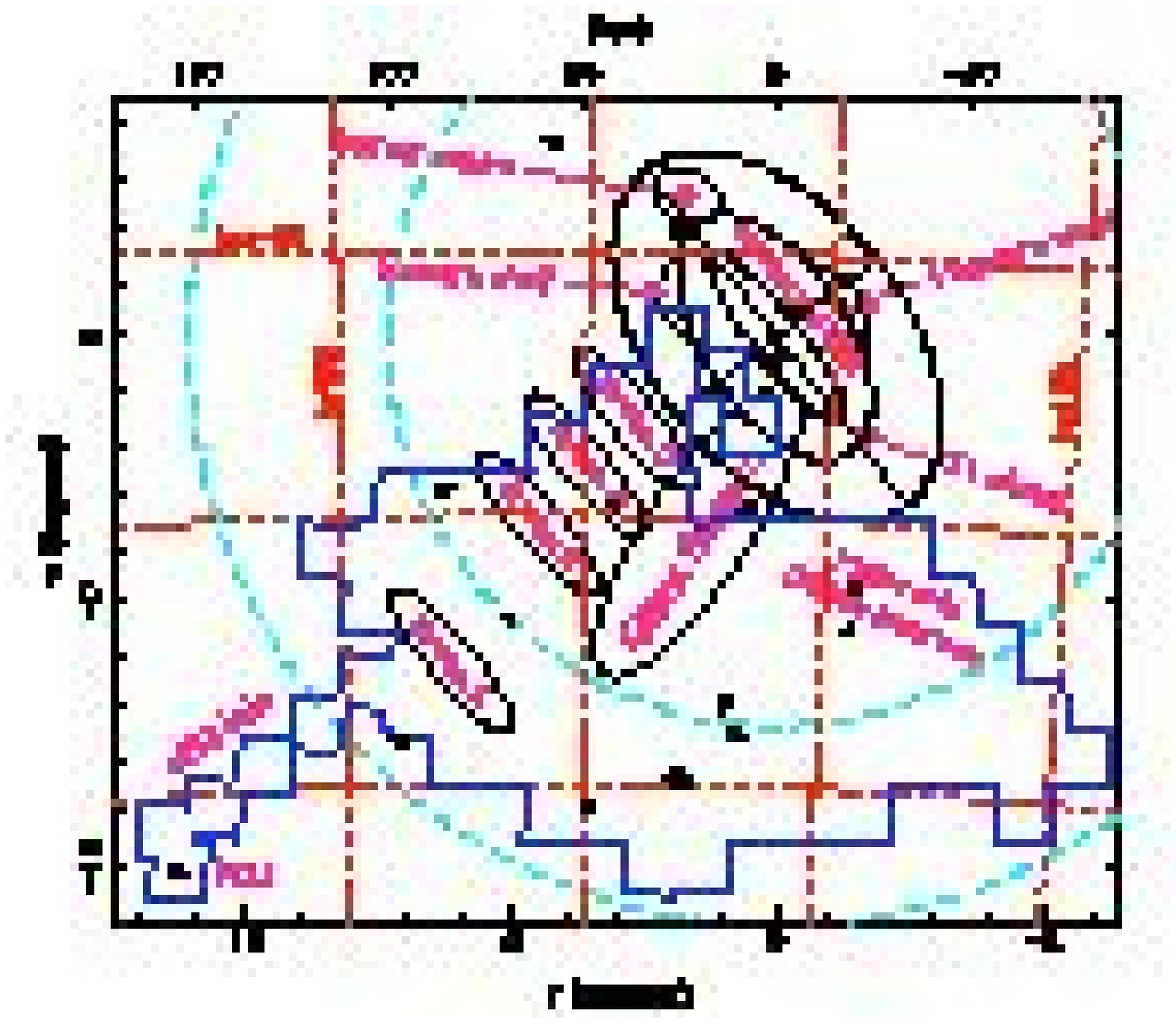}
\end{center}
\caption{Cartoon of the main structures presented in \S5. The circled dots and
`star' markers are reproduced from Fig.~3, and show the positions of dwarf galaxies and selected globular clusters, respectively.}
\end{figure}

The maps displayed in Fig.~22 show the distribution of the matched filter
statistic, so the resulting counts are therefore somewhat difficult to interpret directly. The reason
for this is primarily that the matched filter method relies on a model of the stellar
population that one desires to detect, and the statistic we measure will depend on
the assumed luminosity function and how we choose to weight populations of different 
metallicity. A secondary reason is that, as discussed above, the foreground Galaxy counts do vary
over this vast survey, so the contamination model also varies.
For these reasons we also present in Fig.~23 a straightforward surface density map,
where we have counted up stars in the color-magnitude interval $0.8 < (g-i)_0 < 1.8$ and
$20.5 < i_0 < 23.5$, and have subtracted off the corresponding Besan{\c c}on model counts
over the same area of sky. The main structures previously seen in Fig~21 are nicely
confirmed, and which we highlight in Fig.~23, namely the very extended Giant Stream 
(red polygon), the diffuse major axis structure (green polygon), the minor axis stream-like
structure (blue polygon), the extended outskirts of M33, and the voids elsewhere (pink polygon). 
The advantage of this map is that we can now interpret the physical meaning of the color 
scale, which is shown with the wedge at right-hand edge of the diagram.
Black corresponds to $10^{-4}$ RGB stars per square arcsecond down to $i_0 = 23.5$.
Using the conversion of star-counts to surface brightness discussed above,
the saturated black level translates to ${\rm \Sigma_V = 30.3 \, mag \, arcsec^{-2}}$.

In the next section we discuss in more detail the populations that are highlighted in
Fig.~23. To ease interpretation, in Fig.~24 we show a cartoon of the positions of 
these populations with
respect to the various structures discussed above.

\begin{figure}
\begin{center}
\includegraphics[angle=-90, width=\hsize]{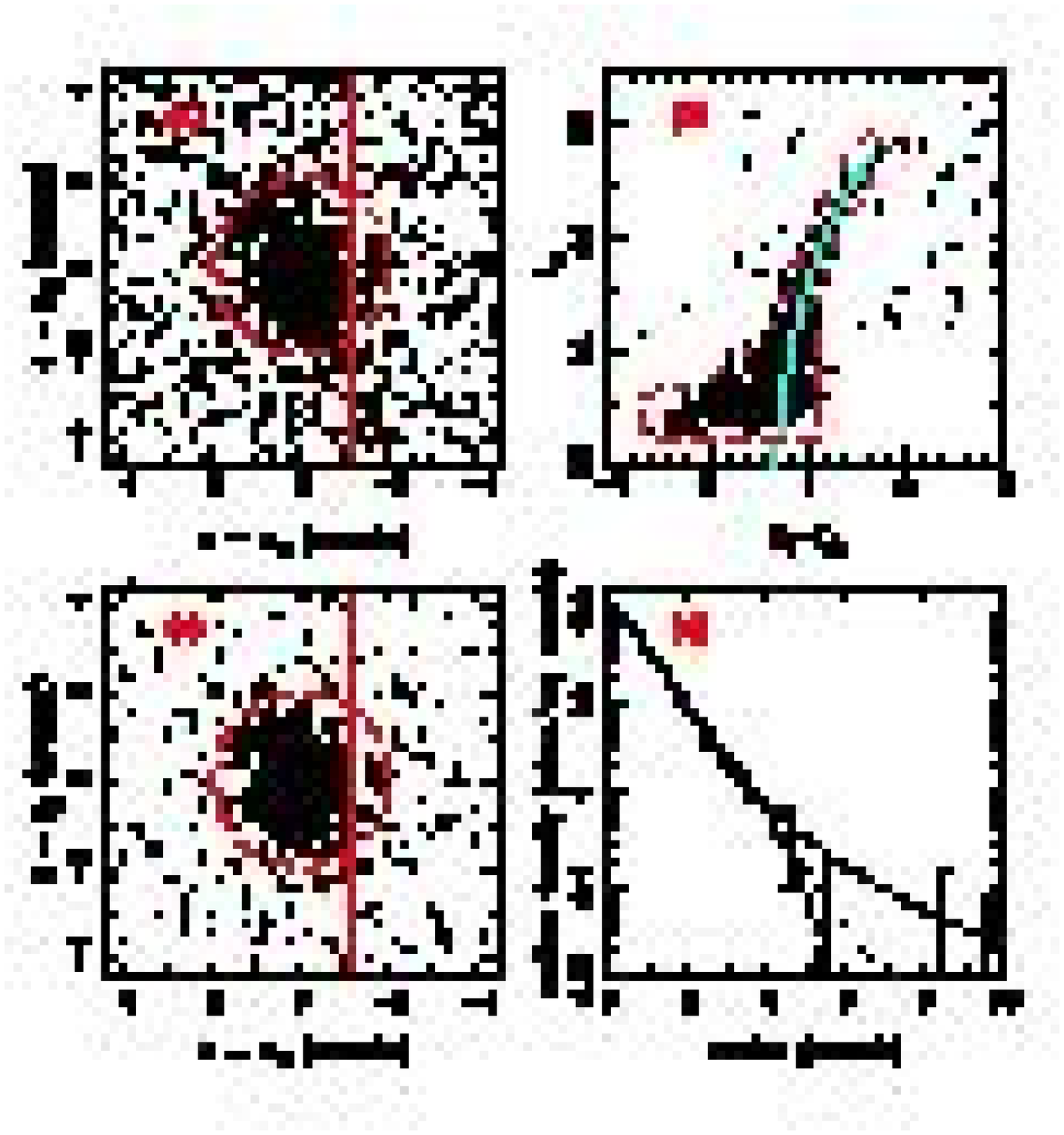}
\end{center}
\caption{The spatial distribution of point-sources in a $9\arcmin\times 9\arcmin$ 
area in the vicinity of And~XV (panel `a'). The parallel red lines mark the 
CCD boundaries, though there is no gap at this location due to the adopted
dithering pattern. The CMD of the stars within the $2\arcmin$ circular
region is shown in panel `b'. Selecting those stars with color and magnitude
within the red dashed polygon, yields the spatial distribution shown in panel 
`c' whose radial profile is given in panel `d'. The continuous, dashed and 
dot-dashed lines in panel `d'
are, respectively, a Plummer model, an exponential model, and a King model 
fit to the profile inside of $5\arcmin$.}
\end{figure}

\section{Spatial subtructures}

\subsection{Discovery of 2 bright satellites}

A thorough analysis of these data regarding the incidence of low mass satellites around M31 and
its implications for galaxy formation theory and cosmology will be presented
in a later publication in this series (Martin et al. 2007b). However, we discuss
briefly here two new dwarf galaxies which were discovered immediately
from simple visual inspection of the starcounts maps. Since the analysis
is identical for both objects we include the results for And~XVI in brackets.

And~XV (XVI), located at $\alpha_0 = 1^h 14^m 18.7^s$, 
$\delta_0 = 38\deg 7\arcmin 3\scnp$
($\alpha_0 = 0^h 59^m 29.8^s$, $\delta_0 = 32\deg 22\arcmin 36\scnp$)
can be noticed as an obvious enhancement in the matched-filter maps
presented previously. In panel `a' of Fig.~25 (Fig.~26), we show the distribution of
all detected point sources in a $9\arcmin\times9\arcmin$ region around
the dwarf galaxy. The color-magnitude distribution of the sources
within the $2\arcmin$ ($1\mcnd5$) radius circle centered at the point of maximum 
density is shown in panel `b'. A very clear and strong RGB is present.
Assuming that the stars outside of the irregular polygon are contaminants,
we proceed to estimate the distance of the structure using the tip of the 
RGB. We adopt ${\rm M_{TRGB} = -4.04 \pm 0.12}$ from \citet{bellazzini01} 
for the absolute I-band magnitude of the RGB tip, and convert into
the Landolt system using the color equations above and those given
by \citet{mcconnachie04a}; this yields a distance modulus
of ${\rm m-M = 24.0 \pm 0.2}$ (${\rm m-M = 23.6 \pm 0.2}$) or
alternatively a distance of $630 \pm 60\kpc$ ($525 \pm 50\kpc$). 
With this distance modulus we find a reasonable visual fit to the RGB
with a Padova isochrone of metallicity ${\rm [Fe/H] = -1.1}$ (${\rm [Fe/H] = -1.7}$). 
Given the angular distance of $6\degg8$ ($9\degg5$) from M31,
the object lies at an M31-centric distance of $170\kpc$ ($270\kpc$).

\begin{figure}
\begin{center}
\includegraphics[angle=-90, width=\hsize]{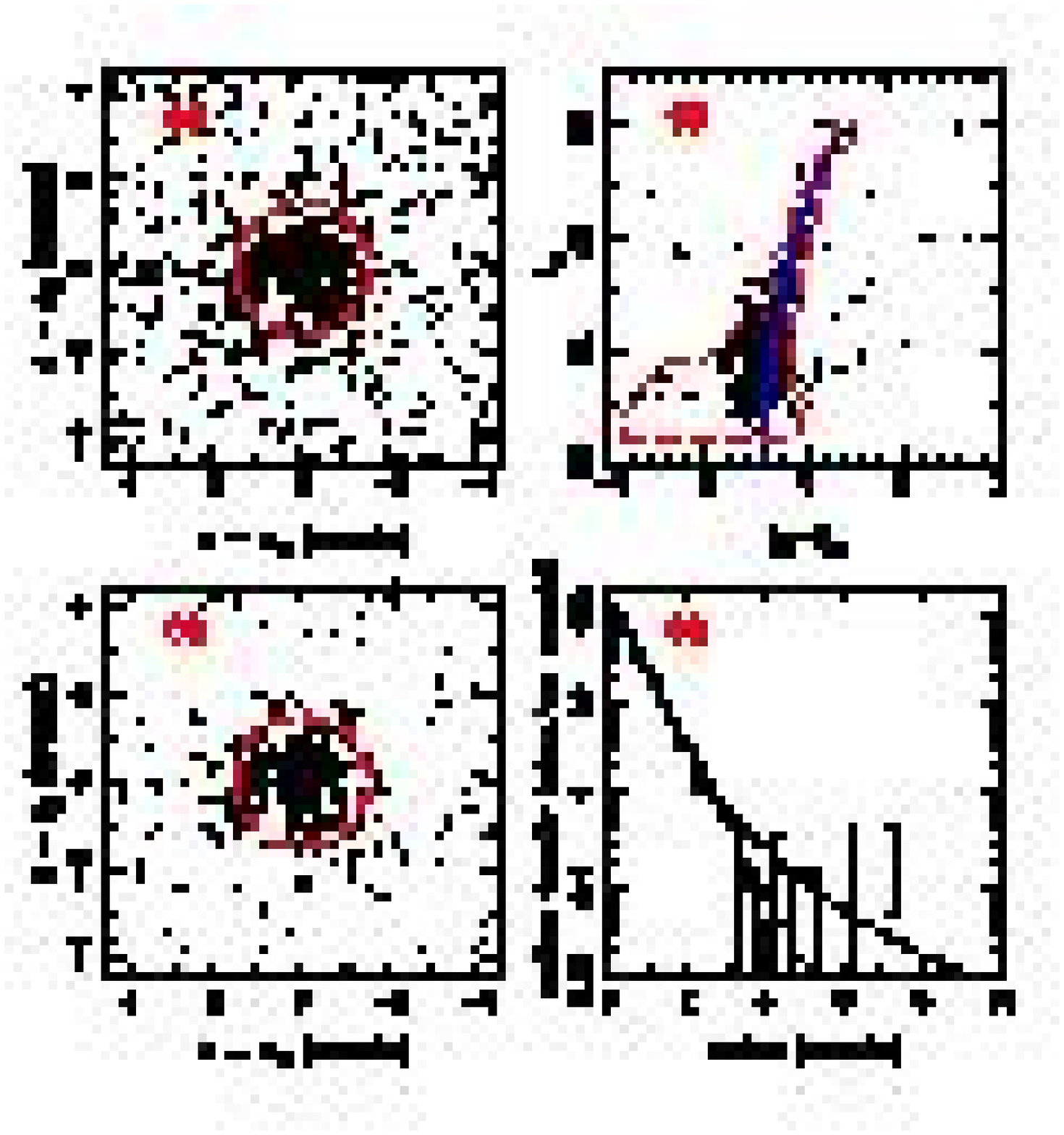}
\end{center}
\caption{As Fig.~25, but for And~XVI. The presence of several bright stars
causes the irregular spatial distribution in the left hand panels.}
\end{figure}

\begin{figure}
\begin{center}
\vbox{
\includegraphics[angle=-90, width=\hsize]{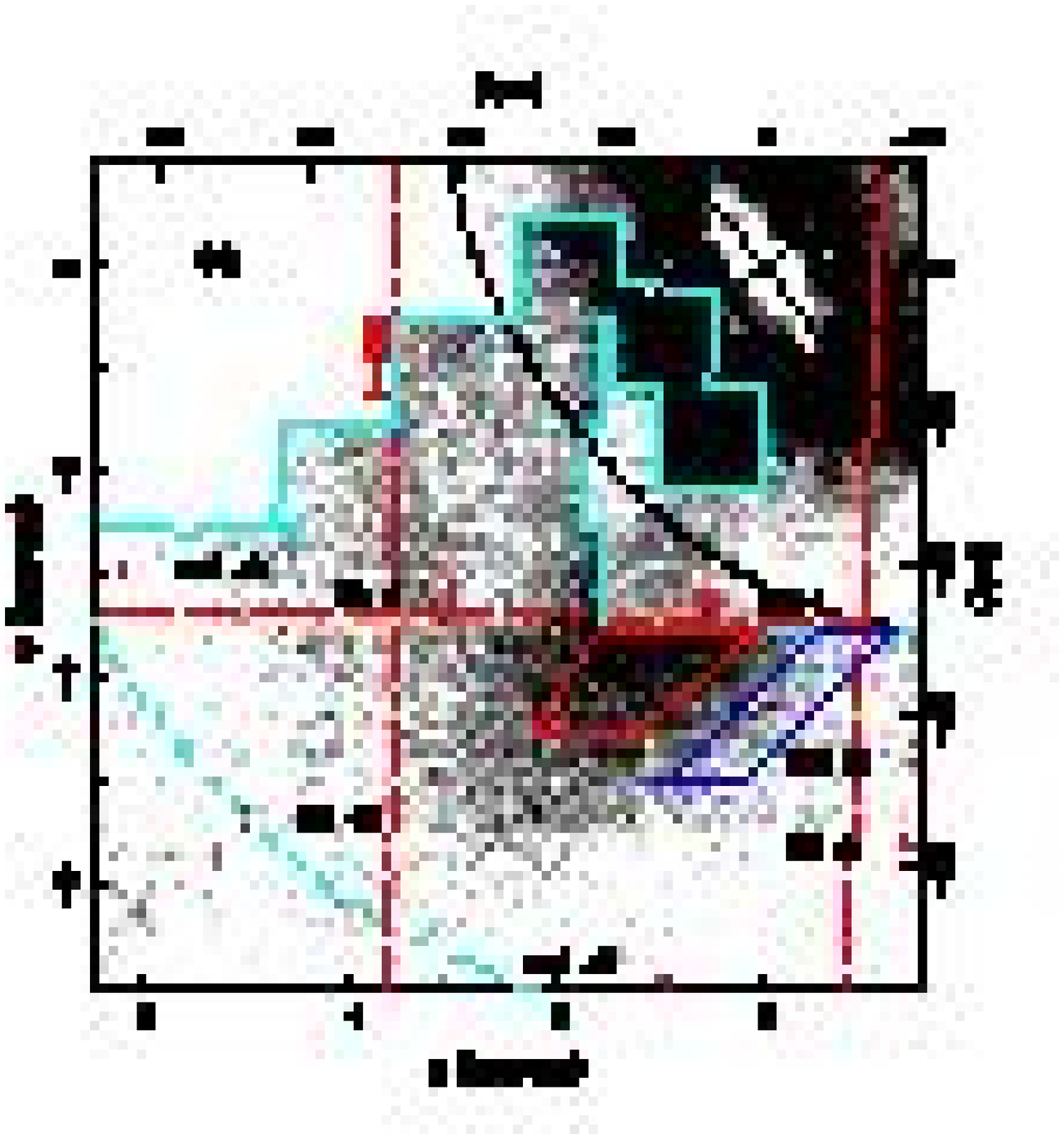}
\includegraphics[angle=0, width=\hsize]{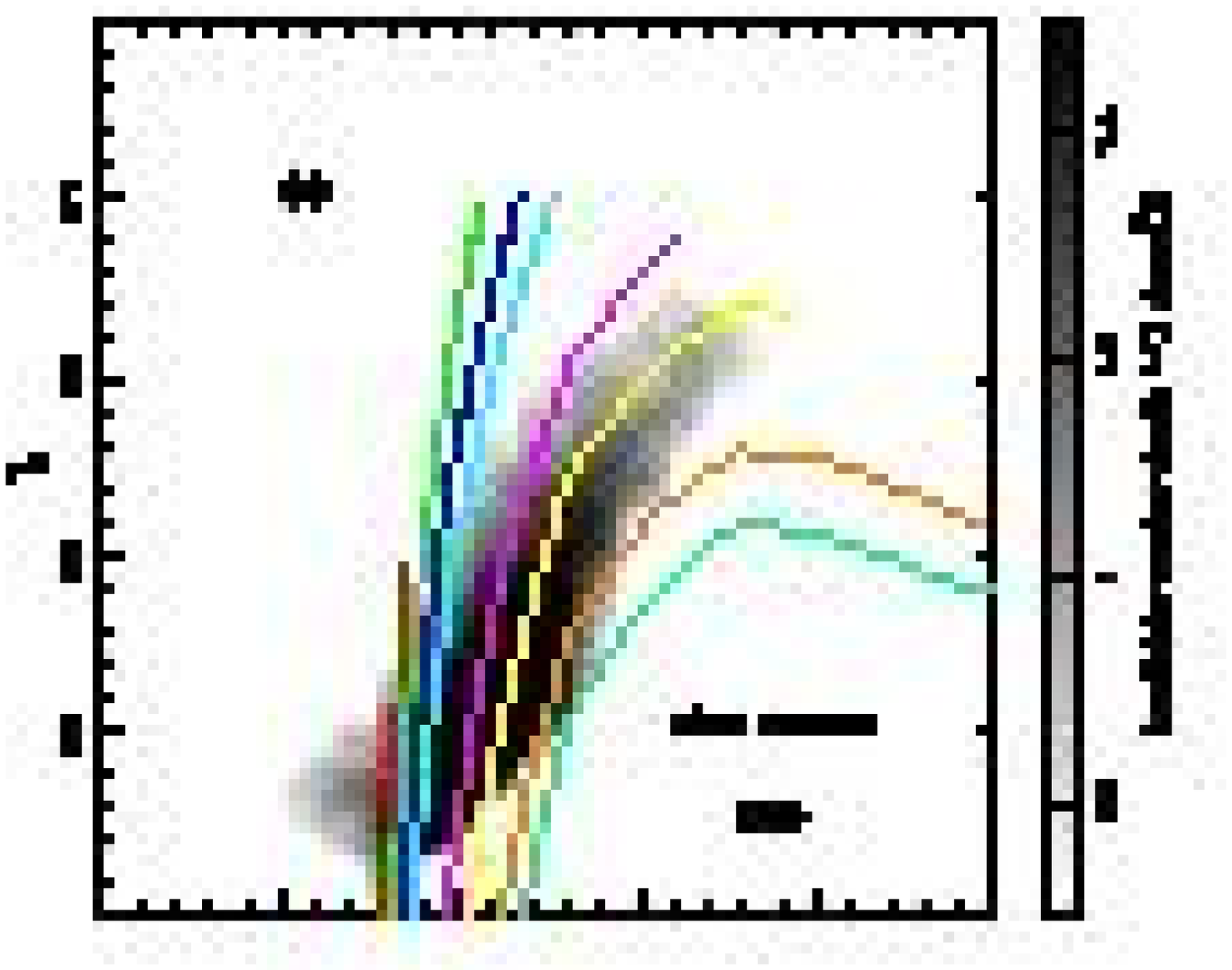}
\includegraphics[angle=0, width=\hsize]{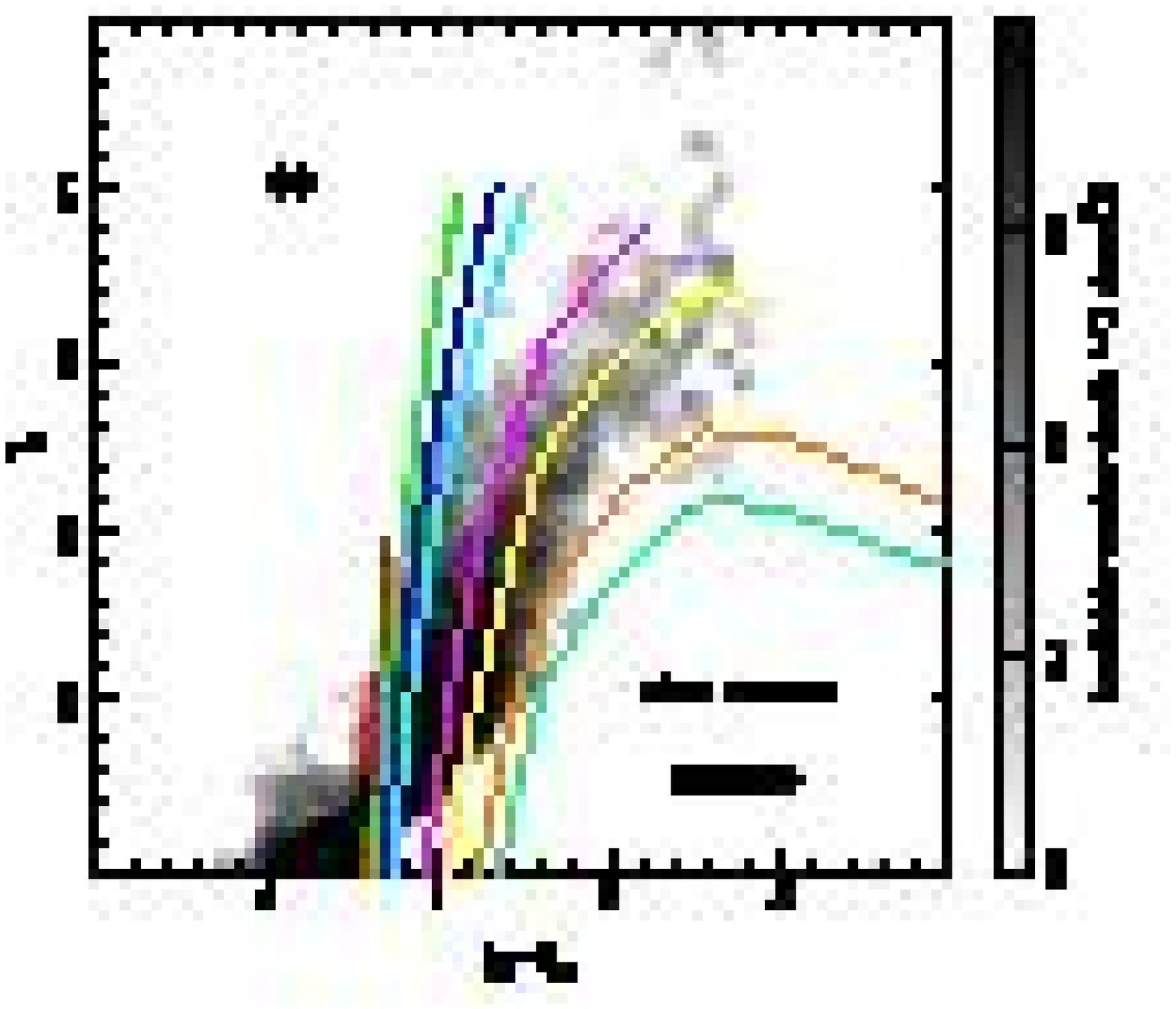}
}
\end{center}
\caption{Panel `b' displays the stellar populations in the core of the Giant Stream 
(sampled in the spatial region shown with a red polygon in panel `a'); while panel
`c' displays those on the periphery of this structure (dark blue polygon
in panel `a'). The foreground contamination has been removed from the
two Hess diagrams.}
\end{figure}

With the CMD selection polygon from panel `b', we filter out foreground
contamination, which gives the distribution shown in panel `c'. The 
corresponding density profile is given in panel `d', where we have subtracted off a
background count determined from an annulus between $10\arcmin$ and $15\arcmin$.
Fitting the distribution with an exponential profile (dashed line), yields a scale length
of $0\mcnd72\pm0\mcnd03$ ($0\mcnd53\pm0\mcnd03$), 
though a Plummer model (solid line) 
of scale size $1\mcnd2$ ($0\mcnd9$) also fits acceptably well,
as does a \citet{king62} model (dot-dashed line) 
with core radius of $0\mcnd91$ ($0\mcnd64$)
and tidal radius of $5\mcnd7$ ($4\mcnd3$).
By integrating the star-counts
up to the half-light radius, and correcting by $2.45$~mag (as above) to account for 
stars below $i_0=23.5$, we estimate a total absolute magnitude of 
${\rm M_V= -9.4}$ (${\rm M_V= -9.2}$).

And~XVI will be a particularly interesting object for further study given its extreme
distance from M31, and its location between M31 and the Milky Way, where
it presumably has felt a non-negligible perturbation from the potential of
our Galaxy. It is also curious that And~XV appears to be
structurally disturbed and elongated, which is suggestive of the action of
galactic tides. Yet how this very distant galaxy might have been
affected by tides is hard to imagine.
(The irregular morphology seen in the distribution
of And~XVI stars in Fig.~26 is an artifact of nearby bright star ``holes'').

\begin{figure}
\begin{center}
\includegraphics[angle=-90, width=\hsize]{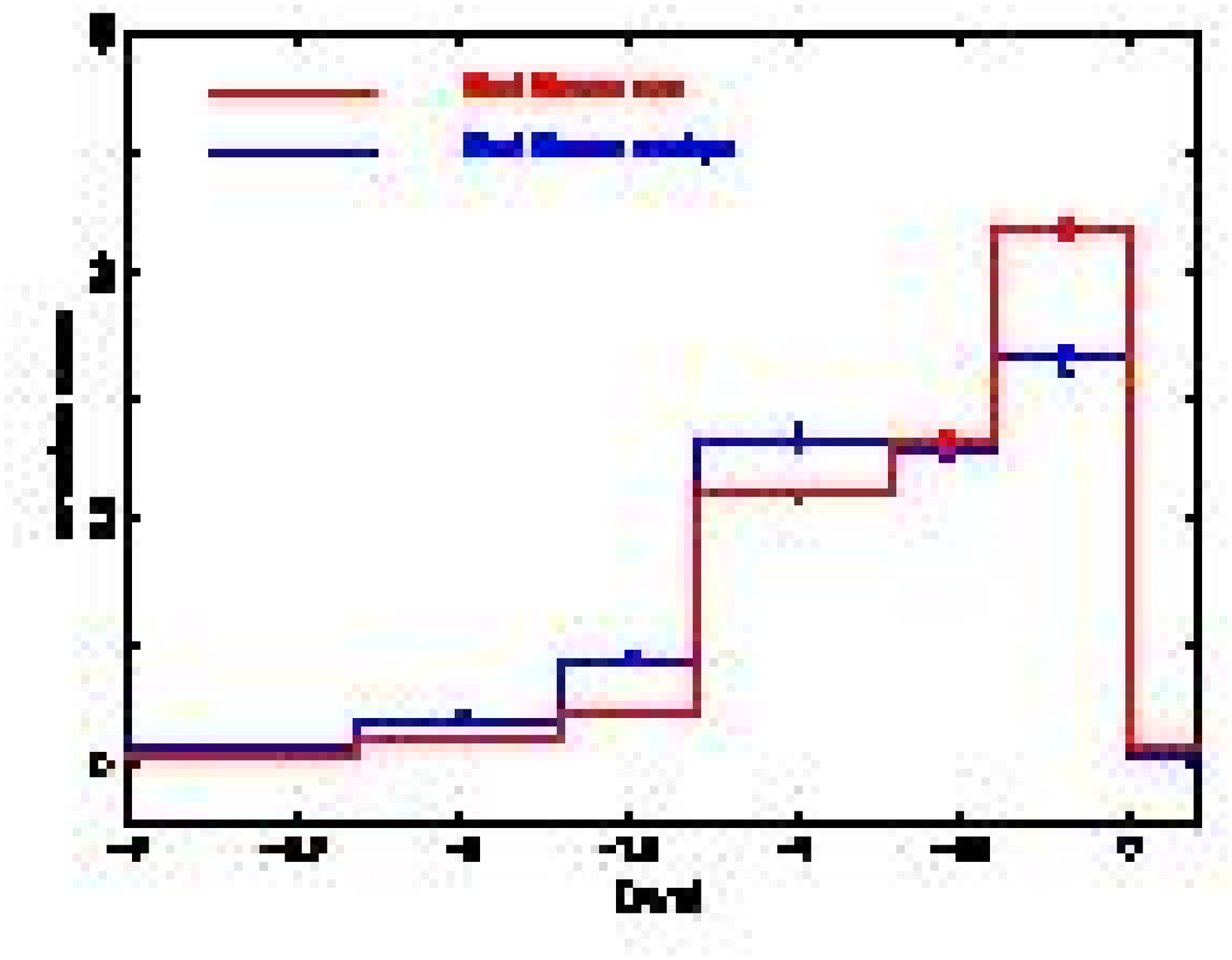}
\end{center}
\caption{The metallicity distribution functions 
(with error bars denoting $1\sigma$ uncertainties) for the Giant Stream core
sample (red) and the envelope sample (blue),
as interpolated from the chosen Padova isochrones. Photometric
limits of $i_0 = 23.5$ and $g_0 = 25.5$ have been imposed. The background fields,
normalized with the Besan{\c c}on model,
have been used to subtract off the expected foreground counts in each 
of the metallicity bins. The two distributions are completely inconsistent
with each other to high confidence.}
\end{figure}

\subsection{Giant Stream}

The Giant Stream around M31 has been the subject of numerous studies,
due to the fact that it is a nearby intermediate mass merging event, and
that it can be used to measure the potential of M31.
The initial discovery in the INT survey \citep{ibata01b} showed the structure to be
(in projection) an approximately linear and radial stream, with a metallicity slightly
higher than that of 47~Tuc (${\rm [Fe/H] -0.71}$), and a total
absolute magnitude of ${\rm M_V \approx -14}$. We probed more fully its extent 
and the line of sight depth with the CFHT12K \citep{mcconnachie03}, a
precursor wide-field camera to MegaCam at the CFHT. These photometric and positional data
were then complemented by radial velocities obtained at 4 locations along the
stream with the DEIMOS multi-object
spectrograph at the Keck Observatory, which allowed a measurement of
the mass of the halo of Andromeda out to $125\kpc$ \citep{ibata04}, and enabled us
to develop a model of the orbital path of the stream progenitor. 
We found the orbit to be highly radial, and predicted that the stream
fans out towards the East after passing very close to the nucleus of M31,
losing its stream-like spatial coherence.
This analysis also posed an interesting puzzle, which is still unsolved: 
since the stream is on
such a highly destructive radial orbit, how did the progenitor survive until
so recently? 

Subsequently, \citet{guhathakurta06} also used Keck/DEIMOS to 
obtain spectra in one stream field, where they measured a mean 
metallicity of ${\rm \langle [Fe/H] \rangle = -0.51}$.
The kinematic data sets were reanalyzed by \citet{font06}, who undertook
N-body simulations to attempt to reproduce the stream morphology. They found that the 
progenitor must have been more massive than $10^8\msun$, and that the time since its dissolution 
is a mere $0.25\Gyr$. Recently, \citet{fardal06} have shown how the fanning-out of the stream
into shells to the East and West can be used to place constraints
on the galaxy potential.
We defer a full re-analysis of the Giant Stream to a subsequent contribution, focussing
here on the salient new features that are revealed in the MegaCam survey.

\begin{figure}
\begin{center}
\includegraphics[angle=0, width=\hsize]{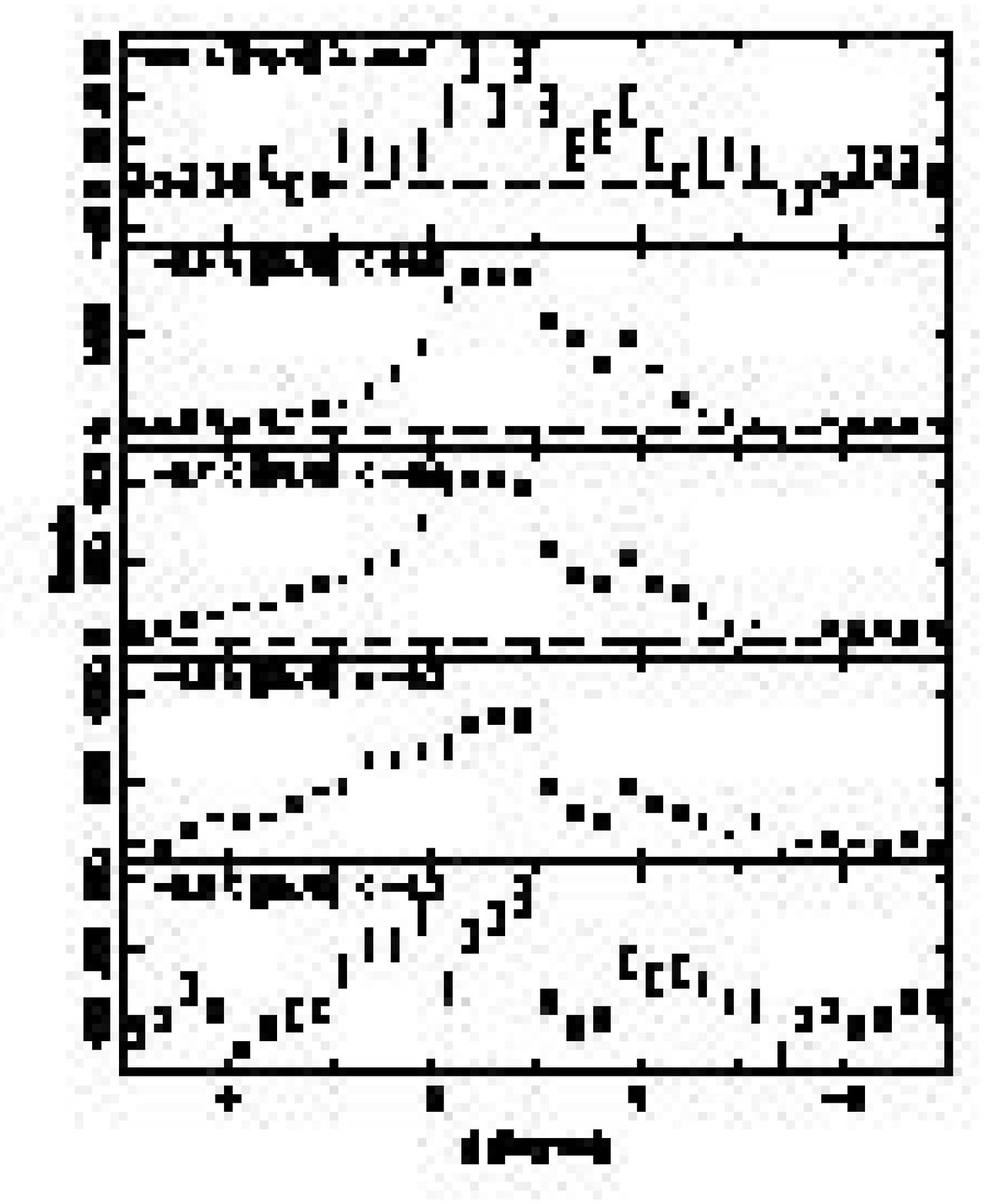}
\end{center}
\caption{Counts in a $1\deg$-wide East-West band between $-4\degg5 < \eta < -3\degg5$
for different metallicity intervals.}
\end{figure}

An inspection of Fig~22, shows that the Giant Stream extends out to a projected radius of
$\sim 100\kpc$ (the inner dashed circle). With the maximum line of sight distance 
to the stream of $886\pm20\kpc$ estimated by \citet{mcconnachie03} (at $\xi =2\deg$, $\eta=-4\deg$),
this corresponds to an apocenter distance of $\sim 140\kpc$.
Though this is further than it had been mapped
out before, the possibility that the stream reaches this projected distance
was anticipated by one of the orbit models presented in 
\citet{ibata04} (cf. Fig.~4 of that paper). That particular orbit model, however, does not
agree well with the measured line of sight distance gradient, though we note that debris
does not exactly follow the orbit of the progenitor. 
Further detailed modeling
is clearly required to understand the dynamics of this stream.

The MegaCam data also shows that there are stellar populations variations in the stream.
We illustrate the evidence for this in Fig.~27, where the 
colour magnitude distribution in the core of the Giant Stream (panel `b') is compared
to a region on the western periphery of the structure.
Both of these spatial selections contain stars over a wide range
of metallicities, and peak at high mean metallicity, consistent with the 
mean photometric metallicity of ${\rm \langle [Fe/H]=-0.51 \rangle}$ measured
from a kinematically-selected sample of stars on the periphery
of the Giant Stream \citep{guhathakurta06}.
It is clear from an inspection of this diagram, however, that
relative to the outer field the core is lacking the blue stellar populations
(around the isochrone with
${\rm [Fe/H] = -1.3}$). The concentration
of very ``metal-rich'' stars to the core of the stream can also be seen in Fig.~22 (compare
panel `a' to panel `c'). 
We stress here that these red stars need not be as metal-rich as they appear from
comparison to these ancient isochrones, due to the well-known
age-metallicity degeneracy. While the majority of other ``halo'' populations
studied in this contribution are very likely old, this is not the case for the Giant Stream.
In the spectral sample of bright stream stars obtained by \citet{ibata04}, many targets
could be identified as Asymptotic Giant Branch (AGB) stars from their spectral
features, which indicates that a fraction of these stars are of intermediate age.
This is consistent also with the deep photometric survey in a Giant Stream field undertaken
by \citet{brown06b} with the Advanced Camera for Surveys (ACS) on board the HST.
They detected a 
dominant population of age $\sim8\Gyr$, as well as a younger $\sim 5\Gyr$ component.
We will continue to label these red stars as ``metal-rich'' for the sake of brevity, though the
above caveat should be kept in mind.

The stellar populations differences can be put on
a more quantitative basis by constructing the metallicity distribution
functions for the ``stream core'' and ``outer stream'' selections; this is displayed in Fig.~28, which
shows the striking difference very clearly. 
The core of the stream clearly has a very large
fraction of red stars. Fig.~29 shows the star-counts
in different metallicity intervals as a function of $\xi$ in a $1\deg$-wide band between 
$-4\degg5 < \eta < -3\degg5$. The distribution, which peaks near $\xi \sim 1\degg5$
for ${\rm [Fe/H] > -0.4}$, becomes broader for the metallicity intervals
${\rm -1.3 < [Fe/H] < -0.4}$.

\subsection{Major axis structure}

\begin{figure}
\begin{center}
\vbox{
\includegraphics[angle=0, width=\hsize]{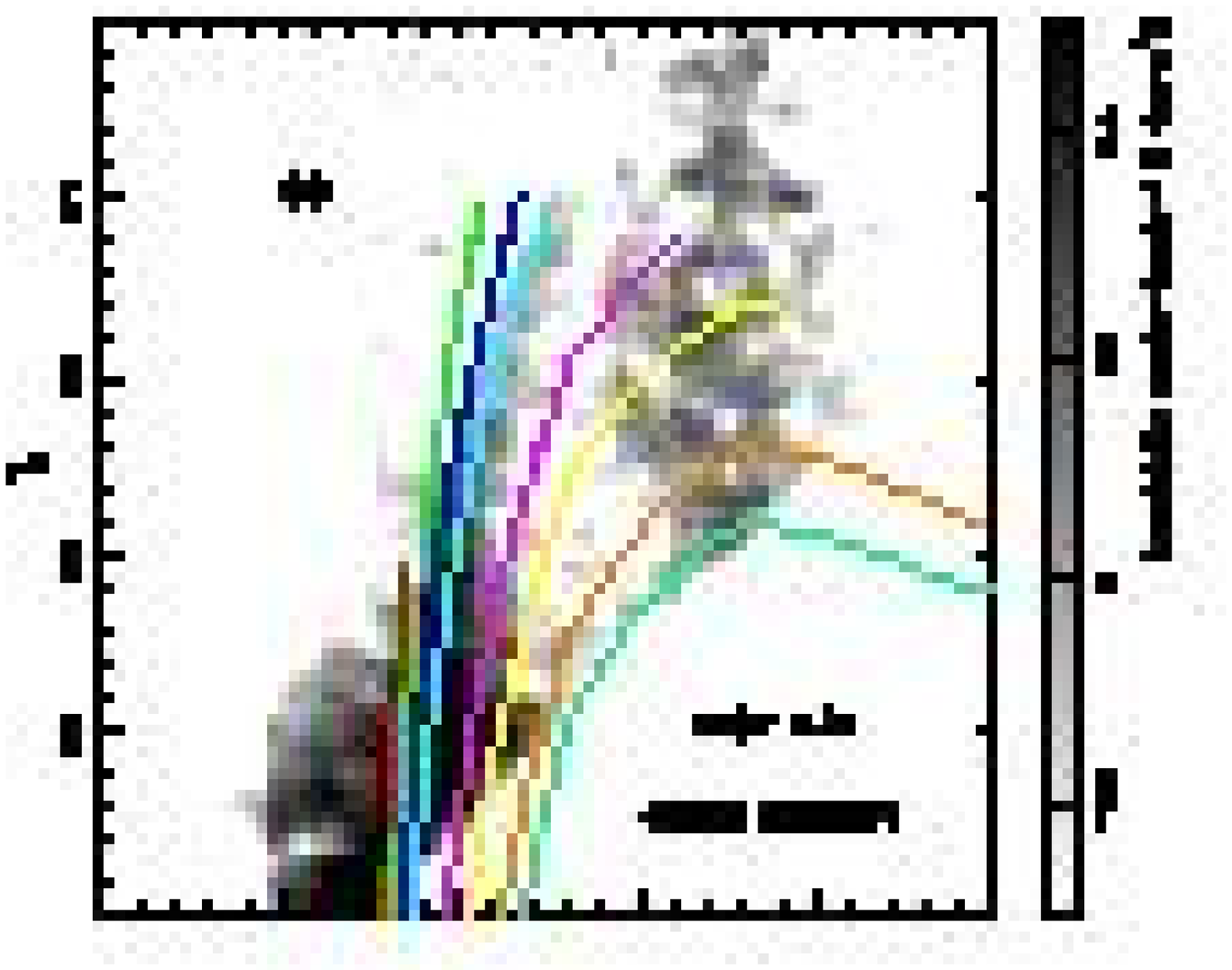}
\includegraphics[angle=0, width=\hsize]{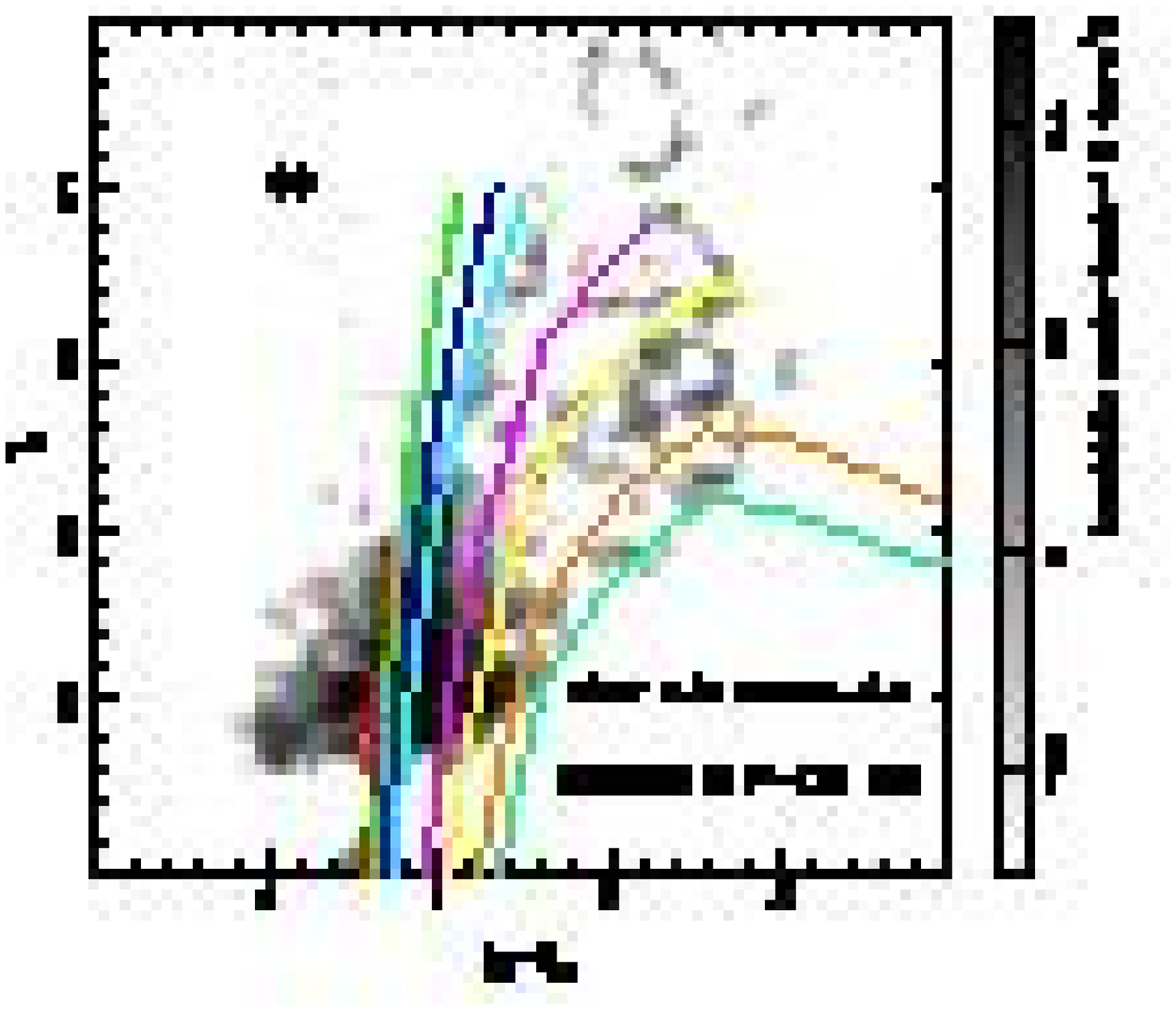}
}
\end{center}
\caption{Panel `a' shows a 
foreground-subtracted Hess diagram of the major axis diffuse population
over the region marked out with the green polygon in Fig.~23, while
panel `b' presents the foreground-subtracted Hess diagram of the minor axis stream population
over the region marked out with the blue polygon in Fig.~23.
The gray scale wedge on the right shows the count level per CMD bin of
size ${\rm 0.05 mag \times0.05 mag}$.}
\end{figure}

The faint diffuse population detected on the major axis between a projected distance of $50\kpc$ 
and $100\kpc$ (delineated with
the green polygon in Fig.~23) is a conspicuous feature of the MegaCam survey.
The average surface brightness
in this region is $\approx 31$~mag~arcsec$^2$.
The dwarf galaxy And~III lies on the edge of this region, so to avoid
contamination we remove the data from a $0\degg5$ radius circle around And~III
for the subsequent analysis.
The color-magnitude distribution of the area is displayed in panel `a' of Fig.~30, which
clearly possesses a well-populated RGB with a dominant population
of color similar to the Padova isochrones
of metallicity ${\rm [Fe/H] \sim -1.3}$. The corresponding MDF 
in Fig.~31 (red line) confirms this visual impression.

Thus despite the visual impression that the ``overexposed''
density map of Fig.~23 gives that the major axis population
merges with the Giant Stream, we find that these two
stellar populations are very different and likely unrelated.
This diffuse low-constrast feature has no clear spatial 
structure as one would expect of a stream. Indeed, it could
be the inner regions of the halo, though it appears not to be a smooth
roughly spherical structure since there is an obvious deficit of
stars at ($\xi \sim -0\degg6$, $\eta \sim -6\deg$) compared to
($\xi \sim -3\deg$, $\eta \sim -5\deg$). We refrain from
estimating the total luminosity of the structure, since we have clearly
only detected a fraction of the entire object.
Additional photometry to the North and West and possibly even
kinematics will be needed to understand this structure further.

\subsection{Distant minor axis stream `A'}

In contrast, the structure on the minor axis (delineated
with the blue polygon in Fig.~23, which covers $1.7$~deg$^2$) 
at $R \sim 120\kpc$ is much more
confined spatially as can be perceived from an inspection of the matched-filter
maps in Fig.~22. Curiously, this population (which we refer to
as stream `A' in the discussion below) has a very similar
color-magnitude distribution to that of the major axis structure,
with a dominant population again just slightly redward
of the ${\rm [Fe/H] = -1.3}$ Padova isochrone, as can
be seen in panel `b' of Fig.~30. The corresponding MDF is
compared to that of the diffuse major axis feature in 
Fig.~31. 

The structure is very faint, with an average surface brightness 
of ${\rm \Sigma_V \sim 31.7 \pm 0.2 \, mag \, arcsec^{-2}}$ . Integrating
over the blue polygon in Fig.~23, and subtracting the average
counts at this radius calculated from the ``outer halo'' region
(contained in the pink polygon), gives a total luminosity
of $L_V \sin 2.3 \times 10^6 \lsun$ ($M_V \sim -11.1$). 
If we are detecting the entirety of the 
stars in the original structure, the progenitor must have been a 
galaxy similar to the Sculptor dwarf spheroidal 
($M_V =-10.7 \pm 0.5$, \citealt{irwin95}).

\begin{figure}
\begin{center}
\includegraphics[angle=-90, width=\hsize]{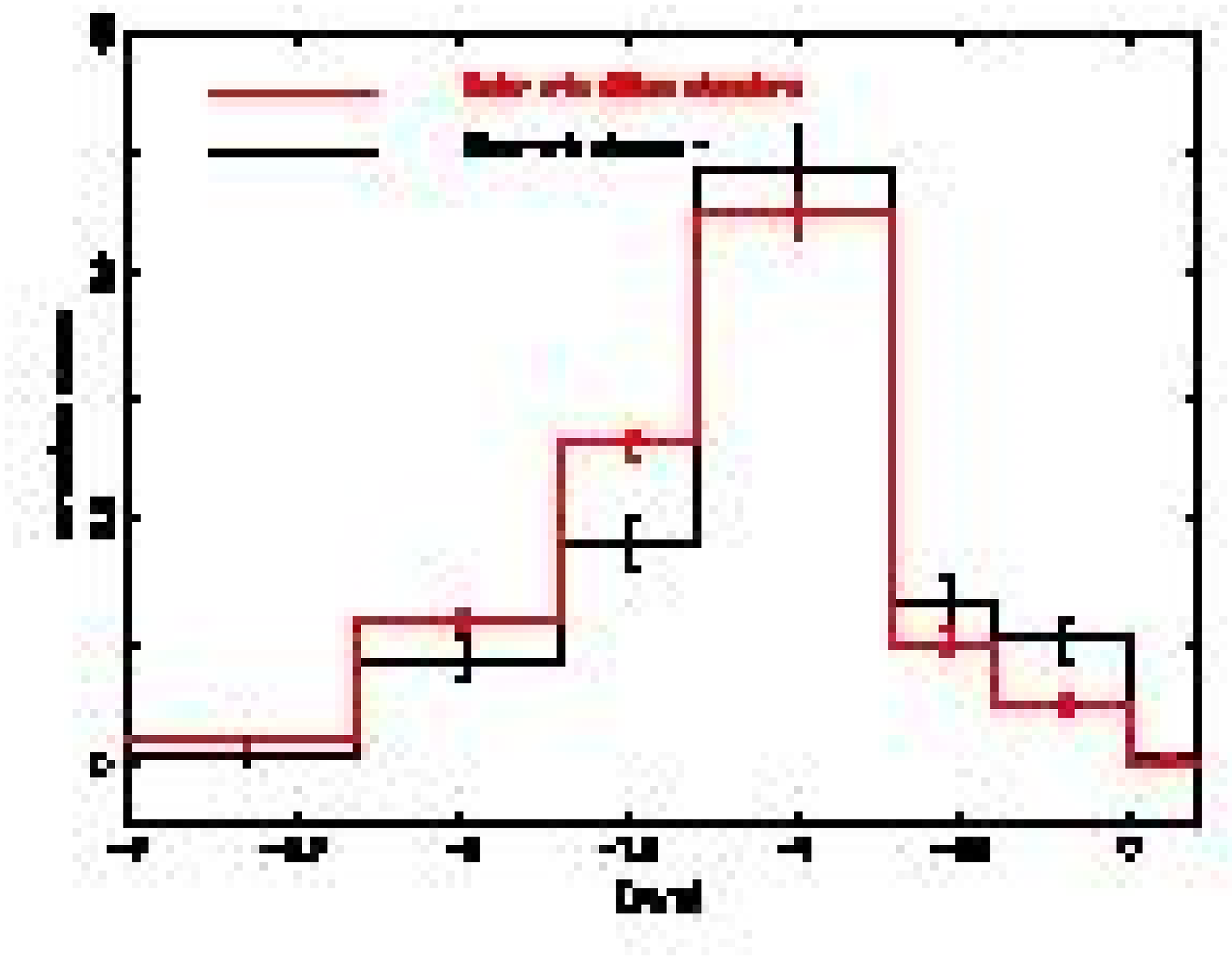}
\end{center}
\caption{The metallicity distribution function of the 
major axis diffuse structure and the
minor axis stream `A' population, as derived from the
data in Fig.~30.}
\end{figure}

\subsection{Minor axis streams at $R<100\kpc$}

Figure~32 shows a close-up map of the minor axis region
in the proximity of M31 and the Giant Stream. Here we
have used the matched-filter technique to detect
structures of metallicity in the range ${\rm -3.0 < [Fe/H] < 0.0}$,
and have chosen a grayscale representation that highlights
the three linear structures that appear almost perpendicular to the minor
axis and merge into the Giant stream. Three arrows have been 
added to the diagram to indicate the approximate location of
these stream-like features, which we denote `B', `C' and `D'
in order of increasing declination.

\begin{figure}
\begin{center}
\includegraphics[angle=-90, width=\hsize]{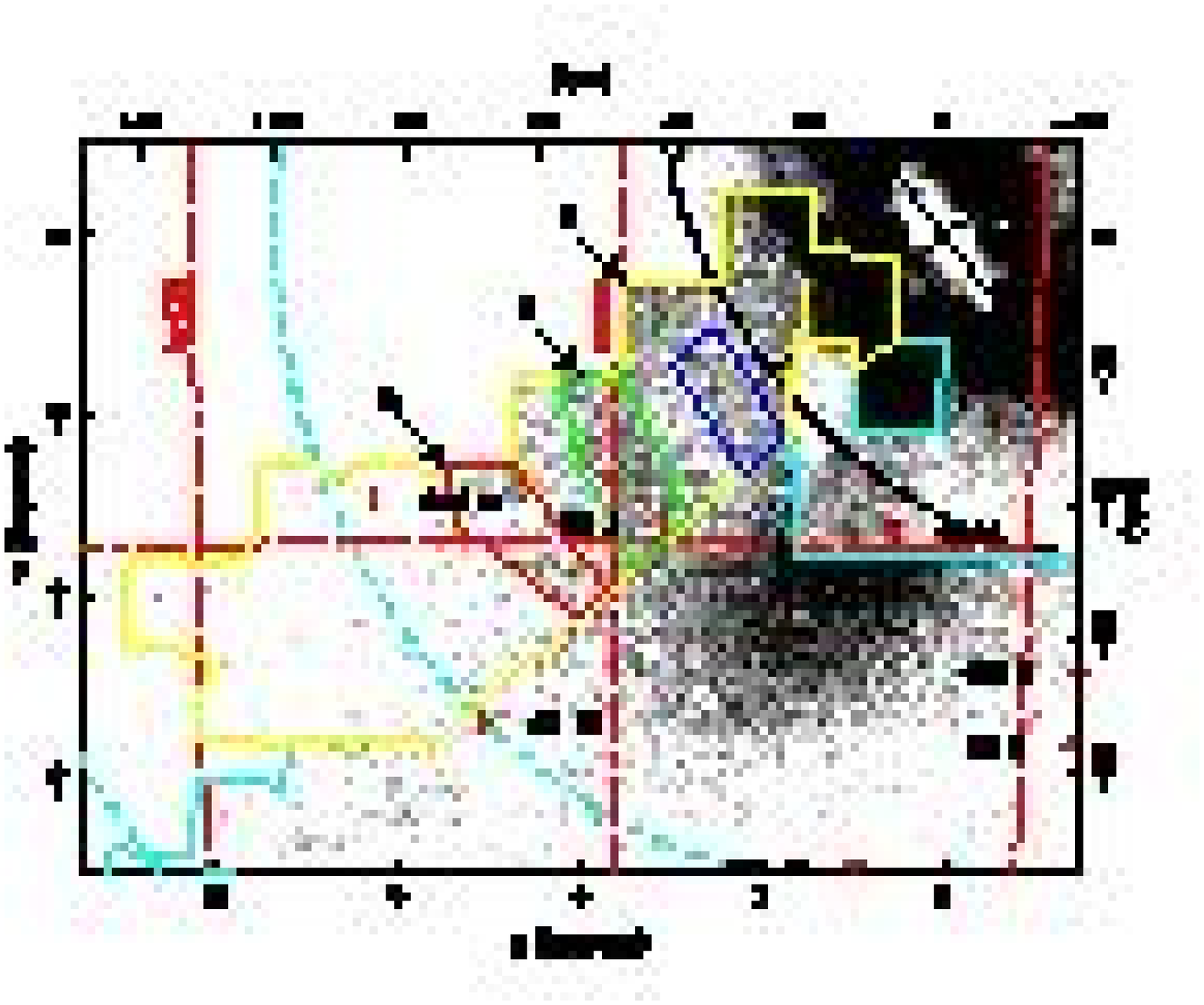}
\end{center}
\caption{Matched-filter map of the minor axis populations
with metallicity in the range ${\rm -3.0 < [Fe/H] < 0.0}$.
The map is, as before, a superposition of MegaCam and
INT photometry, the differences in depth of which account
for the discontinuous density distribution. The region surrounded
by the yellow polygon encloses the MegaCam area used
to investigate the minor axis density profile in Fig.~33.
The red, green and blue polygons enclose the stream-like
structures labeled, respectively, `B', `C' and `D'. (These
structures can be appreciated better in panel `a' of Fig.~27).}
\end{figure}

\begin{figure}
\begin{center}
\includegraphics[angle=-90, width=\hsize]{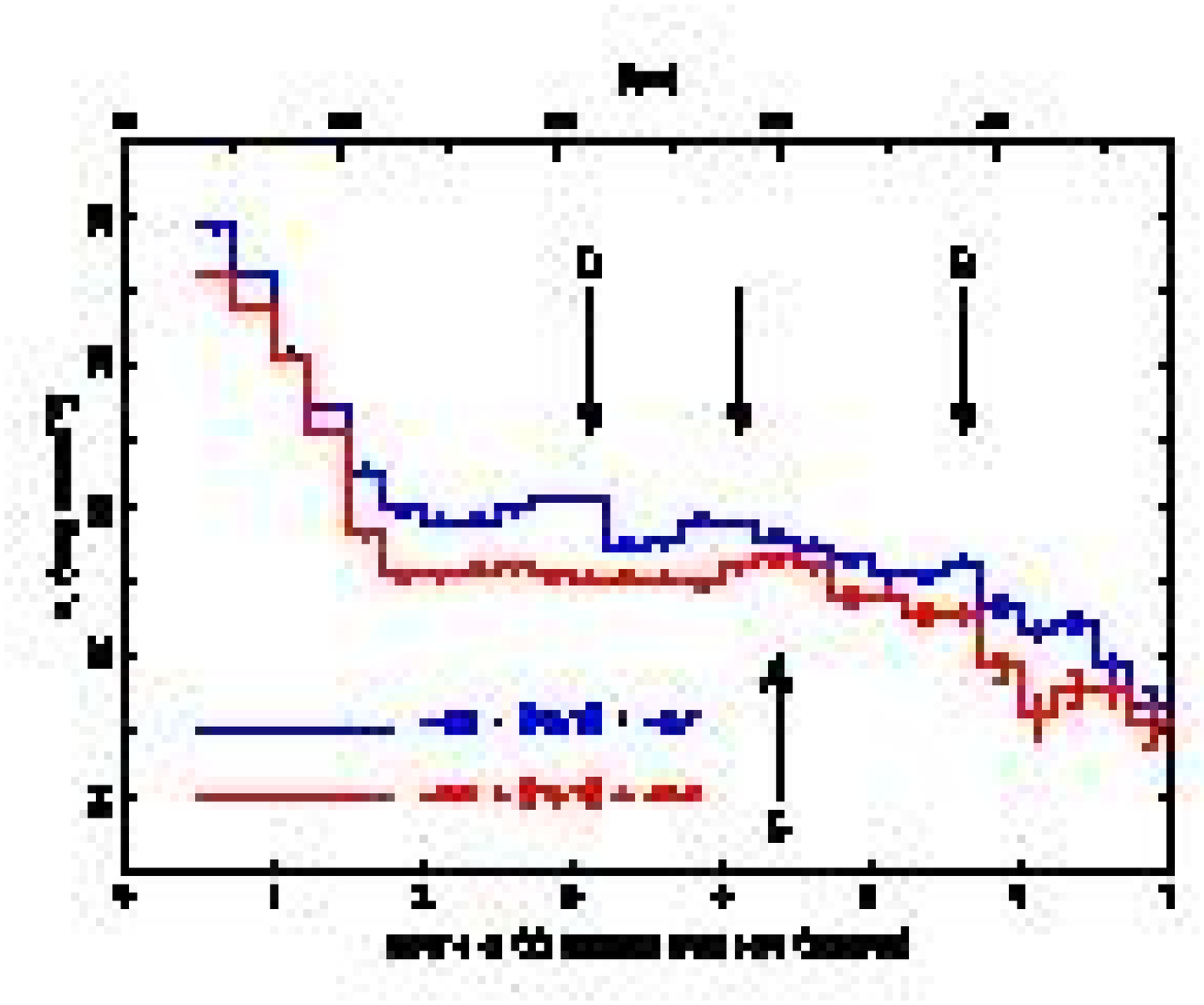}
\end{center}
\caption{The surface density profile along the minor axis,
selected from the region within the yellow polygon
in Fig.~32. The arrows point out significant peaks in the
profile. The positions of the labeled peaks correspond to
the streams seen in Fig.~32.}
\end{figure}

The nature of these streams becomes more apparent
if we investigate the color profile along the minor axis region.
We choose to remove the Giant Stream stars by selecting
only those point-sources within the yellow-line polygon in Fig.~32,
and sum stars perpendicular to the minor axis (rather than
taking radial bins) so as to enhance the density peaks.
The corresponding foreground-subtracted surface
brightness profiles are shown in Fig.~35, where the blue line
shows the metal-poor populations with ${\rm -3.0 < [Fe/H] < -0.7}$
and the red line those with ${\rm -0.7 < [Fe/H] < +0.2}$.
The foreground, as before, is estimated using the Besan{\c c}on model.
As expected, several strong peaks are detected, however,
the locations of the peaks in the metal-poor subsample do not
coincide with those of the metal-rich subsample, suggesting
very strong stellar populations differences between these 
stream-like features. 

\begin{figure}
\begin{center}
\includegraphics[angle=-90, width=\hsize]{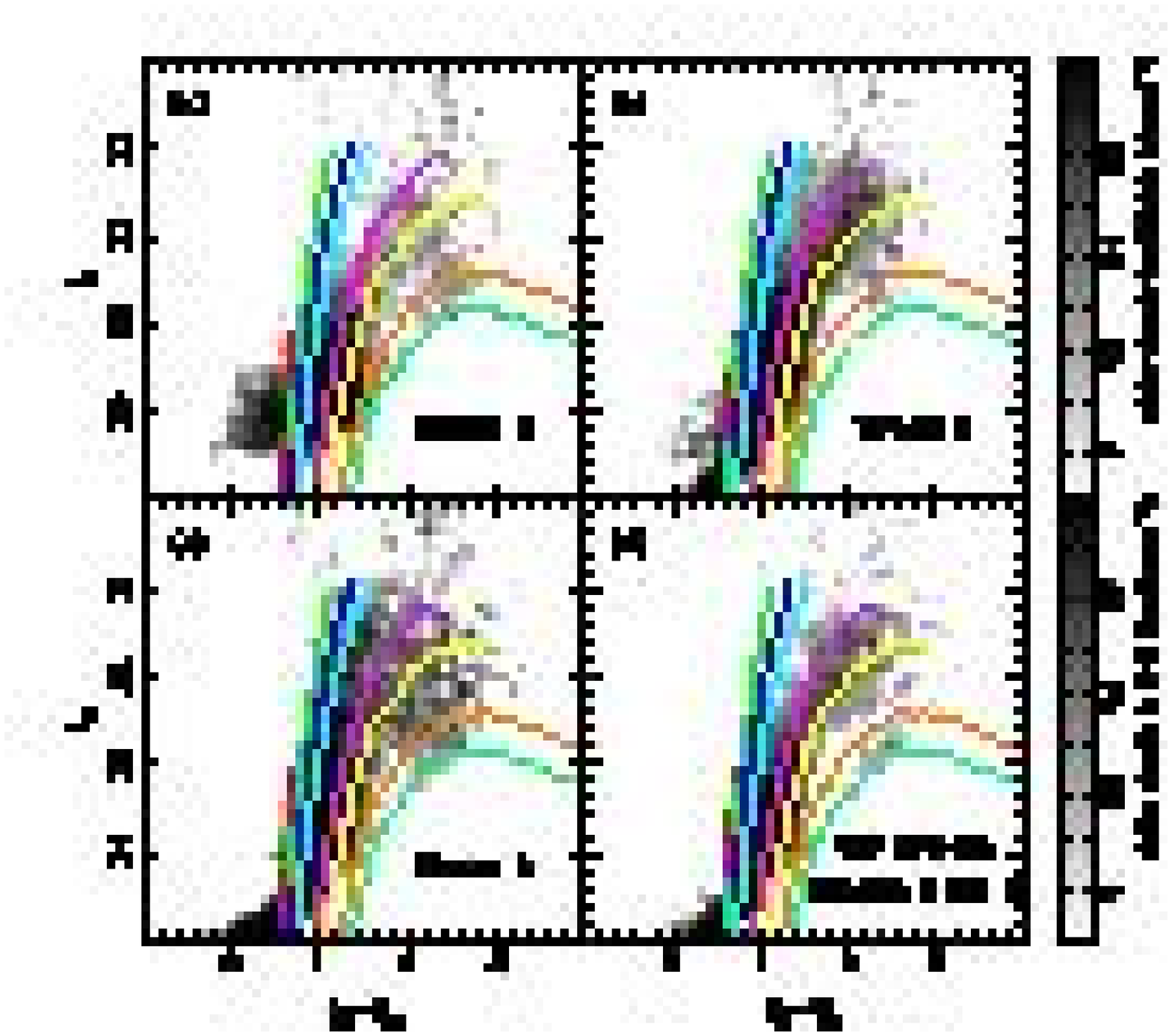}
\end{center}
\caption{Background-subtracted Hess diagrams for four 
adjacent MegaCam fields near the minor axis. 
In panels `a', `b', `c' and `d' we display the data for fields 
13, 14, 23 and 24, respectively.}
\end{figure}

\begin{figure}
\begin{center}
\includegraphics[angle=-90, width=\hsize]{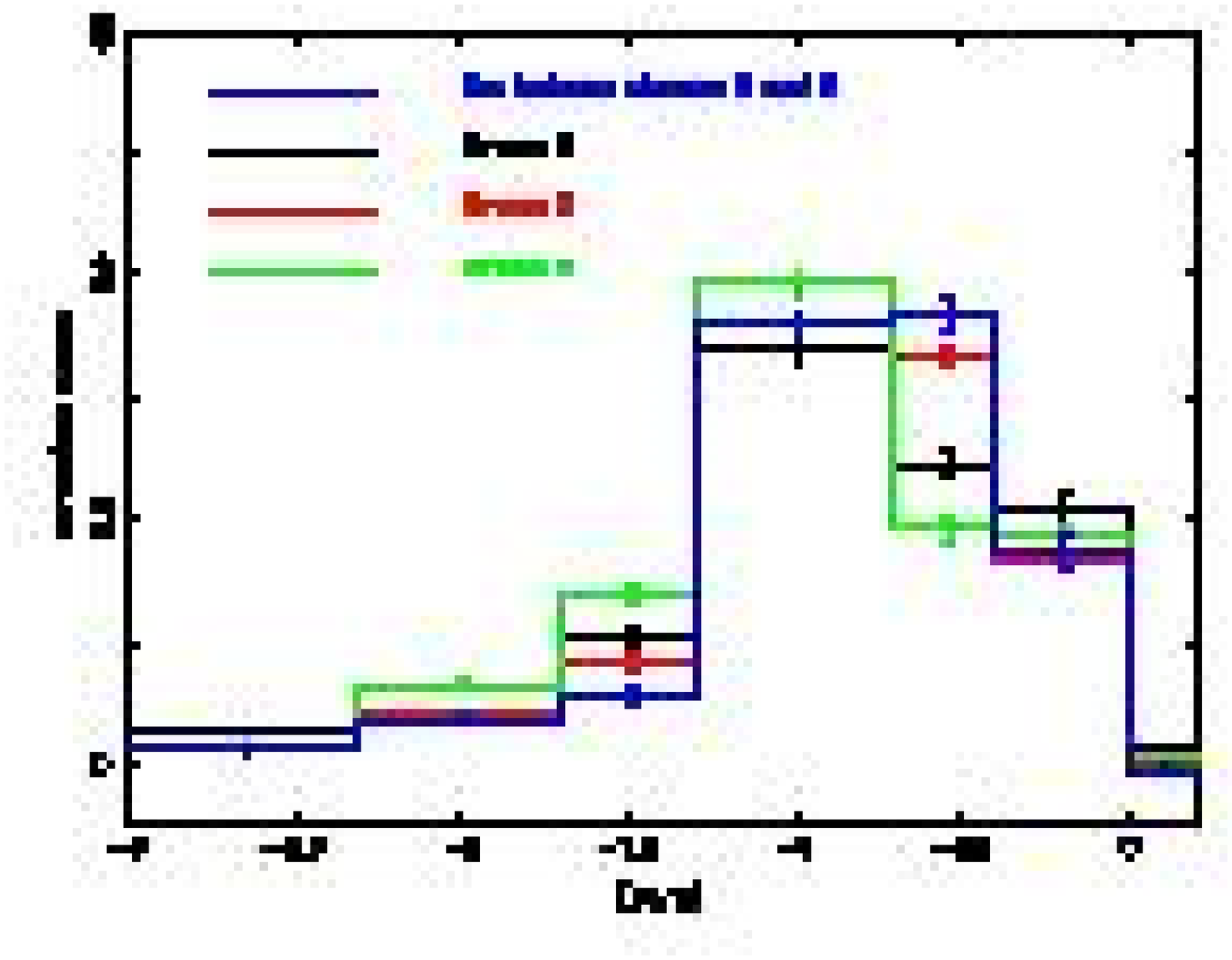}
\end{center}
\caption{The MDF determined from the four fields of Fig.~36:
13 (black), 14 (red), 23 (green) and 24 (blue).}
\end{figure}

This deduction is borne out by the
variations in the color-magnitude distributions in adjacent spatial
locations. In Fig.~34 we display the Hess diagrams
of the stream-like structures enclosed within
the green, red and blue polygons of Fig.~32, and also
show the stellar population between streams `B' and `C'.
The corresponding MDFs are given in Fig.~35.
These data show that stream `D' is a relatively metal-poor
structure, while stream `C' is predominantly metal-rich.
Curiously, the population contained within gap between
streams `B' and `C' has a narrow range of metallicity
and is metal-rich.

These stream-like structures overlap along the line of
sight (which is why we chose not to extend the stream `D'
spatial selection polygon in Fig.~32 up to the northeastern end of
the survey region). A spectacular example of this can be seen
in Fig.~36, which shows the CMD of MegaCam field 14, where streams `C' and `D'
cohabit over essentially the entirety of the field. 

\begin{figure}
\begin{center}
\includegraphics[angle=-90, width=\hsize]{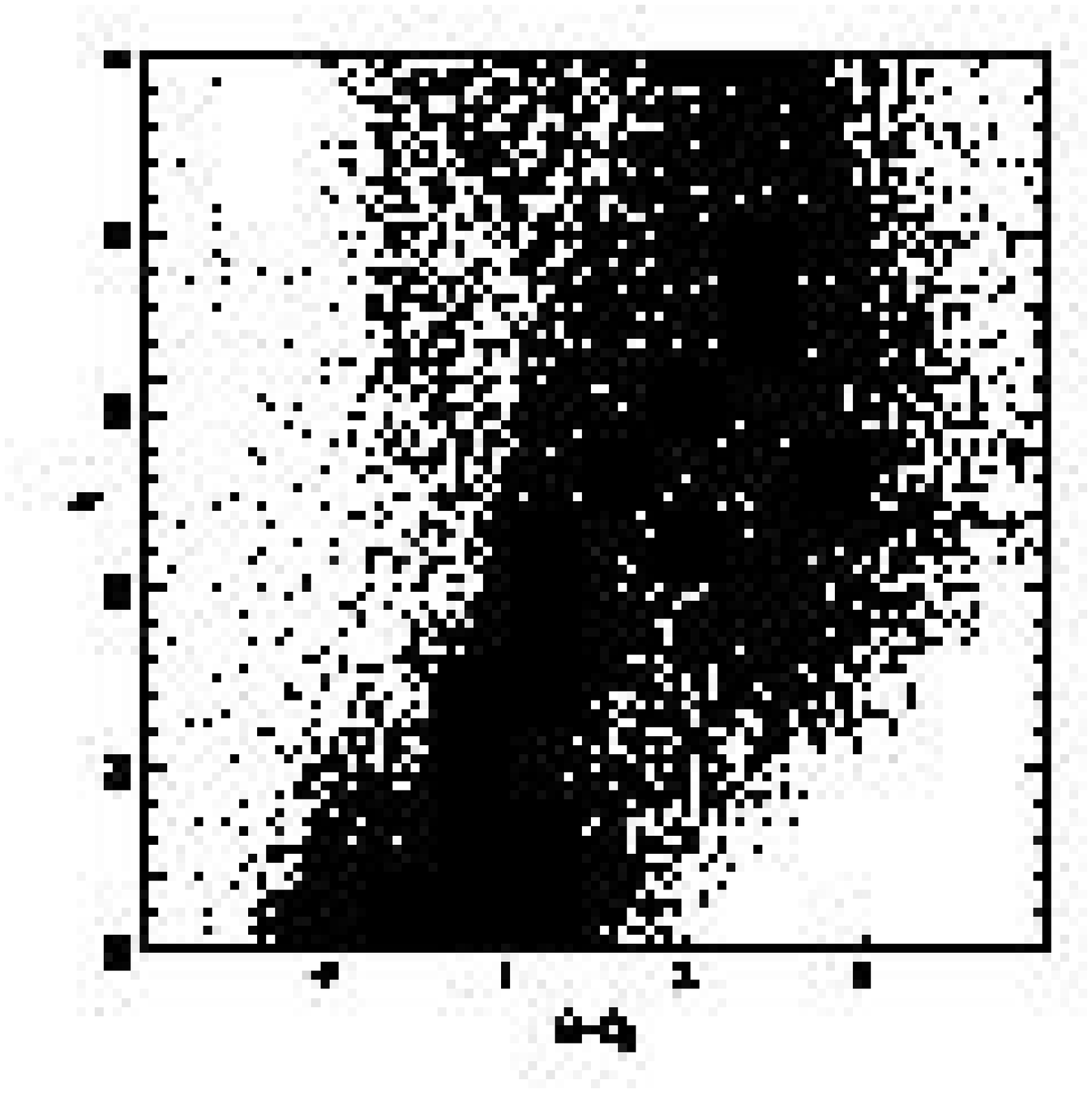}
\end{center}
\caption{The color-magnitude diagram for field 14, showing
the presence of two co-spatial populations with very different
RGB tracks.}
\end{figure}

Thus, although these stream-like structures appear to merge spatially with
the Giant Stream, such that it is tempting at first sight to associate
them to that huge structure, their stellar populations properties are
so different both from each other and from the Giant Stream, that this
proposition is untenable. 

In the present survey these streams or stream-like structures are
clearly truncated at the Eastern edge of the dataset, so it is impossible 
to determine their full extent or nature. Instead, we obtain a first and very 
rough estimate of their luminosities by integrating within the 3 stream
polygons in Fig.~32. In this way we estimate that stream `B',
which lies at $R \sim 80\kpc$, has a luminosity within the red polygon of
$\sim 1.0 \times 10^7 \lsun$; stream C at $R \sim 60\kpc$ has $\sim 1.4 \times 10^7 \lsun$
in the green polygon;
while stream `D' at $R \sim 40\kpc$ has $\sim 9.5 \times 10^6 \lsun$ in the blue polygon.
In estimating these luminosities we have ignored the complex background
in this region. Nevertheless, these estimates indicate that the
progenitors of the streams were sizable dwarf galaxies, likely more
luminous than the Fornax dSph. We note that 
the extended globular cluster \citep{huxor05} EC4 \citep{mackey07} lies within
or superimposed on stream `C'.

\section{The Outer halo}

The primary reason for undertaking this survey was initially to investigate the large-scale
structure of the halos of M31 and M33, and to some extent the substructures
discussed above are a hindrance for this purpose. In particular, we had not expected
the Giant Stream to be as extended and polluting of the inner halo as it turned out
to be, and the various ``contaminating'' streams along the minor axis were
a surprise, as we had chosen those fields from the shallower INT survey to probe
the surface density profile of the ``clean'' inner halo.

However, there is a relatively empty region of the survey free from obvious
substructures towards the South-west. This $\sim 30$~deg$^2$ region previously surrounded
with a pink polygon in Fig.~23, is reproduced in Fig~37, where we have converted
the counts of stars in the various metallicity ranges shown into an equivalent
surface brightness. The four white pixels within the polygon in the diagram are pixels discarded 
from the analysis as they contain the dwarf galaxies And~XI, XII, XIII and XVI.

\begin{figure}
\begin{center}
\vbox{
\includegraphics[angle=-90, width=\hsize]{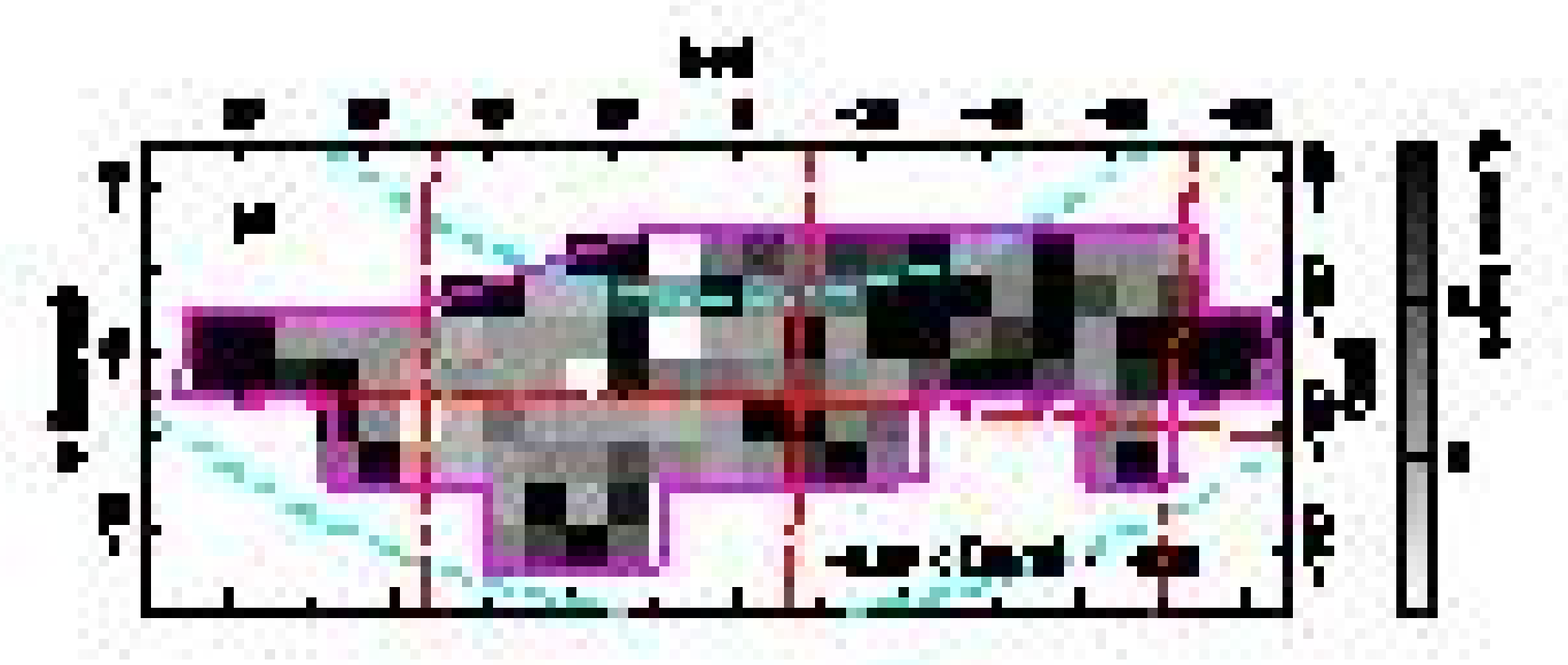}
\includegraphics[angle=-90, width=\hsize]{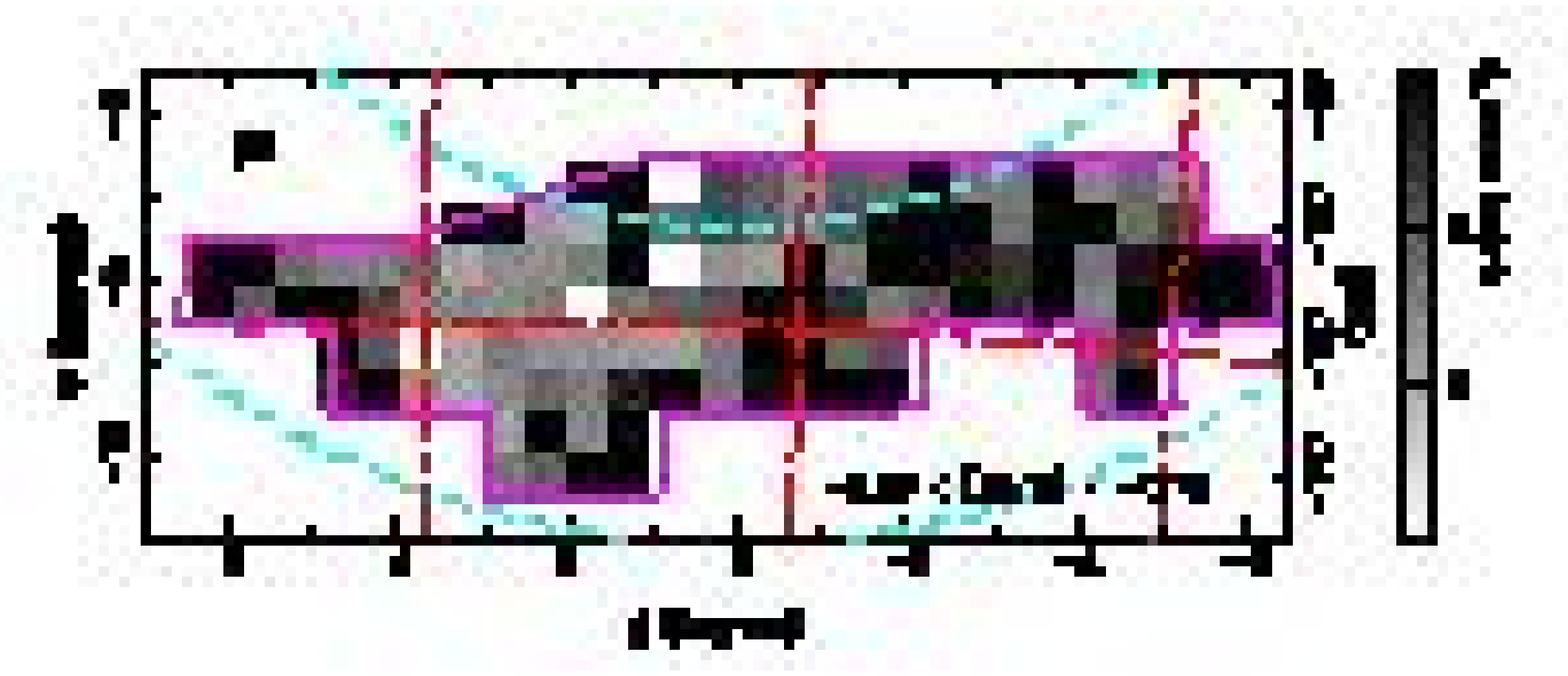}}
\end{center}
\caption{Background-subtracted maps of the equivalent surface brightness in the 
outer halo region. Panel `a' shows stars in the metallicity range
${\rm -3.0 < [Fe/H] < +0.2}$, while panel `b' is restricted to
 ${\rm -3.0 < [Fe/H] < -0.7}$, a range which suffers much less from 
uncertainties in the background correction.}
\end{figure}

The equivalent mean surface brightness of the outer halo stars for the full ${\rm [Fe/H]}$
range given in panel `a' of Fig.~37 is ${\rm \Sigma_V = 33.0 \pm 0.05 \, mag \, arcsec^{-2}}$,
where the uncertainty is calculated using Poisson statistics, assuming no
uncertainty in the background subtraction. 
Note that a 2\% error in the subtraction (the average difference of the residuals
between the Galactic model and observed Galactic disk found in Fig.~14), 
will incur a $0.25$~mag systematic error.
However, the rms scatter
in the pixel values in Fig.~37 (calculated in counts and then converted 
into magnitudes) is $1.1$~mag; for this calculation, 
we only took into account those (128) pixels in Fig.~37 for which the 
surface area correction was less than 10\%. 
The fact that this rms scatter is larger 
than the 0.2~mag random uncertainty expected from Poisson 
uncertainties in the total measured 
star counts, could be due to an intrinsic lumpiness in the star distribution
on the $0\degg5 \times 0\degg5$ scale of the pixels in Fig.~37, but
we consider it likely that it is largely due to slight variations
in observing conditions between fields, and slight variations of image quality
over the camera.
Panel `b' gives the map for the metal-poor range ${\rm -3.0 < [Fe/H] < -0.7}$,
which has the advantage of reducing the amount of residual Galactic contamination.
The equivalent mean surface brightness for this selection is 
${\rm \Sigma_V = 33.7 \pm 0.08 \, mag \, arcsec^{-2}}$. It is pertinent to point
out here that the six fields chosen to probe the background all lie within this outer
halo region, indeed they are the fields closest to the outer dashed circle segment marking
a projected radius of $150\kpc$ (cf. Fig.~18).

The Hess diagram for the outer halo region is shown in Fig.~38, with the foreground
subtracted as before. Though noisy, a low contrast RGB population can identified 
that is strongest between the ${\rm [Fe/H] = -1.7}$ and ${\rm [Fe/H] = -0.7}$ isochrones.
Clearly for ${\rm [Fe/H] > -0.7}$, residual foreground subtraction errors dominate
the data.

\begin{figure}
\begin{center}
\includegraphics[angle=0, width=\hsize]{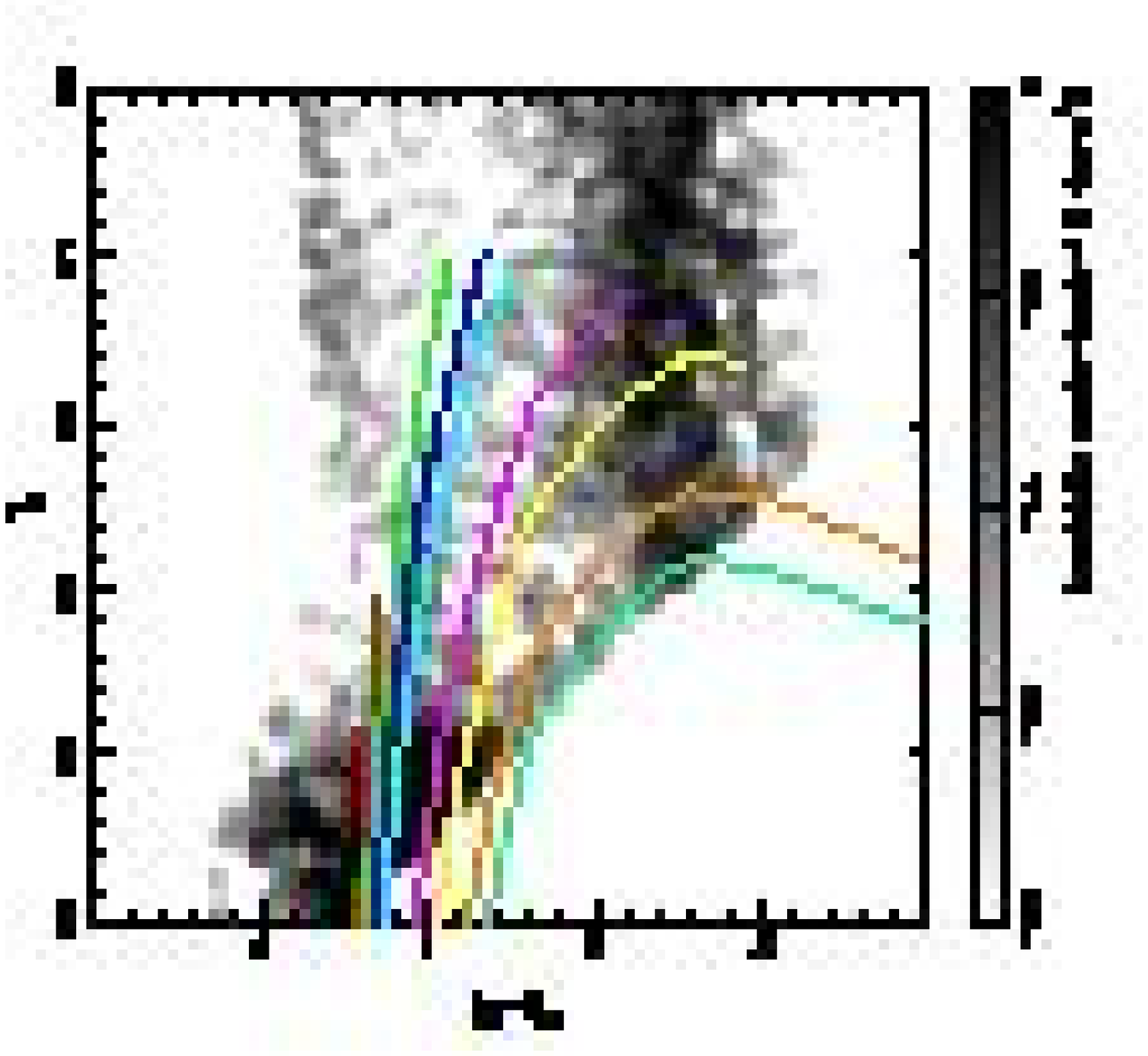}
\end{center}
\caption{Foreground-subtracted Hess diagram of the outer halo region 
shown in Fig.~37.}
\end{figure}

\section{Halo profiles}

Before presenting the radial profiles of the stellar population 
present in the survey,
we first investigate whether our conversion from star-counts
to ``equivalent surface brightness'' (first presented in \S4.1)
yields consistent results with previous studies. To this end 
we compare our measurements to those of \citet{irwin05}
who analyzed the profile along the minor axis of M31. 
We attempted to 
reproduce as closely as possible the spatial selection chosen 
by \citet{irwin05}, a band between $\pm 0\degg5$ of the minor
axis (see their Fig.~1). The MegaCam survey covers 
most, but not all, of this area (there is a small gap near 
$\xi \sim 1\degg5$, $\eta \sim -1\degg5$, as can be seen in Fig.~32, for example).
In panel `a' of Fig~39 the black dots mark the surface brightness 
measurements from integrated light by \citet{irwin05},
while the blue histogram shows the MegaCam profile. 
Though the measurements from integrated light end
at $R=0\degg5$, just before the beginning of the MegaCam
survey, there is good consistency between these two profiles.

\begin{figure}
\begin{center}
\includegraphics[angle=0, width=\hsize]{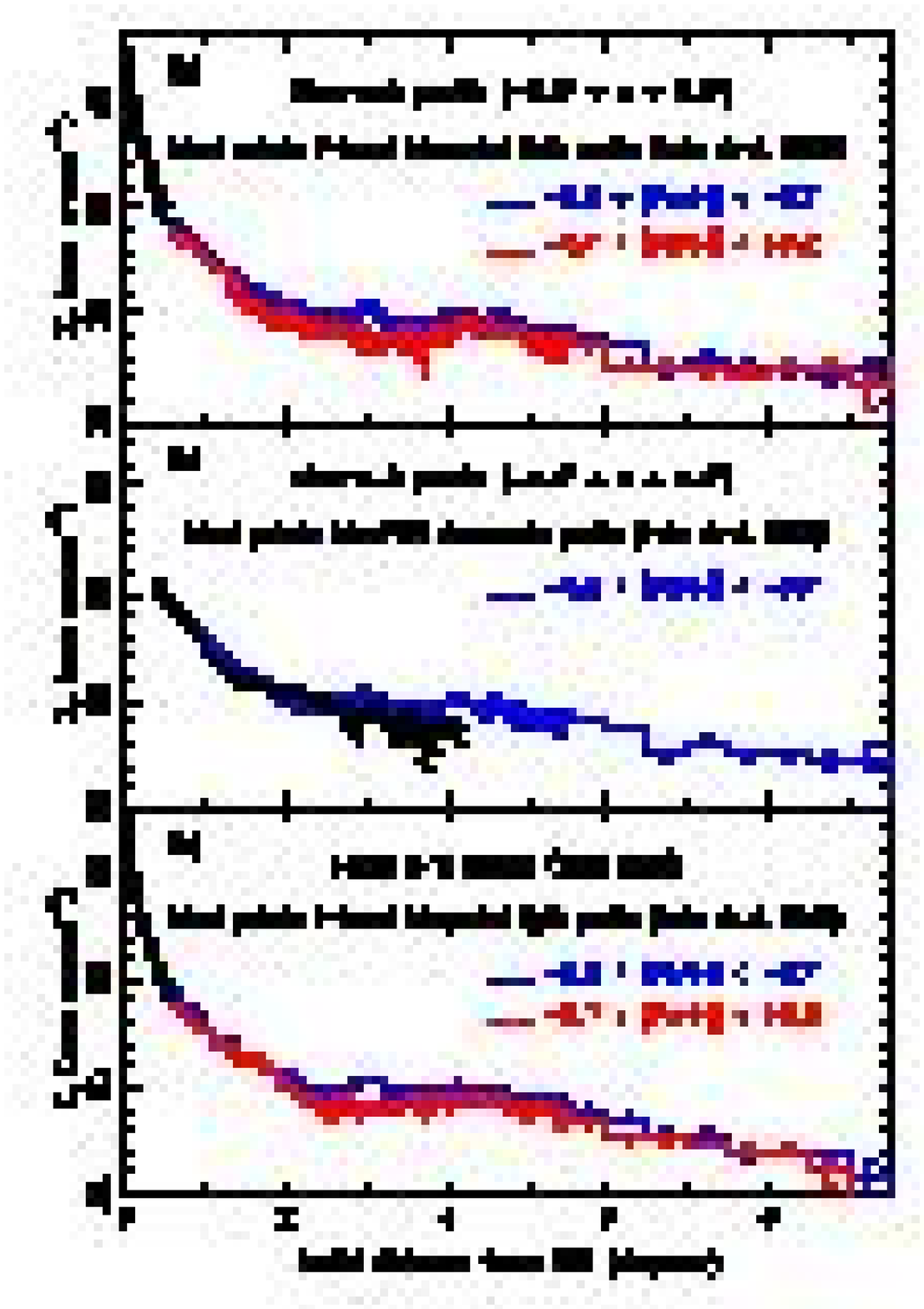}
\end{center}
\caption{Panel `a' compares the surface brightness
profile from integrated light (black points) deduced
from the INT survey by \citet{irwin05}, with the 
converted star-counts derived from the present survey
in a $\pm 0\degg5$ band around the minor axis of M31.
The color variations as a function of radius are attributable
to substructures with different stellar populations intersecting
this area. The blue RGB star-counts profile of \citep{irwin05}
is compared in panel `b' to the metal-poor MegaCam selection in the
same spatial region. The differences between these
curves at $r < 2\degg5$ are likely due to the fact that the
two stellar selections, though similar, are not identical. For
$r > 2\degg5$ the \citet{irwin05} profile decreases sharply
due to over-subtraction of foreground contaminants in that
analysis. Panel `c' is similar to panel `a' but the profile
is derived over a wider minor axis area (contained within the
yellow polygon of Fig.~32).}
\end{figure}

In panel `b' the black dots now show the star-counts profile 
of the blue RGB selection of \citet{irwin05} converted into 
an equivalent surface brightness. 
This V-band profile was determined from a color cut
in the INT (V,i) system, designed to select metal poor stars.
Here we have chosen not to adopt that approach, relying
on interpolation between Padova isochrones. This 
difference in stellar populations must account for some of the 
differences between the two profiles. However, the shape of the 
\citet{irwin05} profile at large radius drops rapidly unlike
the MegaCam profile derived from the same region. 
This effect is due to the foreground
subtraction method chosen by \citet{irwin05}, who
selected fields within $4\deg$ of M31 to probe
and remove the contaminating foreground populations. 
With hindsight this is clearly not appropriate given that the 
present MegaCam data shows that the halo is very extended,
and has a rather flat profile. However, out to $R \sim 2\degg5$
the INT and MegaCam profiles agree very well.

To complement the profiles derived from the narrow
$1\deg$ band shown in panels `a' and `b', we present
in panel `c' the surface brightness profile
derived from data over the wider minor axis area 
enclosed within the yellow polygon in Fig.~32. This is
of course less noisy at large radius. The various peaks
in the profile correspond to the locations of the 
stream-like structures discussed above.

Having shown that the minor axis profile is consistent
with previous measurements in the inner regions (for $R < 2\degg5$),
we now proceed to determine the radial trend of the halo populations
over the full survey area. The large amount of substructure detected
in the maps above means that the result we find will depend
sensitively on what populations we decide to include or reject in the
analysis. We therefore adopt a pragmatic approach, taking in turn various
population selections, which may be
helpful when comparing these data to cosmological simulations.

\begin{figure}
\begin{center}
\includegraphics[angle=-90, width=\hsize]{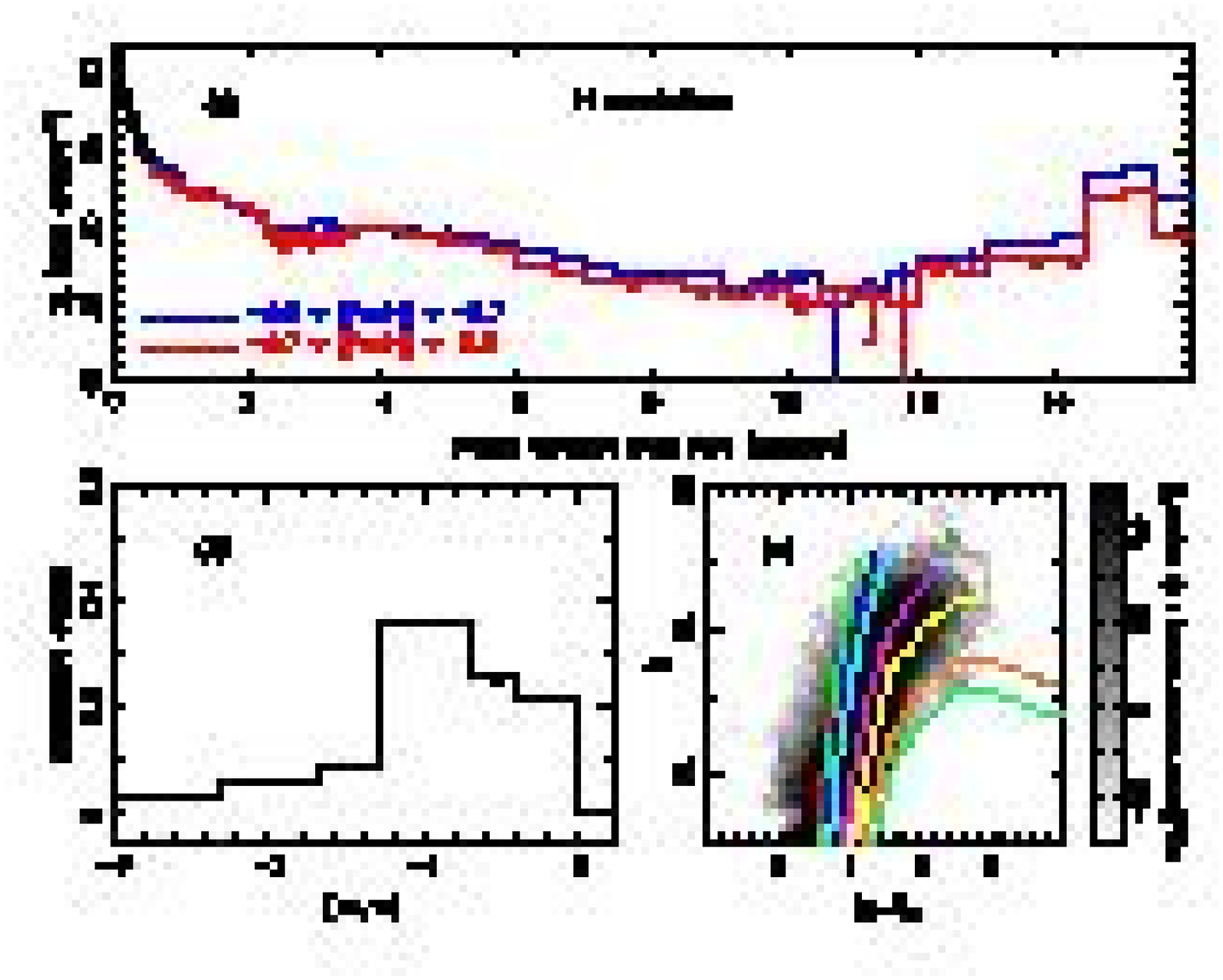}
\end{center}
\caption{Panel `a' shows the radial profile of all the stellar structures
present in the MegaCam survey to a limiting magnitude of
$i_0=23.5$. The points show the V-band
surface brightness profile from integrated light, as derived
by \citet{irwin05}. The profiles colored in blue and red show the
MegaCam data for the metal-poor and metal-rich selections,
respectively. Panel `b' gives the ``metallicity'' distribution of this
entire region (and down to $i_0=23.5$), derived from the
stellar color by comparison to $10\Gyr$ old 
Padova isochrone models. The corresponding background-subtracted
Hess diagram is shown in panel `c'.}
\end{figure}

\begin{figure}
\begin{center}
\includegraphics[angle=-90, width=\hsize]{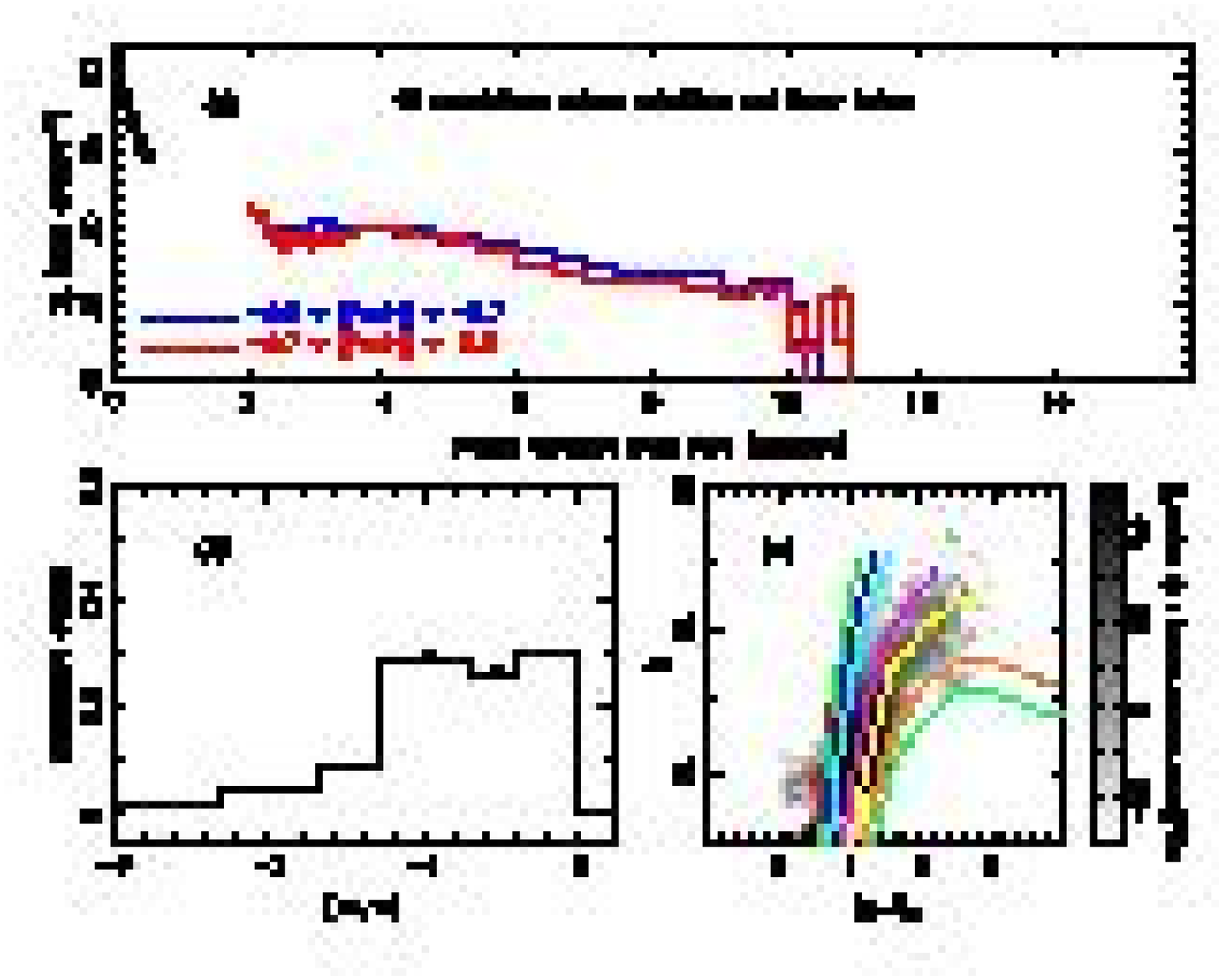}
\end{center}
\caption{As Fig.~40, but removing the inner halo of M31
out to $r=2\deg$ and that of M33 out to $r=5\deg$,
as well as all known satellites. For the satellites
we excised data within $0\degg5$ of And~II and III, and 
within $0\degg2$ for the remaining satellites in the MegaCam
region.}
\end{figure}

We begin by showing the profile of all stellar populations present
in the survey down to a limiting magnitude of $i_0=23.5$
(Fig.~40, panel `a'). The counts in each radial bin are derived
from averaging over the entire azimuthal coverage of the MegaCam survey, 
with the foreground subtracted using the Besan{\c c}on model.
The V-band surface brightness
profile measured from the integrated light in the 
INT data \citep{irwin05} is reproduced here 
with black dots. The profiles measured from the present MegaCam
data are displayed in blue for 
${\rm -3.0 < [Fe/H] < -0.7}$, and in red for
${\rm -0.7 < [Fe/H] < 0.0}$. As discussed above, we expect the metal-rich
selection to be compromised by foreground correction uncertainties,
though this is likely only to be an issue at low surface brightness levels
where the signal is small. 

This sample contains all populations -- including satellites and streams,
so the profile is not obvious to interpret. However, it transpires that the peak 
near $4\deg$ is due to the presence of the Giant Stream at that location.
The metal-rich nature of that structure enhances the metal-rich (red)
profile in the region between $3\degg5 < R < 6\deg$, giving the impression
that the halo becomes more metal poor at large radius. 
This is, however, purely an artifact of the presence of that one stream.

A clear radial decrease is detected in the surface brightness of this
combined population up to a distance of about $R \sim 10\deg$, where
it begins to rise again towards M33. Given that M31 and M33 lie
at approximately the same Heliocentric distance \citep{mcconnachie05},
this is a spectacular demonstration that the stellar halos of the two
galaxies actually pass through each other like ghostly bodies.

Panels `b' and `c' show the MDF and background subtracted Hess 
diagram of these stellar populations. In this situation the distributions
are overwhelmingly dominated by stars close to M31 and in 
the disk of M33.

In Fig.~41 we repeat this analysis, after removing large
areas around the inner halos of the two main galaxies and
their known satellites. Clearly the tiny bound satellites found within 
the MegaCam survey do not have a significant
effect on the global surface brightness profile. However
the Giant Stream does have a large effect, and the MDF
and Hess diagram in panels `b' and `c' are dominated by 
that population (compare to panels `b' and `c' of Fig.~27).

\begin{figure}
\begin{center}
\includegraphics[angle=-90, width=\hsize]{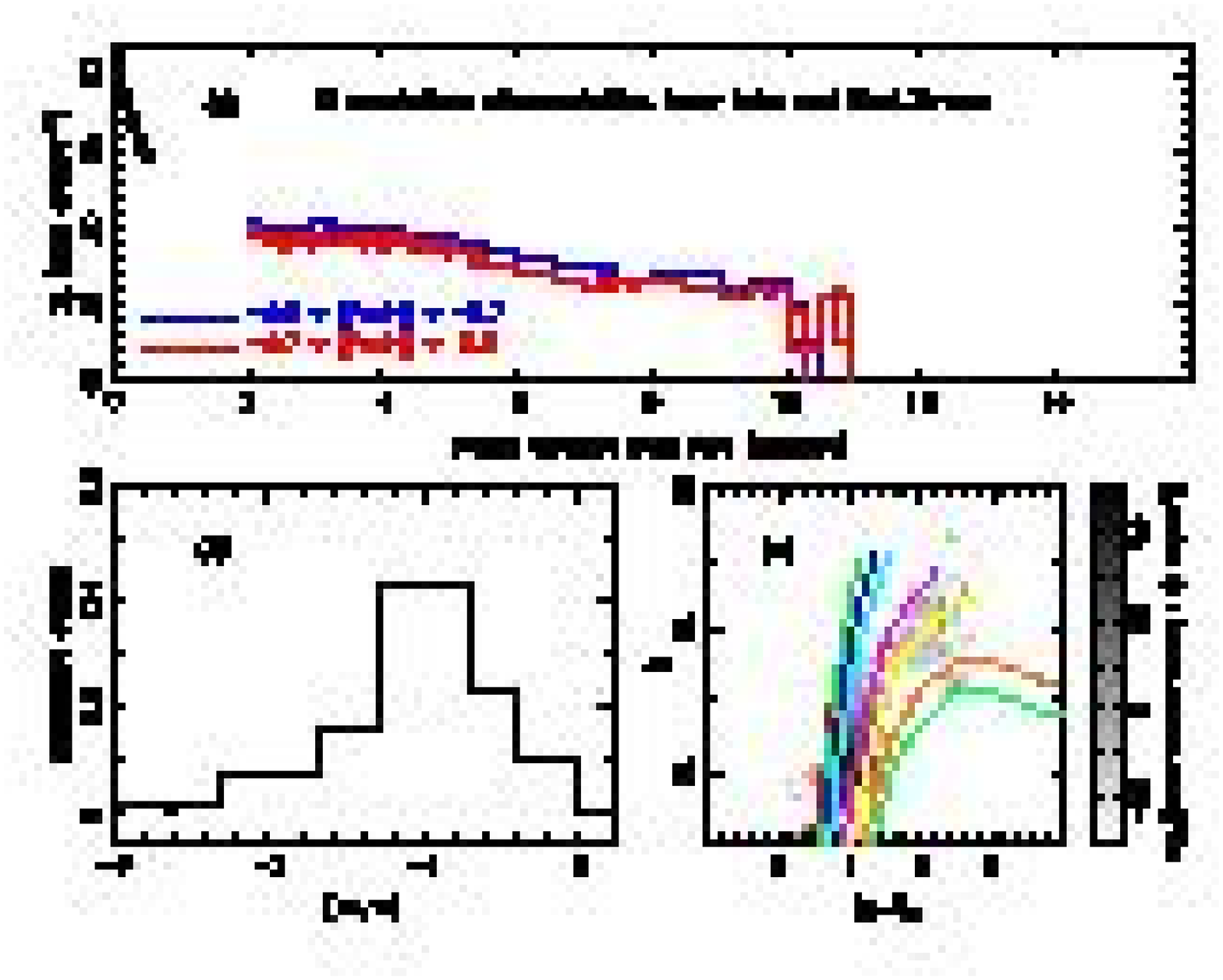}
\end{center}
\caption{As Fig~41, but with the additional removal of the Giant Stream,
as contained within the red polygon of Fig.~23.}
\end{figure}

\begin{figure}
\begin{center}
\includegraphics[angle=-90, width=\hsize]{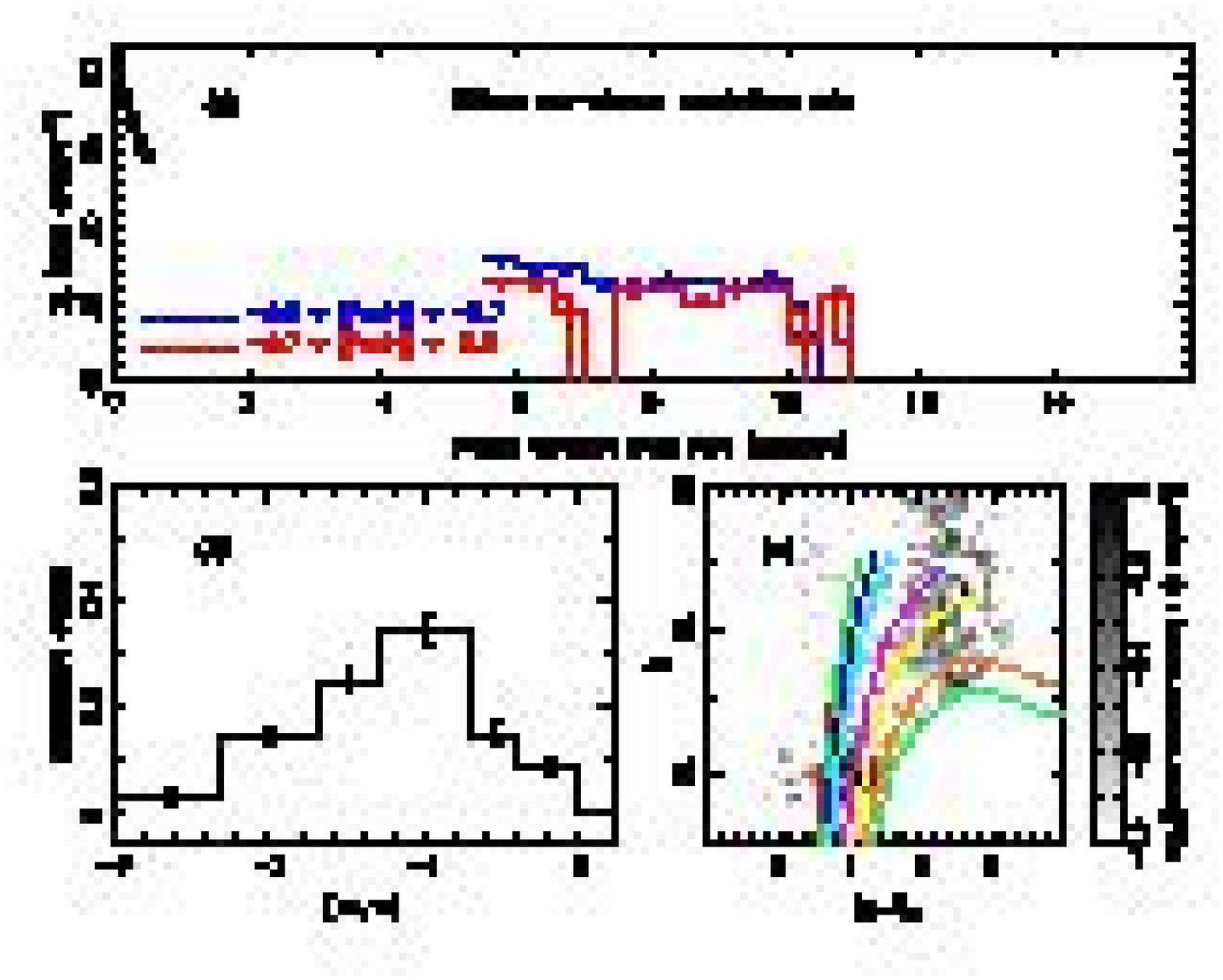}
\end{center}
\caption{As Fig.~42, but containing only diffuse stellar populations
not identified as streams.}
\end{figure}

Removing the Giant Stream in addition to the inner halo and
bound satellites, reveals a fascinating profile (Fig.~42).
We find a very flat decrease as a function of radius, 
visually resembling an exponential profile in the log-linear
diagram of panel `a'.  
Moving outwards from $2\deg$ to $5\degg5$ the offset
between the metal-poor and the metal-rich profile remains approximately
constant. This data comes primarily from the minor axis
area previously presented in Fig~34 (the region within the yellow polygon). 
At a radius of $R \sim 5\degg5$ the metal-rich population drops
significantly, and again appears to mimic the metal-poor
profile out to $R \sim 7\deg$. The fact that the
metal-poor and metal-rich profiles track each other
fairly well in each of these two radial ranges, suggests that the
mix of stellar populations present does not change considerably
over each range.
Whether the drop at $R \sim 5\degg5$ reflects a real change in stellar populations at this radius 
($75\kpc$) remains to be confirmed.

Finally, we show in Fig.~43 the result of removing all of the identified structures
from the survey, leaving only the widely-distributed diffuse population behind. For this
we have excised the inner halos of M31 and M33, as well as the satellites
as detailed previously. We have also removed the areas within the 
red, green and blue polygons in Fig.~23, and the region contained within
the yellow polygon in Fig.~32. 
As can be appreciated from panel `c' of Fig.~43, we cannot have much confidence in the metal-rich 
selection, and correspondingly the red profile of panel `a' is very uncertain.
However, the metal-poor profile appears fairly smooth.

\begin{figure}
\begin{center}
\includegraphics[angle=-90, width=\hsize]{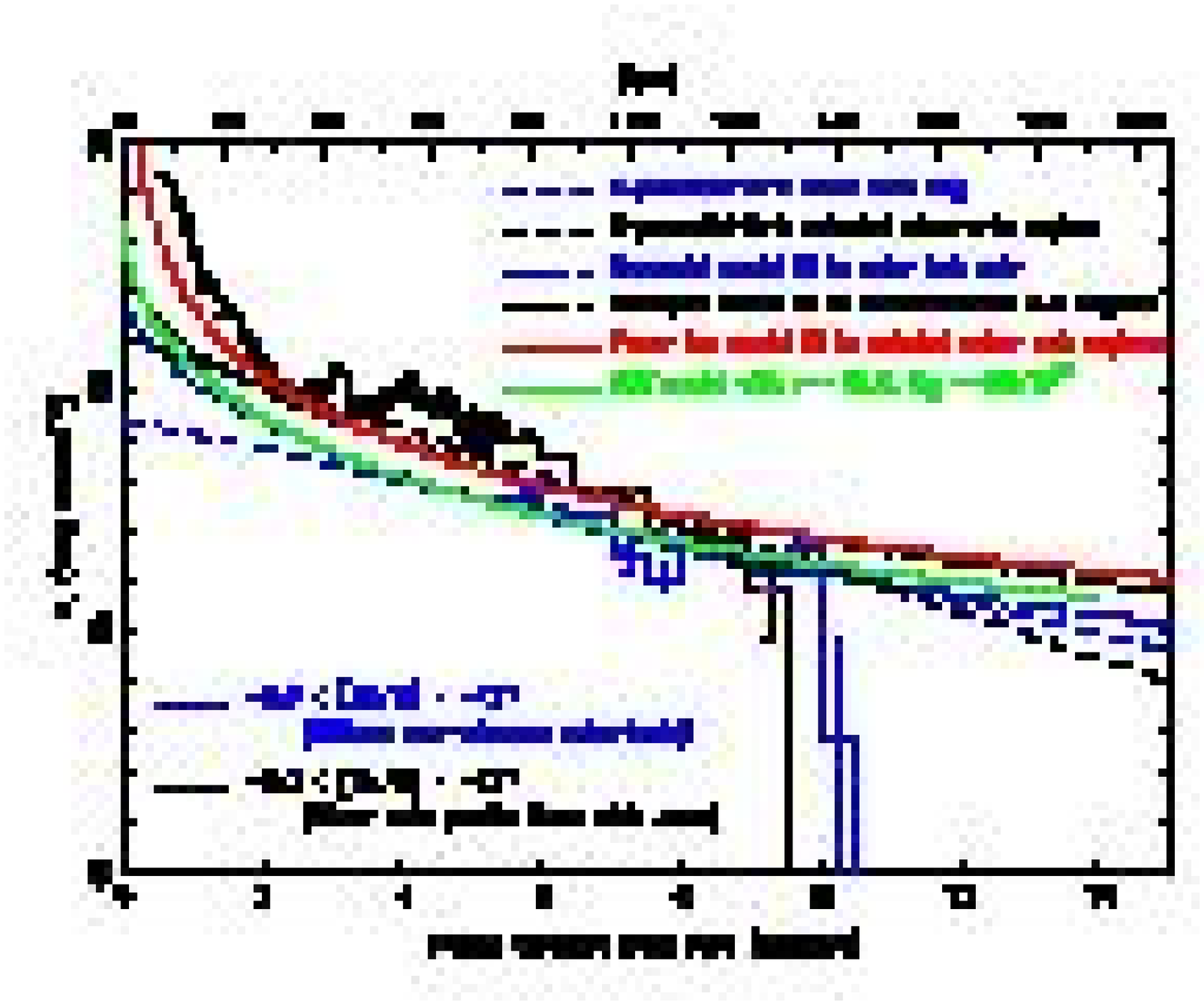}
\end{center}
\caption{The radial profile from the metal-poor selection of
the diffuse outer halo (previously shown in Fig.~43) is displayed
with the blue histogram. The blue dashed line is an
exponential fit to these data with $h_R=46.8 \pm 5.6 \kpc$, while
the blue dot-dashed line is a Hernquist model with
scale length of $53.5 \pm 0.2 \kpc$. The black histogram reproduces
the metal-poor minor axis profile of panel `c' of Fig.~39. We reject the data
below $R=30\kpc$, as the profile is dominated by the inner $R^{1/4}$ 
de Vaucouleurs profile in this region \citep{pritchet94,irwin05}. 
As we have shown, between $35 < R < 90\kpc$ there are copious stream-like
substructures on the minor axis, so we reject these regions as well.
The best exponential model fit to the remaining data (marked with red points) is
shown with a black dashed line, and has $h_R=31.4 \pm 1.0 \kpc$.
The black dot-dashed line shows the best fit Hernquist 
model, which has a scale length of $53.7 \pm 0.1 \kpc$. The red line
shows a power-law model fit to these data, which has an exponent of
$1.91\pm0.11$.
In addition, with the green line, we show the NFW model halo mass profile 
fit by \citet{ibata04} to the kinematics of the Giant Stream, with an
offset (arbitrarily) chosen to fit the outer halo data. The virial radius
of this model is $191\kpc$.
}
\end{figure}

Indeed, the outer halo profile appears remarkably flat in log-linear representation,
essentially an exponential function. The blue dashed line in Fig.~44 
shows an exponential model fit to the outer halo data (blue histogram);
we find an extremely long exponential scale length of $h_R=46.8 \pm 5.6 \kpc$.
We also show a projected Hernquist model fit (blue dot-dashed line) to these data,
a model choice motivated by the simulations of \citep{bullock05}; the
best model has a scale radius of $53.5 \pm 0.2 \kpc$, more that a factor of 
3 larger than predicted by \citet{bullock05}. The black histogram in 
Fig.~44 reproduces the metal-poor minor axis profile from panel `c' of Fig.~39.
Recall that this minor axis selection contains the stream-like structures
`B', `C' and `D', so it does not represent the underlying halo.
Nevertheless, beyond $R = 6\degg5$ there was no 
obvious substructure in that region of the halo, and we see that the profile from the minor
axis agrees reasonably well with that deduced from the ``outer halo''.

For $R > 6\degg5$ the minor axis profile appears slightly higher than
the  ``outer halo'' profile. It is possible that this may reflect
the real geometry of the halo, the difference would be consistent
with the halo being a slightly prolate structure.
We do not favor this interpretation, however.
The copious substructures seen at $R < 80\kpc$ testify to the 
dominance of stochastic accretion events in the halo. Given this, its seems
more natural to postulate that the variation in the profile 
that we see here is another consequence of this messy merging process. 

If such a thing as a smooth dynamically relaxed halo exists underneath all of
the substructure, it cannot have a hole, so the interval $30 < R < 35\kpc$ is a good
place to probe the upper limit to the radial profile in the inner region.
We therefore fit models to the data in that region and also at $R > 90\kpc$ (the 
data points used are marked red in Fig.~44). The best-fit exponential model 
to these minor axis data (black dashed line) has $h_R=31.4 \pm 1.0 \kpc$; 
while the best fit projected Hernquist 
model (black dot-dashed line) has a scale radius of $35.7 \pm 0.1 \kpc$. 
We also fit a power-law model, and find that an exponent of $1.91\pm0.11$
is preferred. 
Thus we find again a similar slow decline and a long scale length.

\begin{figure}
\begin{center}
\includegraphics[angle=-90, width=\hsize]{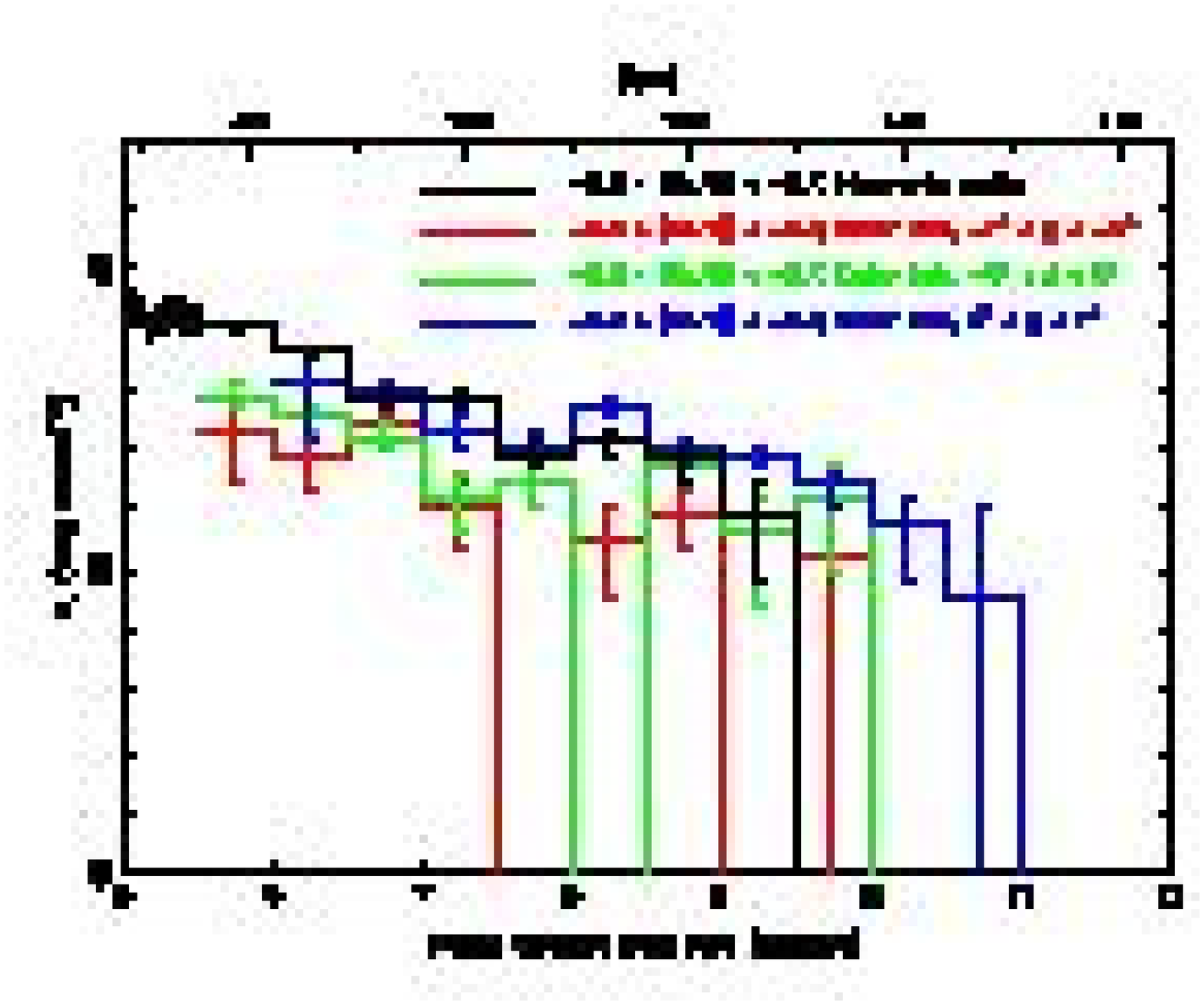}
\end{center}
\caption{The radial surface brightness profile for stars with
${\rm -3.0 < [Fe/H] < -0.7}$ is shown for the minor axis data
(black), and for three sub-samples of the outer halo region:
$-7\deg < \xi < -1\deg$ (in red), $-1\deg < \xi < 2\deg$ (in green),
and $2\deg < \xi < 7\deg$ (in blue). The similar radial decease
indicates that an underlying halo population is present in all these
samples, which are separated by up to $150\kpc$.}
\end{figure}

\begin{figure}
\begin{center}
\includegraphics[angle=-90, width=\hsize]{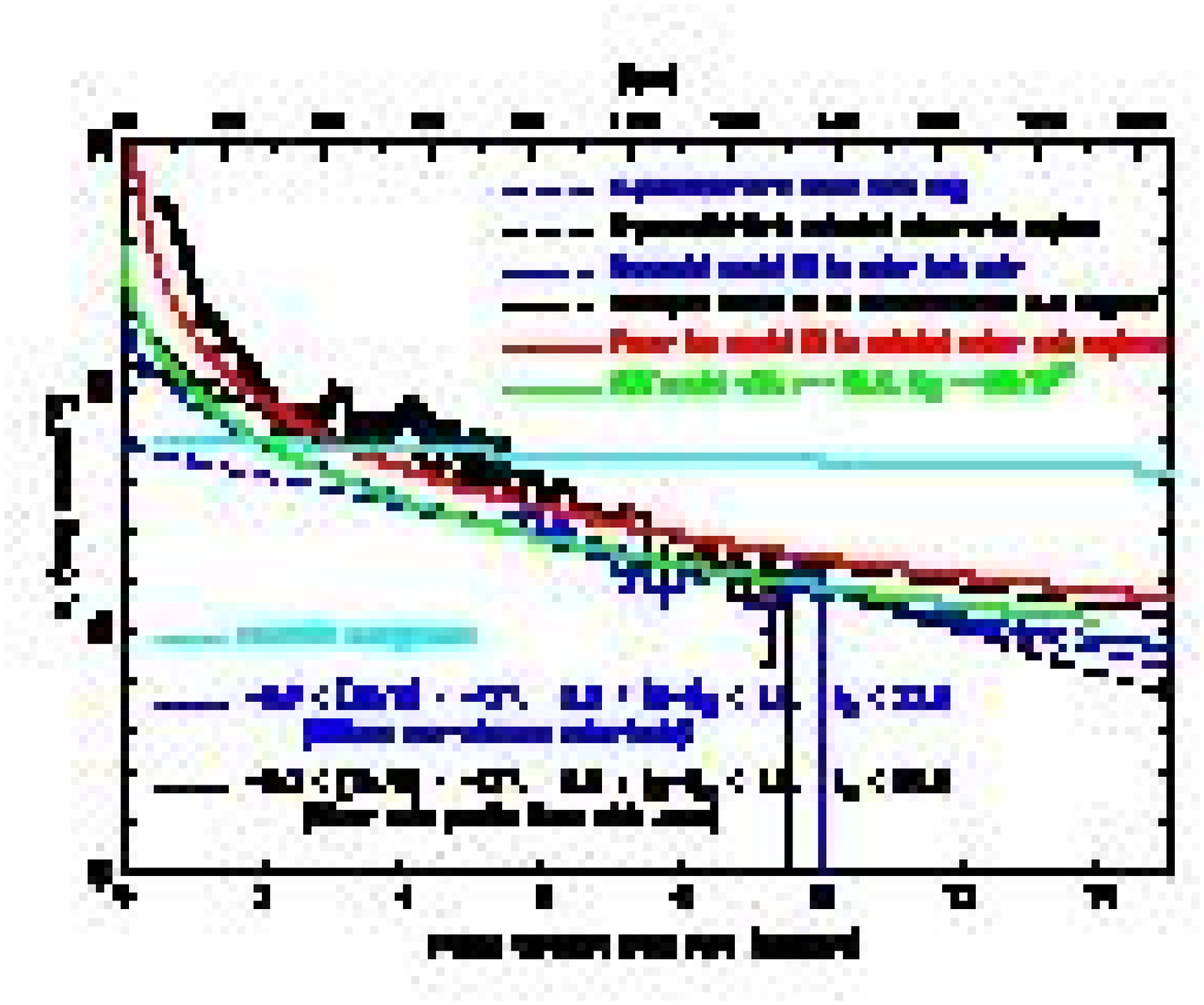}
\end{center}
\caption{As Fig.~44, but for stars 
restricted to the small color-magnitude region ${0.8 < (g-i)_0 < 1.8}$
and ${i_0 < 22.5}$, to ensure a minimal contamination from
the Galactic halo and disk. 
Since this selection is for the purpose of verification only,
we make no attempt to calibrate the absolute surface brightness 
values; hence the ordinate includes an unknown constant.
The exponential fit to the outer halo
(blue dashed line) has $h_R=48.8 \pm 8.8\kpc$, while
the Hernquist fit (blue dot-dashed line)
has a scale radius of $53.6 \pm 0.3 \kpc$. 
The black histogram is the metal-poor minor axis selection, also
constrained to the narrow color-magnitude region.
The exponential fit to these data (black dashed line) has
$h_R=32.5 \pm 1.5 \kpc$, while the Hernquist model 
has a scale radius of $53.9 \pm 0.1 \kpc$. The power-law
fit to these same data (red line) has an exponent of $1.85\pm0.16$.
For comparison, we also show the profile of the Galactic foreground
as predicted by the Besan{\c c}on model (turquoise line). The same color-magnitude selection
is used as for the observed profiles, although we show here the 
model prediction over the entire MegaCam survey area (not just 
the ``outer halo'' or minor axis regions). The model predicts a 
decrease in the foreground contamination with radial distance
over the survey region, but it is essentially flat compared to the
observed decrease in the M31 populations.}
\end{figure}

\begin{figure}
\begin{center}
\includegraphics[angle=-90, width=\hsize]{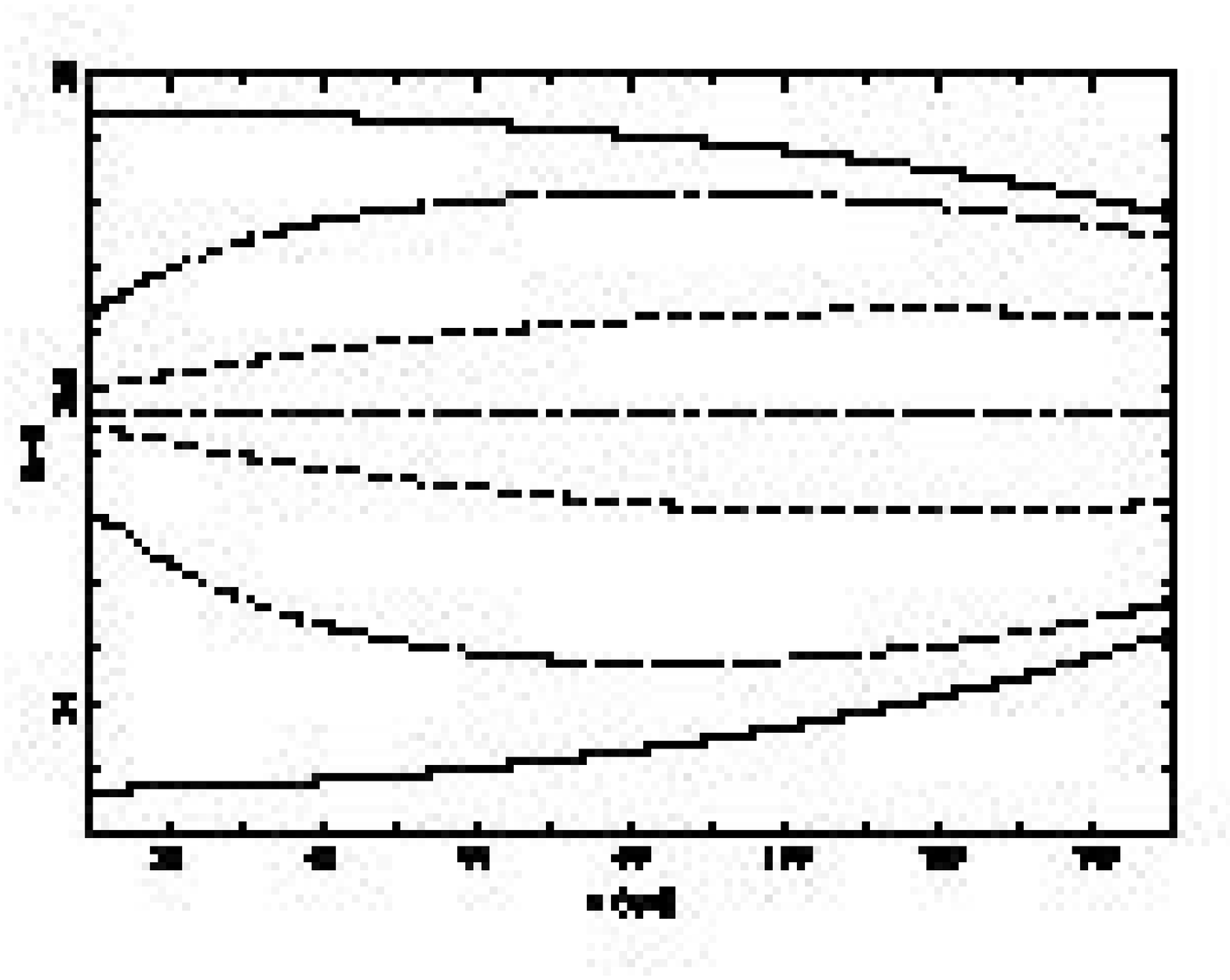}
\end{center}
\caption{The expected spread in distance modulus 
as a function of projected radius if the underlying 
halo component falls off as $\rho(r) \propto r^{-2.91}$.
The dashed line shows the distance modulus to M31, while
the full lines shows the limit of $r=191\kpc$ (the virial radius
estimated by \citealt{ibata04}). The dashed and dot-dashed
lines mark the region enclosing 50\% and 90\% of the stars.}
\end{figure}

This is a very important, and rather unexpected result, and therefore
deserves to be checked carefully. In Fig.~45 we have split the ``outer halo''
sample into three sub-samples (contained within the regions
red: $-7\deg < \xi < -1\deg$, green: $-1\deg < \xi < 2\deg$ and 
blue: $2\deg < \xi < 7\deg$); the same slow decline with radius is
seen in each sub-sample, and in the minor axis sample shown in black, 
indicating that we are not simply detecting the effects of some localized substructure:
approximately $150\kpc$ separate the red and black profiles!
It is possible that the signal
arises from an incorrect subtraction of the Galactic contamination.
Since the density of stars decreases away from the Galactic plane, which also
happens to be the direction away from the centre of M31, an insufficient
subtraction of the contaminants could leave a residual that 
decreases with $R$ as observed. Furthermore, the Galactic disk has 
an exponential profile, which would naturally explain the observed decline.
To examine this possibility we recalculate the surface brightness profiles as before,
selecting on metallicity, but this time in addition using a draconian 
color-magnitude selection. We limit the 
data to $i_0 < 22.5$ and retain stars only in the color interval $0.8 < (g-i)_0 < 1.8$.
An inspection of panel `b' of Fig.~16 reveals that this selection avoids the bulk of
the Galactic disk and halo. The results are shown in Fig.~46, and reassuringly they are
qualitatively and quantitatively identical to the previous selection with deeper data and the full color interval. The predicted behavior of the Galactic
foreground contamination (with this same color-magnitude selection) is also
shown in Fig.~46 (turquoise line). The profile of the contamination is nearly flat 
in this log-linear representation, so contamination cannot account for the observed profile.
Thus a slow decline with an exceeding long scale length for the outer halo
population is a robust result of this survey.

This slow decline has important consequences on the detectability of 
halo populations. In particular one may worry about the distance spread in the
halo, whether we are able to detect stars on the far side of M31, and the 
corresponding spread in the CMD. Assuming an $\rho(r) \propto r^{-2.91}$ profile,
we display in Fig.~47 the expected spread as a function of projected radius.
We see that even with this extended profile, the distance spread should be relatively
modest, $\sim 0.5$~mag.

\begin{figure}
\begin{center}
\includegraphics[angle=-90, width=\hsize]{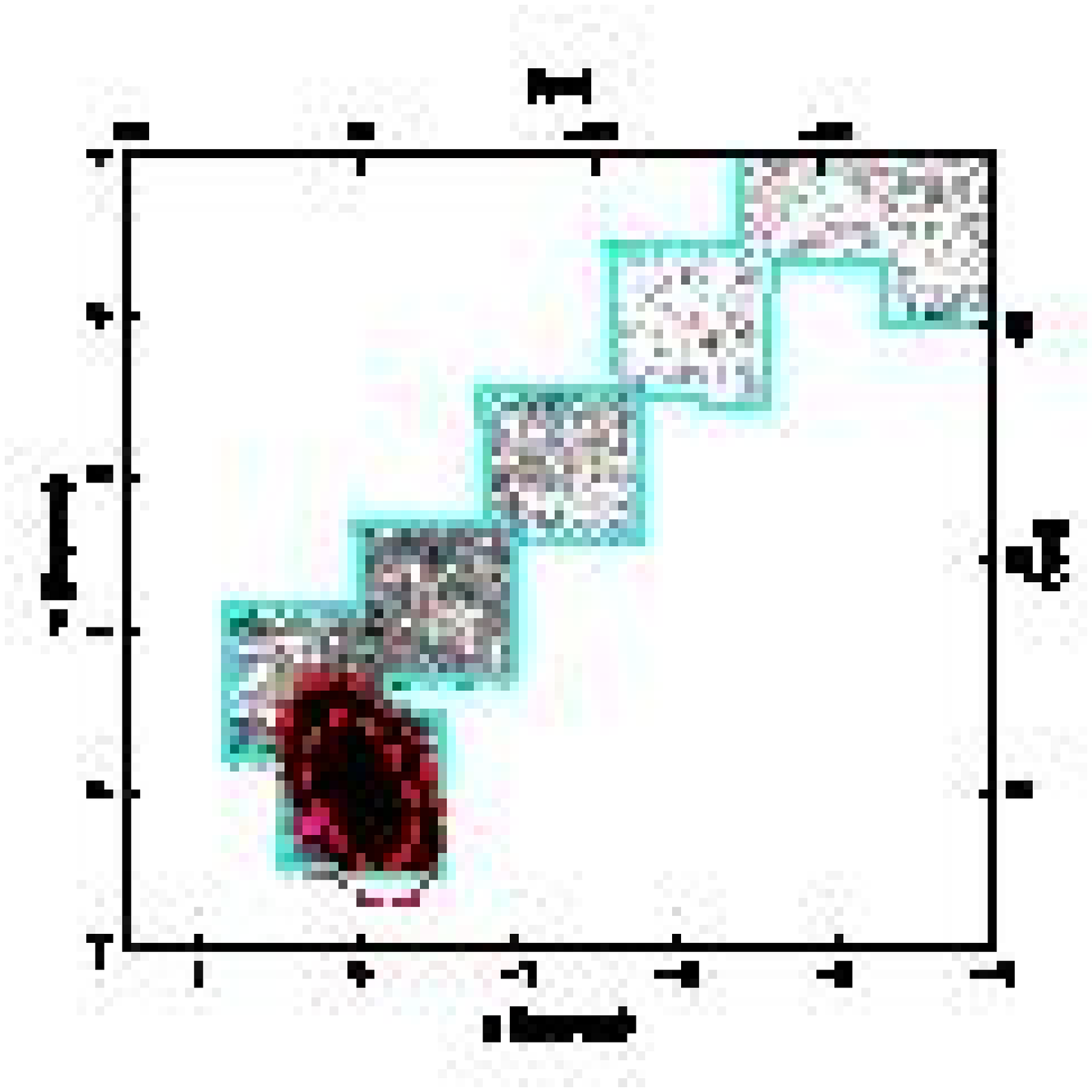}
\end{center}
\caption{Matched filter map (logarithmic representation) to search for structures
around M33 constituted of stars with metallicity in the range ${\rm -3.0 < [Fe/H] < 0.0}$.
A limiting magnitude of $i_0=23.5$ was used.
The two red ellipses mark elliptical radii of $s=0\degg5$ and $s=0\degg75$ around M33. 
The pink square at $\xi=0\degg30$, $\eta=-0\degg24$ marks the location of
the ``halo'' field of \citet{mould86}, which is in fact clearly probing the disk of the galaxy.}
\end{figure}

\section{M33}

The South-eastern corner of the survey extends out to the Triangulum
galaxy, M33. The motivation for this part of the study was to
attempt to investigate the interface region between 
the halos of M33 and M31. 
Four fields were positioned along the extension
of the minor axis of M31, as shown in Fig.~48, connecting to the archival 
data centered on the disk of M33. The map reveals clearly the very regular
outer disk of M33, as well as the presence of an extended component
out to $\sim 3\deg$, possibly the stellar ``halo'' of this galaxy.
A more detailed discussion of the structural and stellar populations properties of M33
based upon a much wider survey conducted with the INT will be presented in a
companion paper (Ferguson et al. 2007, in prep.).
We note here that a previous claimed detection of the stellar halo component
of this galaxy \citep{mould86}, was in reality studying the outer disk
(their field is marked with a pink square in Fig.~48).

We adopted the geometry of the model of \citet{mcconnachie06b} for the
disk of M33, namely a position angle of $23\deg$ and an inclination of
$53\degg8$. The outer red dashed ellipse in Fig.~49 shows the corresponding 
elliptical radius $s=0\degg75$, approximately where the disk appears to truncate in 
this diagram.

\begin{figure}
\begin{center}
\includegraphics[angle=-90, width=\hsize]{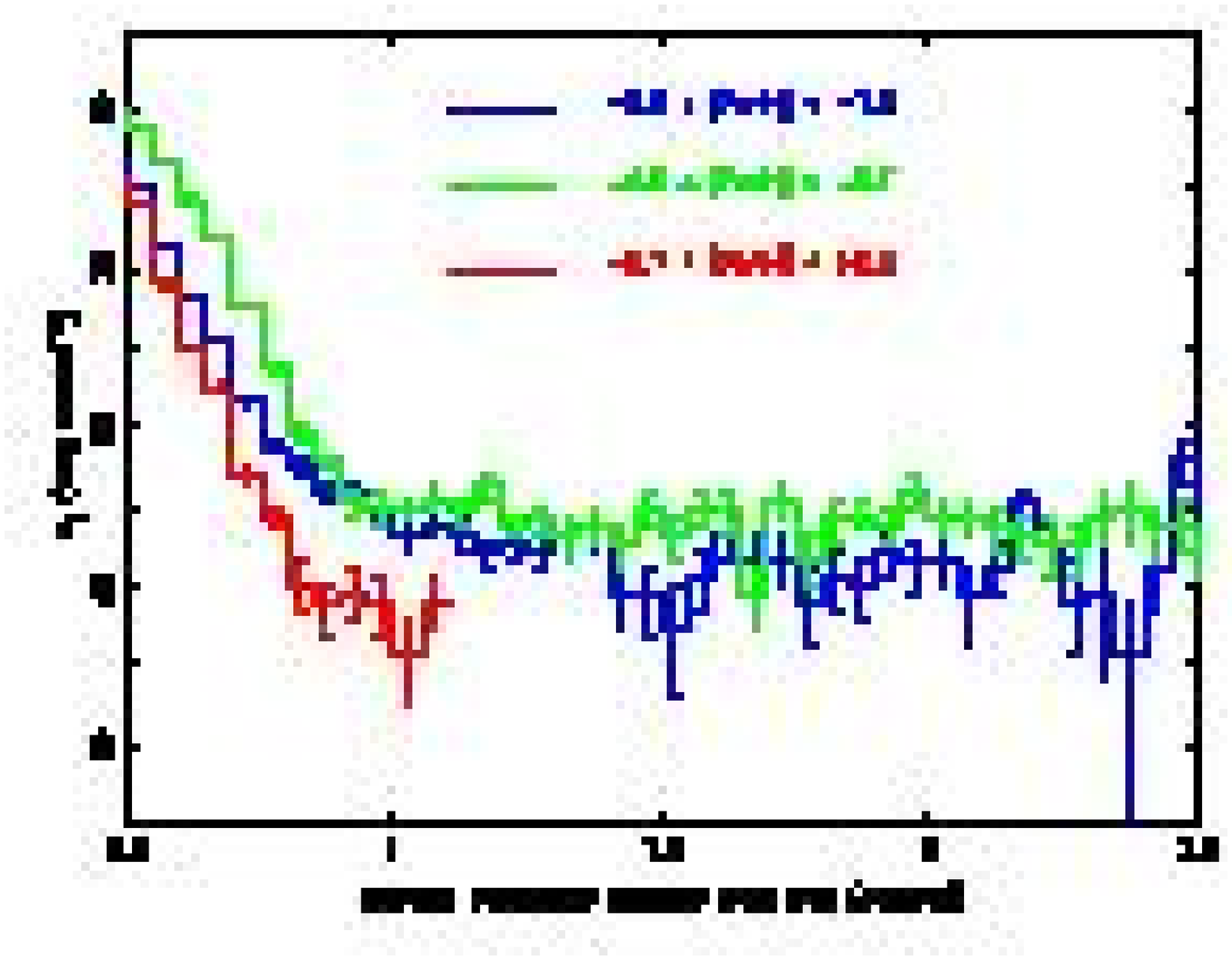}
\end{center}
\caption{The radial profile as a function of elliptical coordinate
distance from M33, in 3 color-magnitude selection
regions corresponding to locations between Padova isochrones.
We truncate the ``metal-rich'' profile (which is more heavily affected
by Galactic foreground contamination), where the noise begins
to dominate.}
\end{figure}

As we have mentioned before, the applicability of the isochrones 
to estimate metallicity is only justified in regions composed of 
old stars, so the ``metallicity'' profiles displayed in Fig.~49
must be interpreted with extreme caution. Here we show the
trends as a function of elliptical coordinate $s$ for 
three different CMD bins, as shown. The data interior to
$s=0\degg5$ is severely affected by crowding, and we therefore neglect 
that region. In the region 
to $0\degg75 < s < 1\deg$, the blue selection becomes more pronounced 
with increasing radius relative to the other two selections,
indicating strong radial variations in the stellar populations.
The exponential profile of the inner disk ends changes abruptly
at $s\sim 0\degg9$ into an apparently flat distribution for 
$1\deg < s < 2\degg5$. 
Fitting the profiles in the interval $1\deg < s < 2\degg5$ with
an exponential function gives exceedingly long scale-lengths,
or even rising profiles.

\begin{figure}
\begin{center}
\includegraphics[angle=-90, width=\hsize]{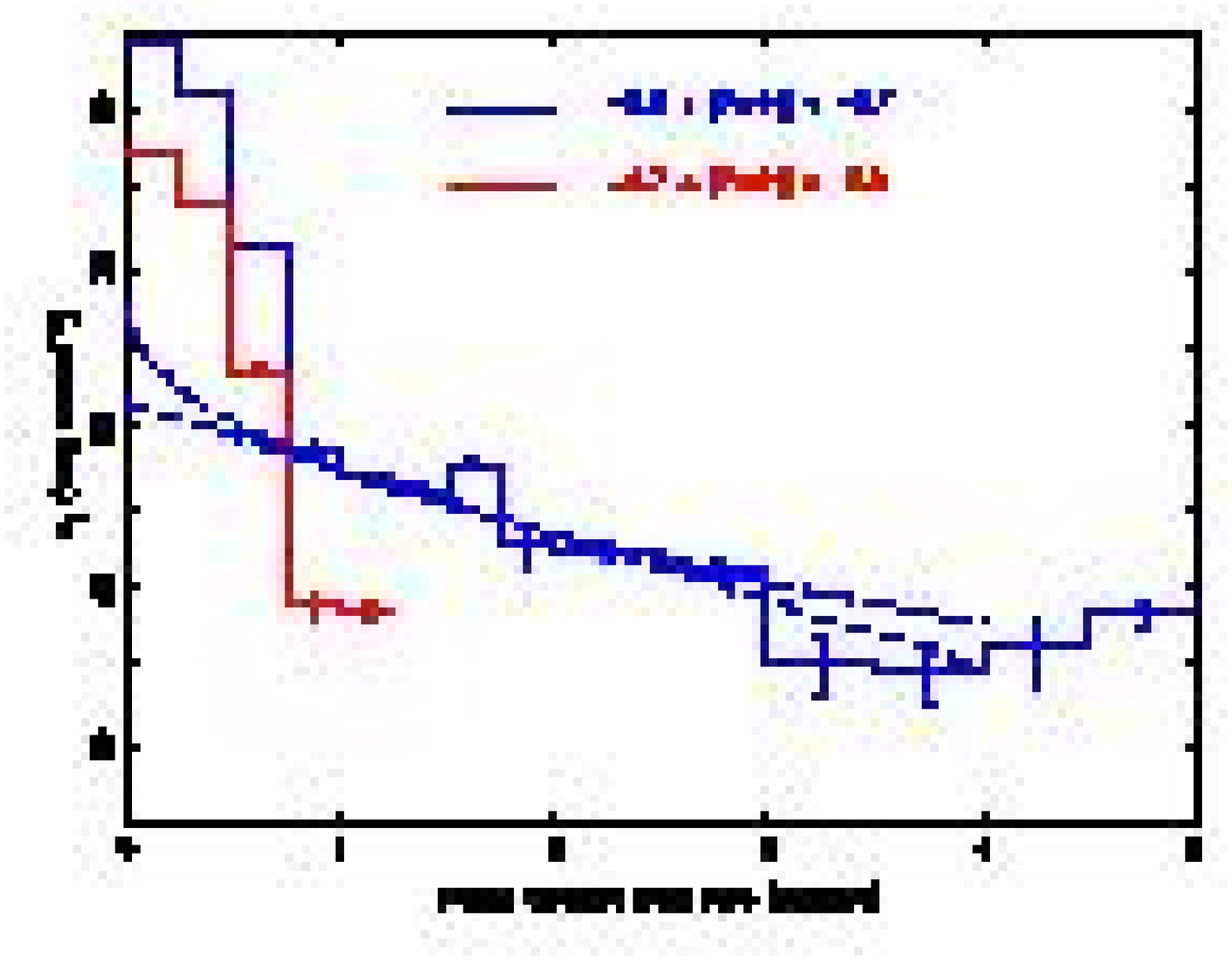}
\end{center}
\caption{The radial light profile in M33 
as a function of the radial coordinate $r$.
We display a fitted exponential model with scale length
$18 \pm 1\kpc$ and a projected Hernquist model with 
scale radius $55 \pm 2\kpc$.}
\end{figure}

The spatial extent of the MegaCam survey around
M33 is very limited, so it is impossible to construct a global
model for the extended outer component. Thus
it is not clear whether the appropriate geometry 
for calculating the profiles is spherical
or ellipsoidal. If we adopt a spherical coordinate as in Fig.~50,
the profile of the extended component
for the selection ${\rm -3.0 < [Fe/H] < -0.7}$ 
seems more reasonable, as it descends monotonically 
apart from a bump at $1\degg6$.

Fitting the data between $1\deg < R < 4\deg$ 
(but rejecting the bin at $1\degg6$) yields a scale length
of $18 \pm 1\kpc$ for an exponential model, 
or alternatively a scale radius of $55 \pm 2\kpc$ 
for a projected Hernquist model.
These scale lengths are surprisingly large, reminiscent of the large
values measured above for the outer halo of M31. Curiously, the central
surface brightness of the extrapolated exponential models
are rather similar too. In M33 the model has $\Sigma_V(0)=29.7\pm0.1$,
while in M31 the two exponentials fit in Fig~44 bracket this value with
$\Sigma_V(0)=30.6\pm0.3$ and $\Sigma_V(0)=29.0\pm0.06$ (taking
the metallicity selection ${\rm -3.0 < [Fe/H] < -0.7}$ for both objects).
We stress here that the detection of a halo component around M33
gives further confirmation that the M31 detection is not due to
errors in the foreground subtraction, since the 
foreground contamination profile
has the opposite slope as a function of galactic radial distance in the
M33 survey fields compared to the M31 fields.

The bump in the surface brightness profile at $1\degg6$ is (just)
visible as a faint arc on the map in Fig.~48, but
we are unsure of the reality of the structure, since it is a very faint 
feature and only extends over one field. Further imaging is required to
determine whether this is a substructure in the halo of M33 or not.

\begin{figure}
\begin{center}
\includegraphics[angle=-90, width=\hsize]{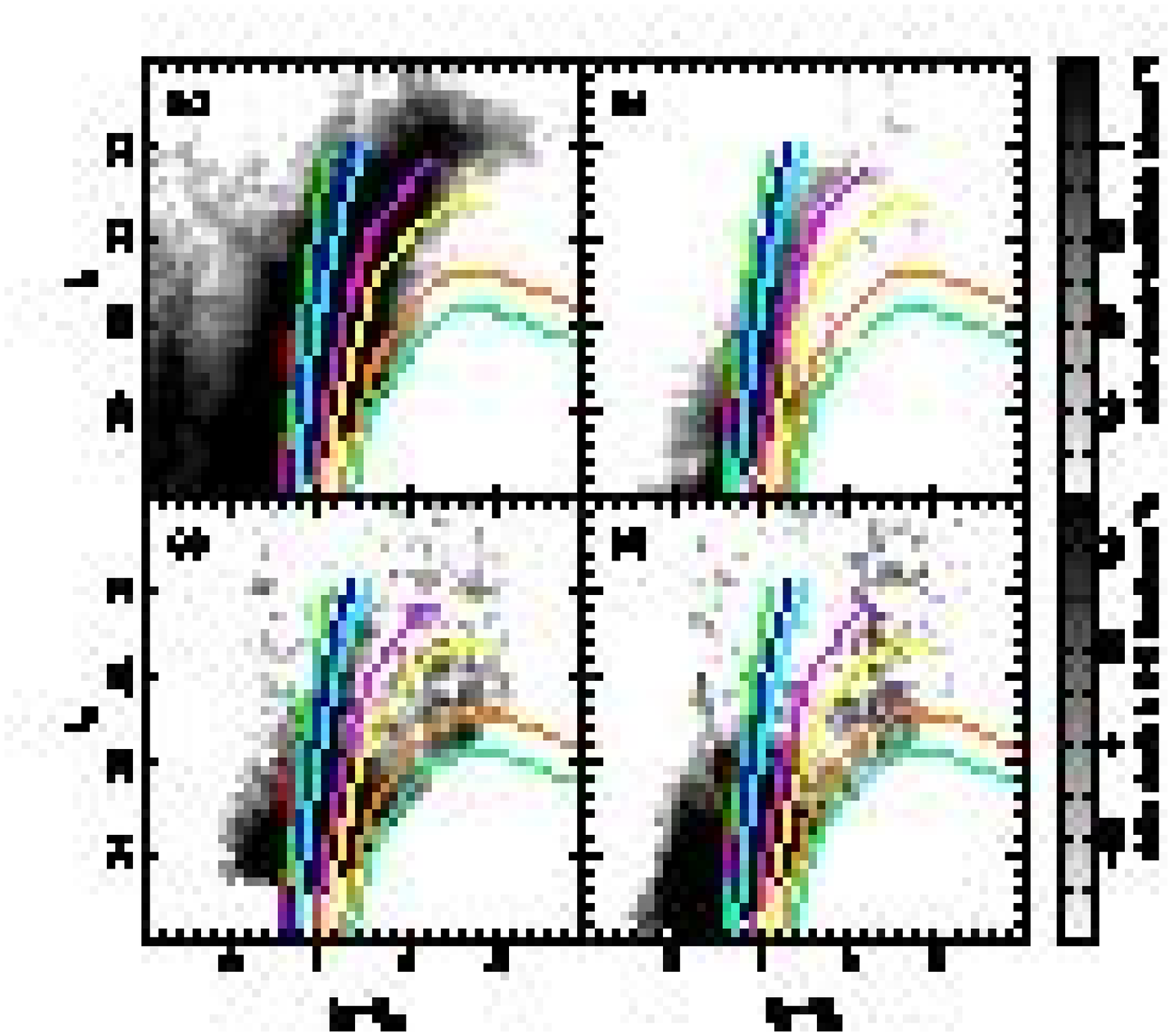}
\end{center}
\caption{Background-subtracted Hess diagrams in four selected
regions near M33. In panel `a', we show the region $0\degg5 < s < 0\degg75$,
panel `b' is for $0\degg75 < s < 1\deg$, panel `c' is for $1\deg < r < 2\deg$ and
panel `d' is for $2\deg < r < 3\deg$.}
\end{figure}

Finally, we show in Fig.~51 the progression in the stellar populations as we 
move from the outer disk into the halo component. Of particular interest is the
difference between panels `b' at the disk edge ($0\degg75 < s < 1\deg$)
with panel `c' ($1\deg < R < 2\deg$) in the halo. The ``halo'' component
contains a higher proportion of blue stars, compared to the broader distribution
in panel `b'.

\section{Discussion}

\subsection{The underlying halo}

The analysis presented above in \S8 indicates that underneath
the many substructures that we have uncovered in M31 lurks an 
apparently smooth and extremely extended halo. A similar structure
is also detected in M33. By ``smooth''
what we mean here is not necessarily that the component is
perfectly spatially smooth, but instead that any substructures
that may be present are below detectability with the current
survey. The detectability threshold
is a function of radius, but it corresponds to approximately
1~mag~arcsec$^{-2}$ brighter than the smooth background
over spatial scales $\simgt 1$~deg$^2$.

The existence of a stellar halo component which appears smooth 
at these surface brightness levels is completely unexpected
given recent numerical models that 
implement recipes for star-formation in merging 
CDM subhalos \citep{bullock05, abadi06}. Those models 
predict that the light at large radius is
confined to arcs, shells and streams, with essentially no
smoothly-distributed stars beyond $\sim 50\kpc$ in a Milky-Way (or M31)
analogue. The reason for this is that dynamical times at large
distances from the galaxy are extremely long, so material has
not had anywhere near enough time to mix. The more recent
the accretion, in general the more spatially confined the stars should be.

Given these considerations, one would expect a smooth 
component to be made in the early violent phases of galaxy formation,
and since the disk is a fragile structure \citep{toth92}, the formation
of the structure would have had to have occurred before the
formation of the thin disk. This scenario still poses problems
however, since the proto-Andromeda at $z\sim 2$ would have been much 
less massive than it is today, so the extreme distances of these
halo stars --- most likely beyond the virial radius of the galaxy at that redshift ---
are hard to explain.

Interestingly, the radial profile of this smooth halo component in
M31 is similar to what is deduced for the Milky Way. As we have reviewed
in \S1.2, in the case of the
Milky Way, current data probe the halo well up to $r\sim 20\kpc$,
we have reasonable constraints up to $r \sim 50\kpc$,
but beyond that distance the information
is very scanty indeed. However, at least up to $r=50\kpc$, and given
variations from study to study (which are probably due to halo
substructures) the density
can be approximated by $\rho(r) \propto r^{-3}$. 
For instance, the study of \citet{siegel02}, which made use of good distance estimates
to halo stars found $\rho(r) \propto r^{-2.75\pm0.3}$.
Similarly, analysis of the RRLyrae sample of \citet{vivas06} yielded
$\rho(r) \propto r^{-2.7\pm0.1}$ or $\rho(r) \propto r^{-3.1\pm0.1}$, depending
on model assumptions of the shape of the halo.
This is completely
consistent with the present $\Sigma(R) \propto R^{-1.91\pm0.11}$ fit to
the minor axis selection in M31.

In modern galaxy formation simulations stars are formed only within the
most massive sub-haloes that merge to form a galaxy. 
This is because star-formation recipes used in the simulations impose a threshold in gas density
below which stars cannot form, basing this condition on observed correlations
between H$\alpha$ emission and gas surface density in galaxy disks \citep{kennicutt89}.
Furthermore, those satellites that were not massive enough to accrete sufficient gas before
the epoch of reionization are expected not to have been able to form stars subsequent to
that epoch \citep{bullock00}.
Dynamical friction acts more strongly upon the most massive subhalos, making them fall
rapidly into the potential well, where they become disrupted and their contents
mixed into the evolving galaxy. 
Because of this, stars accreted from subhalos are expected to 
have a more rapidly falling profile than the dark matter, as we have reviewed in \S1.5,
with the light profile falling as $r^{-4}$ or steeper.
Nevertheless, this prediction does not appear to hold out. 
If dark matter is distributed according to the ``Universal'' NFW profile 
\citep{navarro97}, the density profile in the outer
regions of the halo will be $\rho(r) \propto r^{-3}$, consistent with what we 
have measured from the stars. This suggests that stars in these tenuous
outer reaches of giant galaxies trace the dark matter.

We stress here that the present analysis of M31 is based on a dataset
that is much more spatially extensive than has been possible for the Milky Way.
We have covered substantially more than a quarter of the halo of M31. In comparison,
even the SDSS studies of \citet{yanny00, ivezic00} or \citet{chen01} covered only 1\% of the sky.

Another measure of the halos of these two galaxies that we may now compare is
their total luminosity. Integrating the lower of the two exponential
profiles shown in Fig.~44 out to $140\kpc$, gives a conservative lower
limit to the smooth halo of $L_V \sim 2.2 \times 10^8 L_\odot$. We estimate an upper limit
by integrating the power-law up to the virial radius (which we take to be $191\kpc$),
assuming that the halo density inside $0.5\kpc$ is constant; this yields a value of
$L_V \sim 1.3 \times 10^9 L_\odot$. For the Milky Way, we estimate the
total luminosity by assuming a
Solar Neighborhood V-band luminosity of halo stars of $22300 L_\odot/\kpc^3$ 
\citep{morrison93}; for a density law $\rho(r) \propto r^{-3}$, integration
out to $50\kpc$ gives  $L_V \sim 7 \times 10^8 L_\odot$ or alternatively 
$L_V \sim 1.2 \times 10^9 L_\odot$ for $\rho(r) \propto r^{-3.5}$ (following
\citealt{robin03} we also assume that the density of the halo is constant
in the inner $0.5\kpc$). 
These estimates both for M31 and the Milky Way are very crude, 
but taken at face value they indicate that the stellar halo of M31
is very similar in total luminosity to that of the Milky Way.
Thus it appears that previous estimates (e.g. \citealt{reitzel98})
who reported that the halo in M31 is $\sim 10$ times denser
than that of the Milky Way apply only to the inner regions
of the galaxy, where contamination from the large bulge, extended disk and intervening
substructures are clearly a concern.

As reviewed above,
\citet{chapman06} were able to detect the true inner halo of M31
by observing mostly major axis fields where halo stars
have a very different kinematic signature to other components. 
At radii between 10 and $70\kpc$, the halo component was found
to have a mean metallicity of ${\rm [Fe/H \sim -1.4}$. This
is consistent with the photometric estimate derived for the
outer halo component in Fig.~43 over the radial range
$75 < R < 140\kpc$, and suggests that the halo has a small
or negligible metallicity gradient. This result provides further support for the
case of a smooth monolithic halo formed in a single merging event.

\subsection{Comparison to \citet{kalirai06b}}

Our discovery of a smooth very extended halo component covering the entire
southern quadrant of Andromeda was anticipated by the kinematic study
of \citet{kalirai06b}. These authors used the Keck/DEIMOS spectrograph to
survey a number of fields in this region
of the sky, targeting known dwarf galaxies as well as ``empty'' halo fields. 
The position of the fields presented in \citet{kalirai06b} are shown with red
dots in Fig.~52, green dots mark the positions of fields observed with this
instrument by our own group \citep{ibata04, ibata05, chapman06}.

\begin{figure}
\begin{center}
\includegraphics[angle=-90, width=\hsize]{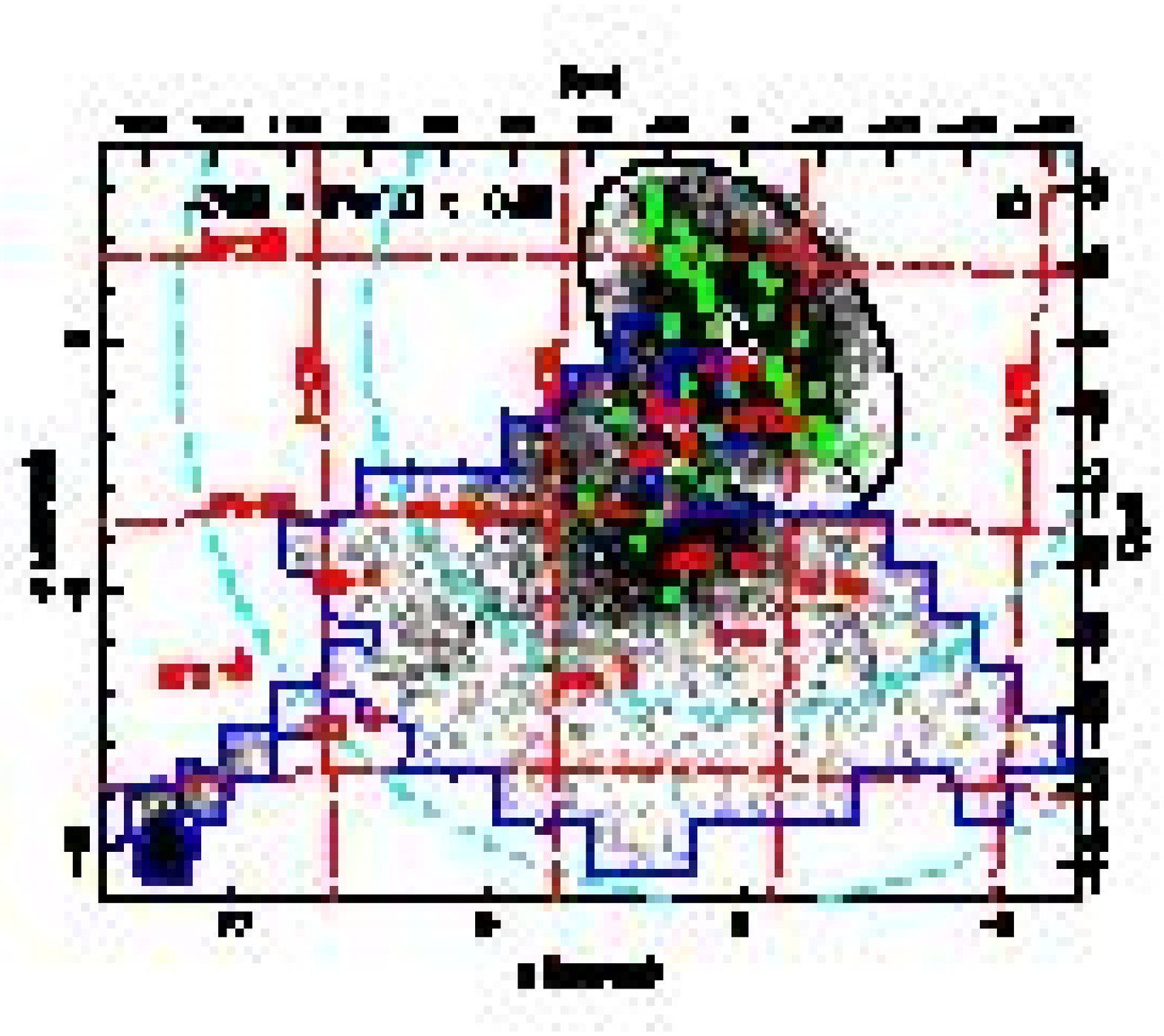}
\end{center}
\caption{Spectroscopically-observed fields. \citet{kalirai06b} fields are shown 
in red, \citet{chapman06} fields in green. Many of these pointing were chosen
without knowledge of the underlying populations, so only now is it possible to
properly interpret the spectroscopic results.}
\end{figure}

The \citet{kalirai06b} fields marked `d2' and `d3' being located on the 
satellites And~II and III, are not of relevance to the current discussion.
But for many of the remaining of their fields our present panoramic survey 
is invaluable, as it allows one to identify the stellar populations
that study actually targeted. In particular, their fields ``m6'' were placed on the edge of
stream `B', while their fields `a13' and `b15' lie on the extended 
cocoon of the Giant Stream. Likewise, in \citet{chapman06} we serendipitously
targeted streams `C' (fields F25 and F26) and `D' (field F7).

Thus we see that only fields `m8' and `a19' were targeted in regions where we can be sure 
that no substructure was present, while field `m11' lies outside of the current
survey region. In these fields, \citet{kalirai06b} report 1 probable M31 halo star
in `m8', 4 stars in `a19', and 3 stars in `m11'.

Are these counts consistent with our results? We normalize with respect
to the \citet{kalirai06b} field `a0' at $30\kpc$, where
we deduce ${\rm \Sigma_V \sim 30 \, mag \, arcsec^{-2}}$. In that field
67 halo stars were detected 
in observations over 3 spectroscopic masks (i.e. 3 subfields were observed). 
Whereas in their field `m11' at $165\kpc$, where a 
mild extrapolation from our survey region 
gives ${\rm \Sigma_V \sim 34}$ -- ${\rm 35 \, mag \,arcsec^{-2}}$, 3 halo stars were
detected using 4 spectroscopic masks.
We therefore expect 40 to 100 times lower
stellar density in `m11' compared to `a0', that is, we expect 0-2 stars to be detected in the 
4 masks observed in field `m11' (taking the best-case scenario that all available halo
stars were observed and correctly classified).
This is then consistent with the sample of 3 halo 
stars that were reported by \citet{kalirai06b}
in field `m11'. We note however, that their field lies $\sim 4\deg$ from 
M33, where we have found that the halos of M31 and M33 overlap, and are approximately
of equal surface brightness.
Though it is dangerous to draw conclusions from such a minuscule sample,
one out of the 3 halo stars in m11 has a velocity of $-150\kms$, and is 
highly unlikely to belong to M31, but could be perfectly consistent with being
a member of the halo of M33.
Likewise, in field `m8' we expect 2.5 stars, while in field `a19' we expect 2.2 stars,
consistent with the number of stars detected spectroscopically.

In summary, despite the very small number of stars in their sample, and despite
the probable contamination from M33 in their most distant (and interesting) field, 
we take the results of \citet{kalirai06b} as confirmation that a smooth extended
stellar halo is present in M31 out to at least $150\kpc$.
We note in passing that \citet{kalirai06b} estimate the photometric metallicity of their outer halo sample
($R > 60\kpc$) to be ${\rm \langle [Fe/H] \rangle = -1.26 \pm 0.1}$.
Although this is apparently consistent with the MDF shown in Fig.~43, 
their sample is almost entirely 
dominated by ``contamination'' from substructure, which as we have shown above,
in predominantly metal-poor.

\begin{figure*}
\begin{center}
\includegraphics[angle=0, bb= 70 1 965 765, clip, width=\hsize]{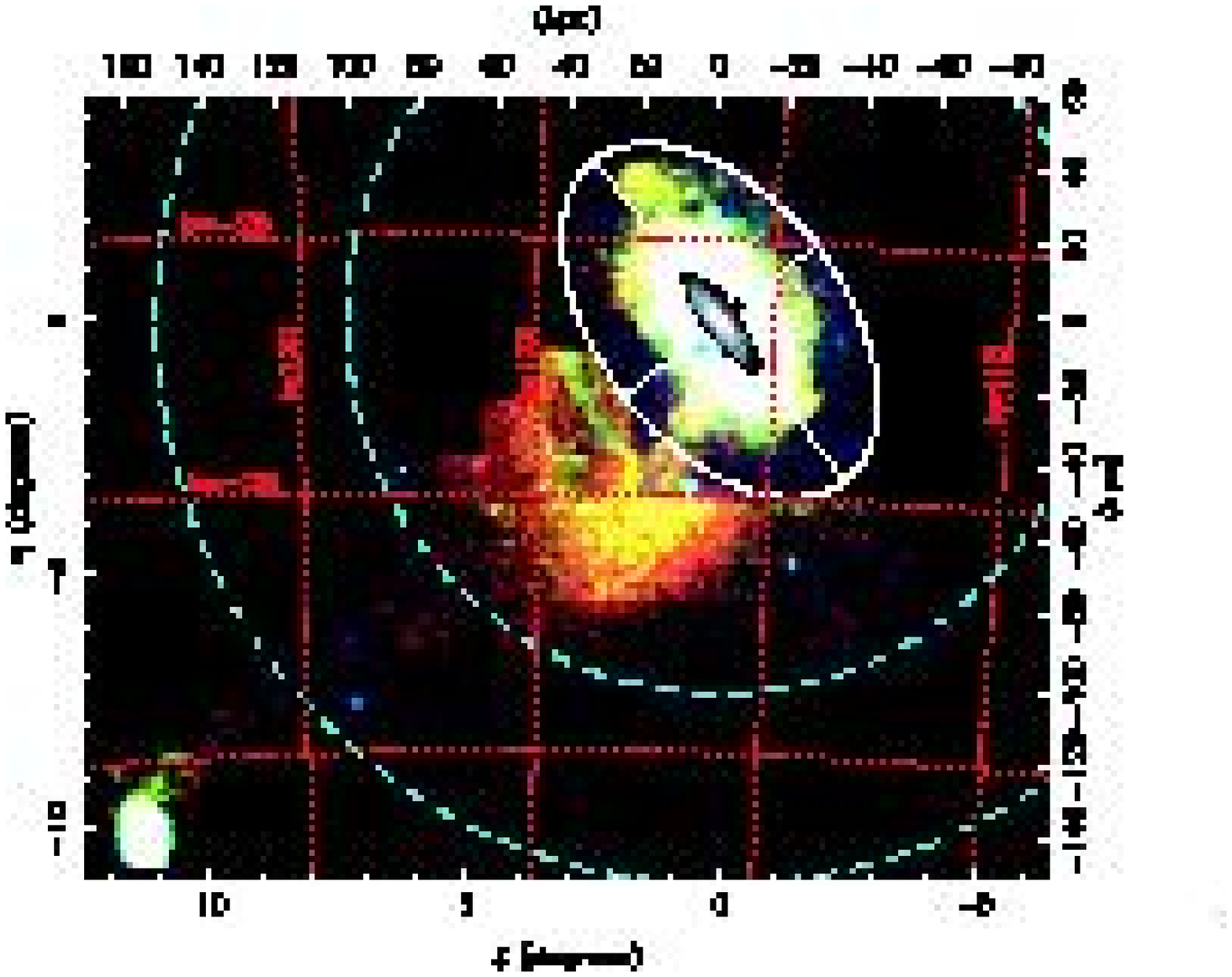}
\end{center}
\caption{RGB color composite map, in which red shows stars with
${\rm -0.7 < [Fe/H] < 0.0}$, green shows ${\rm -1.7 < [Fe/H] < -0.7}$ and
blue shows ${\rm -3.0 < [Fe/H] < -1.7}$. To render the inner region within
the $4\deg$ ellipse easier to interpret, we have removed the MegaCam data
from that region. Dwarf satellite galaxies, being essentially the only structures
with a strong metal-poor population appear blue on this map. The differences
in stellar populations between the Giant Stream and the several minor
axis streams can be seen as striking differences in color. At the center
of the galaxy we have added to scale an image of the central regions of M31 
constructed from Palomar sky survey plates.}
\end{figure*}

\subsection{Shape of the smooth stellar halo}

As reviewed in \S1.2, most studies of the halo of the Milky Way find
that this component is oblate interior to $r \sim 20\kpc$, with flattening $b/a \sim 0.6$.
Studies of the halo component in external galaxies, be it from a medianed stack of
edge-on spirals \citep{zibetti04a}, or from an individual edge-on galaxy \citep{zibetti04b} 
find an identical measurement of $b/a \sim 0.6$, within roughly the same radius.
The data we have presented on M31 do not allow us to make any statement
about the halo flattening in the same volume, and it is very hard to imagine that
such a measurement will be possible in the foreseeable future given the
difficulty of disentangling bulge, disk and halo in the inner regions of M31.
Previous measurements of the flattening of M31 in this region (e.g.
\citealt{pritchet94}: $a/b=0.55\pm0.05$ at $10\kpc$), give an indication 
of the shape of the total light distribution, but do not constrain the shape of the halo.

However, we believe we have been able to identify the main substructures beyond 
a distance of $R=6\degg5$, giving a relatively uncontaminated measurement of
the density profile beyond that radius. We find, however, that the minor axis profile 
is higher than the profile from the broad region we have termed ``outer halo'' and which
lies closer to the major axis. This allows us to firmly reject an oblate halo with 
$b/a \sim 0.6$ at these distances, and suggests instead the possibility that the halo is 
prolate, with $c/a \simgt 1.3$. Further data in other quadrants is required to 
assess the reliability of this estimate. However, in any case, the shape of the outer halo
of M31 is manifestly different to that of the inner halos of other galaxies observed
to date.

\subsection{Substructures}

Every step we have taken in obtaining a wider view of Andromeda has awarded
us with new discoveries in the form of previously unknown substructure.
The large area surveyed with MegaCam in the present contribution has 
continued this trend showing new dwarf galaxies, and several diffuse
stellar populations in the form of arcs, streams or shell segments. These
structures testify that accretion and therefore galaxy
buildup is still continuing to the present time.

Of the substructures that are present in the survey region the Giant Stream
is by far the most significant.
The data presented in \S6.2 shows that
the Giant Stream is a long cigar-shaped structure made up
of metal rich, or young, stars with a metal-poor envelope or cocoon, possibly $\sim 3\deg$ wide.
This lack of homogeneity of the stellar populations in the Giant Stream indicates
that so far the system has not been fully mixed during the course of the tidal disruption process, so it is 
likely a dynamically very young stream. The requirement that the center and the cocoon remain
spatially distinct will likely provide very useful additional constraints for the modeling of the system.

We count up the Giant Stream stars to $i_0=23.5$, and as before use And~III to normalize
the total luminosity. (We caution the reader again that using And~III as a 
reference introduces a large uncertainty into the luminosity estimate).
Integrating within the red polygon shown in Fig~23 (and removing
a $0\degg5$ circle around both And~I and And~III), and subtracting off the expected 
foreground from the Besan{\c c}on model, we find 
$L_V \sim 1.5 \times 10^8 L_\odot$ ($M_V \sim -15.6$) 
over this region. This corresponds to approximately
a tenth of the luminosity of M33, and given that the MegaCam region only probes a 
fraction of the total stream, it is plausible that the progenitor of the Giant Stream
was initially a galaxy of similar luminosity to M33. The width of the stream appears
consistent with this possibility, though of course it must have been broadened 
in the merging process. 
The core and cocoon dichotomy support further the analogy with a dwarf disk galaxy
like M33. Indeed, the metal-poor cocoon may be the remnant of a vestigial halo. It will be interesting 
to conduct new simulations in which a small disk galaxy is accreted by M31. 

This luminosity of the Giant Stream, measured from the southern quadrant,
is between a factor of 1 and a factor of 10 less luminous than that of the 
total smooth halo component estimated above.
This indicates that the Giant Stream is a very significant, probably the largest,
merging event into the halo that has ever taken place in Andromeda.
If merging dwarf galaxies are responsible for contributing globular clusters 
into halos, one should therefore expect to find a commensurate number
of halo globular clusters with kinematics compatible the Giant Stream and its
extension.

\begin{figure}
\begin{center}
\includegraphics[angle=0, bb= 120 1 650 550, clip, width=\hsize]{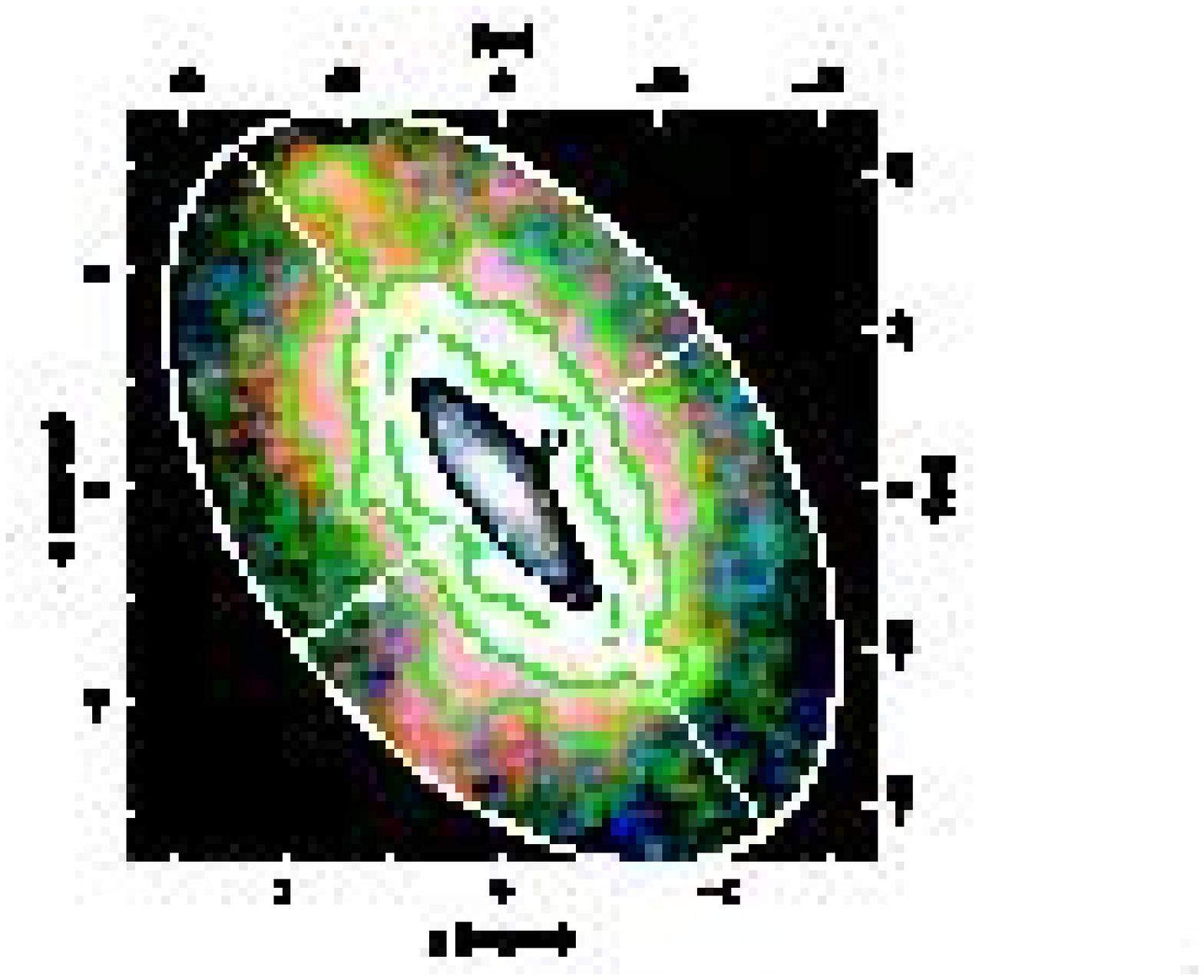}
\end{center}
\caption{RGB color composite map, as Fig.~53, but for the INT data
within the $4\deg$ ellipse (the blue area around the center of the image
is an artifact of crowding in certain central fields). 
We again see the presence of many streams
and structures that have been discussed in earlier articles by our group.
This RGB image, however, shows vividly the differences and
similarities in the stellar populations of these structures. In particular,
one notices that the color of the Giant Stream is similar to
that of the two ``shelves'' (at $\xi \sim 2\deg$, $\eta \sim 0\degg5$ 
and $\xi \sim -1\degg5$, $\eta \sim 0\degg5$) that appear on this map
(see \citealt{ferguson02}).
Other structures, such as the diffuse NE structure ($\xi \sim 1\degg5$, $\eta \sim 2\degg5$)
and the G1 clump ($\xi \sim -1\degg5$, $\eta \sim -1\degg7$) possess a
different distribution of stellar populations. This diagram also allows one
to understand the nature of populations seen at various distances along the 
minor axis. It is clear that at $R\sim 10\kpc$ on the minor axis the 
dominant stellar population is that of the extended messy ellipsoidal 
structure that we have shown previous is a giant rotating component \citep{ibata05}.
Beyond that radius out to $R\sim 20\kpc$ we discern a stellar population
with the same color as the Giant Stream. The contours show the approximate location
of $\Sigma_V=27$, 28 and ${\rm 29~mag/arcsec^2}$.
The locations of the ACS fields of \citet{brown03, brown06a, brown06b, brown07}
are indicated with purple squares (the ACS field sizes have been exaggerated for display purposes).}
\end{figure}

In Fig.~53 we present an RGB
image of the survey region, in which the red, green and blue channels
contain, respectively, the matched filter maps for metal-rich (${\rm -0.7 < [Fe/H] < 0.0}$), 
intermediate (${\rm -1.7 < [Fe/H] < -0.7}$) and metal-poor (${\rm -3.0 < [Fe/H] < -1.7}$)
stars. This image shows the striking differences in stellar populations of the 
halo substructures we have identified in this survey. Even though the Giant Stream
remains the most significant accretion, many more smaller systems are being accreted.
M31 is evidently still leading a colorful life assimilating its small neighbors.

We see also that halo formation is evidently a stochastic process. 
The halo profile and detailed properties of the halo
can therefore be expected to differ from galaxy to galaxy depending
on the amount of substructure and merging debris that is 
present. This makes it all the more surprising that the
profile of the smooth halo discussed above resembles well that of the Milky Way,
suggesting that the reason for this is an underlying similarity in the mass
distributions, which is independent of the detailed assembly history.

\subsection{The inner minor axis}

The several streams detected on the minor axis from $\sim 6\degg5$ all the way
into the edge of the disk are particularly important in that they shed light on
the numerous previous studies (reviewed in \S1) made in this region because
it has been 
considered ``clean halo'' for many years. Indeed, it is not obvious that there 
exists a region of ``clean halo'' in the inner galaxy. This is demonstrated in Fig.~54,
which shows a RGB color composite similar to Fig.~53, but using only INT data
and with a smaller pixel scale. The variations in stellar populations are apparent
as color differences, and one can readily see that the G1 clump and NE structure
have a different distribution of stellar populations to the Giant Stream and the
two ``shelves'' to the East and West (the figure caption states their location).

\begin{figure}
\begin{center}
\includegraphics[angle=0, width=\hsize]{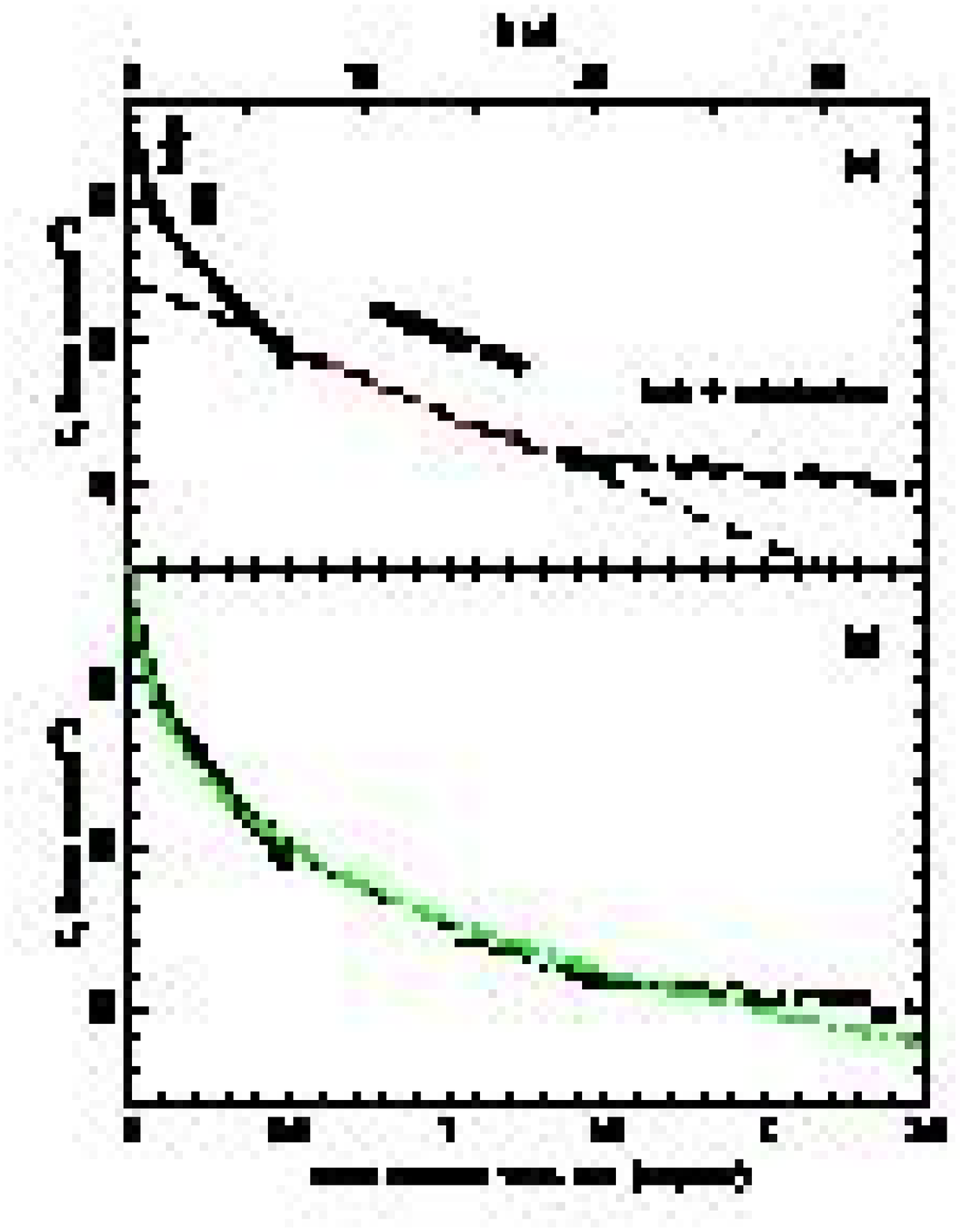}
\end{center}
\caption{The black points in both panels 
reproduce the V-band minor axis surface brightness
profile from \citet{irwin05}. The radial interval $8 < R < 18\kpc$ (marked
with red points in panel `a') 
is clearly almost straight in this log-linear representation. Fitting the data in
this region with an exponential function (dashed line), yields
a scale length of $3.22 \pm 0.02\kpc$ (where the uncertainty is the formal
error on the fit). 
We also indicate the regions where the various components
are dominant. Recent analysis of the 2MASS 6X imaging data of
M31 shows a high-contrast bulge that dominates the near infrared
light out to $\sim 2.6\kpc$ on the major axis \citep{beaton07}.
The bumps in the disk-dominated region (at projected
radii between $2\kpc \simlt R \simlt 6\kpc$ on the minor axis) are due to spiral arms
and the star-forming ring.
The dashed line in panel `b' shows a de Vaucouleurs model fit 
using an effective radius of $R_e=0\degg1$
as found by \citet{pritchet94}, equivalent to $1.4\kpc$ \citep{irwin05},
which overestimates the starcounts between $1\deg < R < 1\degg5$.}
\end{figure}

The contours in Fig.~54 show the iso-luminosity surfaces derived from star-counts
for stars with ${\rm -3.0 < [Fe/H] < 0.0}$, with contour separation of
${\rm 1~mag/arcsec^2}$ (the levels correspond approximately to $\Sigma_V = 27$, 28 and
${\rm 29~mag/arcsec^2}$). It is immediately 
apparent from this diagram that at a projected radius from $R\sim 10\kpc$
to $R \sim 20\kpc$ on the minor axis,
the dominant component is a large irregular ellipsoidal structure whose major
axis size extends out to $R \sim 40\kpc$. We
have shown previously from kinematics in many fields around the galaxy that this
is an extended rotating disk-like component \citep{ibata05}.
Thus, although the surface brightness profile on the minor axis 
follows approximately an $R^{1/4}$ law out to 
$1\degg4$, or $19\kpc$ \citep{pritchet94, irwin05}, it is unlikely that the bulge
itself extends out to those radii. Indeed the bulge in near
infrared wavelengths is a relatively compact structure that dominates
out to $\sim 2.6\kpc$ on the {\it major axis} \citep{beaton07}.
It is therefore pertinent in the current context to 
review the evidence for the $R^{1/4}$ law profile. In Fig.~55 we reproduce the 
V-band minor axis profile from \citet{irwin05}; in the interval $8 < R < 18\kpc$ 
the light profile is actually remarkably similar to an exponential function with
a scale length of $3.22\kpc$.
We stress that this exponential behavior is not confined to the minor axis data alone: it is
present with the same density profile (and normalization) at all azimuth angles (see Fig.~3 of \citealt{ibata05}).
In contrast, the de Vaucouleurs profile of \citet{pritchet94}, shown in panel `b' of Fig.~55,
over-predicts the counts in the radial range $1\deg < R < 1\degg5$.

The ``extended disk'' component
was found to have an intrinsic scale length of $6.6\pm0.4\kpc$ \citep{ibata05},
and to follow an exponential profile out to $\sim 40\kpc$ (after which the profile
flattens out). For the minor axis scale length of $3.22\kpc$ to be consistent
with that intrinsic scale length, the inclination of the outer disk would have to be $60\degg8$,
very close to the value of $64\degg7$ estimated by \citet{ibata05}.
Furthermore, the intrinsic break at $40\kpc$ (deprojected) 
would correspond to $1\degg4$ on the minor axis, 
exactly where it is seen.

If one wishes to adhere to the previously-held
assumption that the minor axis is dominated out to $R \sim 20\kpc$ by an immense
$R^{1/4}$-law ``bulge'' or ``spheroid'', it requires a considerable stretch of credibility. It means
that this ``spheroid'' has to be substantially flattened to be consistent with
the contours of Fig.~54; the ``spheroid'' must have
an exponential-like profile between (deprojected) radii
of $15 \simlt R \simlt 40\kpc$ at all azimuth angles; and it must be rotationally-supported, but with
a rotation rate almost as fast as that of the \ion{H}{1} disk. 
We therefore judge that the ``extended disk'' picture is a far more likely and less contrived
model. This confirms the visual impression of Fig~54: 
in the distance range $10 \simlt R \simlt 15\kpc$ the minor axis profile is dominated by
a disk-like population, with only minor contribution from the bulge or spheroid. 

Since we now understand the kinematic and chemical behavior of the ``extended disk'' from observations
close to the major axis (where stars of different components may be more easily 
distinguished by their differences in kinematics), we can use these insights
to interpret the radial variation in the
properties of the stellar populations on the minor axis. Interior to $\sim 0\degg2$ on the minor
axis the dominant population will clearly be the bulge; further out
between $0\degg2 < R < 0\degg4$, the normal disk contributes
in a non-negligible fashion to the profile, as noted by \citet{irwin05};
then from $0\degg5 < R < 1\degg3$ the extended disk component
becomes dominant; finally beyond $1\degg5$ the 
underlying smooth halo becomes important, though spatially confined
streams dominate at various locations. 

Consequently, one should also expect strong radial variations in metallicity and
kinematics. The kinematics on the minor axis in particular will be complex, and difficult
to disentangle, since all populations have the same mean velocity and 
their velocity distributions overlap. Going out from the center
one should therefore expect to find the
bulge, with high metallicity and high velocity dispersion; then in the bulge plus disk region,
a wide metallicity range, but a narrower velocity dispersion;
then with the addition of the extended disk, the mean metallicity should
decrease towards ${\rm [Fe/H] \sim -0.9 \pm 0.2}$, and the velocity
distribution should contain a significant fraction of stars in a peak with
dispersion in the range $20\kms$ to $50\kms$ \citep{ibata05}; then the
halo component should appear with ${\rm [Fe/H] \sim -1.4}$ and
with a large velocity dispersion of $\sigma_v \sim 140\kms$ at $R=20\kpc$, 
decreasing outwards \citep{chapman06}. In addition to these smooth structures
one will find the multiple streams detected (and not yet detected!) in this area, which as we have shown
can have quite different stellar populations, but which are likely to be dominated
by the metal-rich Giant Stream. The velocity distribution of these streams 
in a small field will in general be a narrow velocity spike of
dispersion $\sim 10\kms$.
However, we stress that the minor axis is a very complex region interior to $\sim 30\kpc$, 
with a complex mix of many stellar populations, each component overlapping considerably
with the others in terms of radial velocity, metallicity, spatial location, color-magnitude
structure, etc.

This finding that the minor axis region between $8 \simlt R \simlt 20\kpc$ is dominated by
the extended disk, and not bulge, halo or spheroid  as has been assumed in numerous
earlier articles, goes a long way towards clarifying the diverse and confusing results
that have been deduced from observations in this region.
In particular, it helps interpret the findings of \citet{brown03, brown06a, brown06b, brown07}. 
These authors obtained ultra-deep
HST/ACS photometry in two minor axis fields, a Giant stream field, and a field 
at the edge of the NE disk, in order to determine
ages of the underlying populations via main-sequence turnoff fitting (field locations are
shown with purple squares in Fig.~54).
Their two minor axis fields lie at projected radii of $R=11$ and $21\kpc$.
Due to the reasons detailed above, their ``spheroid'' field at $R=11\kpc$ 
probes a location which is dominated by the extended disk population. 
From their photometry in this region they deduce a best fitting stellar populations 
model that has ${\rm \langle [Fe/H] \rangle = -0.6}$
and ${\rm \langle age \rangle = 9.7 \Gyr}$. 
\citet{brown06b} dismiss the possibility that the field is related to the extended disk
partly on the grounds that the field lies at a de-projected distance of $51\kpc$,
yet any small warping of the plane of the galaxy, such as we deduced in \citep{ibata04}, invalidates this argument.
The remaining argument is the velocity dispersion measurement of $\sim 80\kms$,
which appears high for the extended disk ($\sigma_v \simlt 50\kms$), 
until one considers the mix of components that must be present at this location.

Further out on the minor axis at $R\sim 20\kpc$ one can discern a diffuse 
component that appears of the same red hue as the Giant Stream with this
color representation. This is clearly a metal-rich region, and possibly related to
the extension of the ``NE shelf'' of \citet{ferguson02}, itself the likely continuation of
the orbit of the Giant Stream \citep{ibata04}. Indeed, \citet{ferguson05} showed that
the Giant Stream and NE shelf are connected on the basis of near identical stellar
populations to 3 magnitudes below the horizontal branch.
With hindsight it is therefore not surprising that the $R=21\kpc$ field
of \citet{brown06b} contains intermediate age stars that have a distribution of stellar
populations essentially identical to that of their Giant Stream field (which is itself
on the outskirts of the ``extended disk'' region). Fig.~53 also
suggests that their NE disk field is also a complex mixture of disk, extended disk,
and possibly metal-rich debris from the Giant Stream.

We note also in passing that the geometry of the minor axis populations
has important consequences for
microlensing studies in M31 (e.g., \citealt{calchi05}). With most of the stellar
populations previously assumed to lie in the spheroid, being confined primarily
in a disk, we predict a much lower self-lensing rate.

\subsection{Kinematics of substructures}

The above discussion also clarifies some previous claims for 
the existence of kinematic substructure around M31.
In a field at $R=19\kpc$, \citet{reitzel02} find four metal-rich stars in their sample
with similar radial velocity of $\approx -340\kms$, 
which they interpreted as evidence for accretion debris. This position
lies within the diffuse region that has stellar populations similar to Giant Stream
(Fig.~54), so the kinematic substructure in the \citet{reitzel02} sample 
is likely related to that structure.

Further kinematic substructure in this region was found by \citet{kalirai06a}
who in studying the kinematics of the Giant Stream find a secondary
kinematic peak $R=20\kpc$ with $\overline{v} = -417\kms$ and $\sigma_v \approx 16\kms$.
The location of this field (H13s) lies at $\xi=0\degg29$,$\eta=-1\degg53$, clearly within
the ellipsoidal contours in Fig.~54, and furthermore the expected mean velocity
of the ``extended disk'' model of \citet{ibata05} predicts $v=-381 \pm 22\kms$ in this field.
The velocity dispersion of the cold component is also similar to what has been found in 
certain regions of the extended disk (e.g., $17\kms$ in field F3 of \citealt{ibata05}).
We speculate therefore that the cold kinematic structure in field H13s is
clumpy structure of the edge of the ``extended disk''.

Most recently \citet{gilbert07} have presented a kinematic survey of several
fields along the minor axis of M31. They detect kinematic substructure in 
three fields, with dispersions of $55.5^{+15.6}_{-12.7} \kms$ ($R=12\kpc$)
$51.2^{+24.4}_{-15.0} \kms$  ($R=13\kpc$) and $10.6^{+6.9}_{-5.0} \kms$
($R=18\kpc$). It is probable that the two structures of velocity 
dispersion $\sim 50\kms$ are also related to the ``extended disk'' component. 
The large de-projected distances
they deduce along the minor
axis ($51$ -- $83\kpc$) are acutely dependent on the assumption 
of constant inclination of the disk, which as we have shown is not
supported by the data \citep{ibata05}. In particular, the $R=12$ and 
$13\kpc$ fields of \citet{gilbert07} lie in the 
distance regime where the extended disk is dominant in Fig.~55.
The cold kinematic component observed in their $R=18\kpc$ field is likely 
related to the Giant Stream for the same reason as is the cold kinematic
structure in the \citet{reitzel02} sample.

\section{Conclusions}

This article has presented a deep panoramic view of the Andromeda galaxy
and part of the Triangulum galaxy. Though it is not the deepest external 
galaxy survey ever undertaken, nor the most extended, we have for the first time
covered a substantial fraction of a galaxy out to a substantial fraction of
the virial radius to sufficient depth 
to detect several magnitudes of the red giant branch and with sufficient photometric accuracy
to estimate stellar metallicity.
To our knowledge this is the first deep wide-field view of the outermost regions of galaxies.

The new CFHT data presented here are combined with an earlier survey of the inner regions
of M31 ($s \simlt 55\kpc$) taken with the INT \citep{ibata01b, ferguson02, irwin05}.
We summarize below the main findings from these surveys.

\begin{itemize}

\item A huge amount of confusion in the literature has arisen from
assuming that the minor axis region between projected radii of 
$0\degg5 < R < 1\degg3$ ($7\kpc < R < 18\kpc$) is representative of the spheroid. 
We have shown here that it is not. Instead it is likely to be a complex
mix of stellar populations, dominated over much of this radial range
by the ``extended disk''. Many of the previous claims that the 
spheroid or stellar halo of M31 is very different to that of the Milky Way 
were based upon a comparison of the properties of genuine Milky Way halo stars
to those of stars in M31 in quite different components.

\item Beyond the inner ($\sim 20\kpc$) disk, Andromeda contains a multitude of streams,
arcs, shells and other irregular structures. Some of these structures
appear to be related 
(they have a similar mix of stellar populations) others are manifestly due to separate
accretion events.

\item The largest of these structures, the Giant Stream, is very luminous, possessing
$L_V \sim 1.5 \times 10^8 L_\odot$ over the region surveyed with MegaCam. This body dominates
the luminosity budget of the inner halo, and once it becomes fully mixed, may double
the luminosity of the smooth underlying halo. This ongoing accretion event must be 
among the most significant the halo has suffered since its initial formation.

\item Ignoring regions with obvious substructure, we find that the remaining
area of the survey exhibits a smooth metal-poor stellar halo component. This structure
need not be perfectly spatially smooth, but the intrinsic
inhomogeneities are below the sensitivity of this study. The smooth halo
is vast, extending out to the radial limit of the survey,
at $150\kpc$. The profile of this component can be modeled with a Hernquist profile
as suggested by simulations, but the resulting scale radius of $\sim 55\kpc$ is almost
a factor of 4 larger than modern halo formation simulations predict. A power-law
profile with $\Sigma(R) \propto R^{-1.91 \pm 0.11}$ (i.e. $\rho(r) \propto r^{-2.91 \pm 0.11}$)
can also be fit to the data. Simulations predicted a sharp decline in the power law
exponent beyond the central regions of the galaxy to $\rho(r) \propto r^{-4}$ or 
$\rho(r) \propto r^{-5}$. This is not observed. Instead, and unexpectedly, 
the stellar profile mirrors closely the expected profile of the dark matter.

\item Since dynamically young accretion events give rise to arcs and streams, 
and because dynamical times are very long in the outer reaches of the halo,
the smoothness of the component over huge areas of the outskirts of the
galaxy suggests that the component is very old. It therefore seems plausible
that the structure was formed in a cataclysmic merging event early in the
history of the galaxy, probably before the formation of the fragile disk.

\item The outer halo of M31 ($R \simgt 80\kpc$) is not oblate. 
On the contrary, the stellar distribution
appears to be slightly prolate with $c/a \simgt 1.3$, though we judge
that a reliable measurement of this parameter will require further data
in other quadrants.

\item Both the density profile of the smooth halo in M31 and its total luminosity
($\sim 10^9 \lsun$) are very similar to the Milky Way. Their
metallicity and kinematic properties also resemble each other
closely \citep{chapman06, kalirai06a}.
This is somewht surprising if halo formation is a stochastic process as suggested by 
simulations (see, e.g. the discussion in \citealt{renda05}).

\item Lumping all stellar populations together, we detect a stellar population
gradient in the survey
such that the more metal-rich populations are more centrally concentrated,
consistent with the predictions of \citet{bullock05}. However,
this is almost entirely due to the presence of the metal-rich Giant Stream ``contaminating''
the inner halo.

\item An extended slowly-decreasing halo is also detected around M33. Fitting this distribution
with a Hernquist model gives a scale radius of $\sim 55\kpc$, essentially
identical to that of M31, though we caution that the poor azimuthal coverage
of the survey around M33 makes this result sensitive to unidentified substructures
and to assumptions about the geometry of the halo.

\item The stellar halos of M31 and M33 touch in projection, and are probably passing through each
other. The kinematics of stars in this overlap region will be fascinating to analyze, though
large samples will probably be needed to disentangle the structures.

\item Two new dwarf satellite galaxies of M31, And~XV and And~XVI, are presented,
which together with those reported in a previous contribution \citep{martin06},
brings the number of new satellites detected in the MegaCam survey region up to five.
Follow-up studies are currently underway to understand the nature of these objects and
those of lower S/N satellite candidates found in the survey.

\end{itemize}

Many questions remain open.
What is the radial dependence of the metallicity and stellar populations in the smooth component?
Is there a discontinuity in properties between the inner halo and the outer halo
similar to the simulations of \citet{abadi03, abadi06}, reflecting native and immigrant stars?

It will be very interesting to extend the survey out to the virial radius of
the Galaxy and verify whether the correlation between the observed
stellar profile and the expected dark matter surface density continues to that radius.
Further photometric coverage to the East of the minor axis will also
be helpful to study fully the morphology and extent of the stream-like
structures detected from $R=30$ to $\sim 120\kpc$ and 
to determine whether these objects are indeed streams, and so make plausible
judgements about their origin and evolution and compare them to theoretical
predictions of the formation of the outer halo. 

This panorama of the outer fringes of Andromeda and Triangulum
has shown that halos are truly misnamed: they are in reality dark
galactic graveyards, full of the ghosts of galaxies 
dismembered in violent clashes long ago.
Other, even more ancient remnants, have lost all 
memory of their original form, and in filling these
haunted halos with the faintest shadow of their former brilliance, they 
follow faithfully the dark forces to which they first succumbed.
The true nature of this most sombre of galactic recesses is finally beginning to be revealed.

\section*{Acknowledgments}

This study would not have been possible without the excellent support of
staff at the CFHT telescope, and the careful and meticulous observations
performed in queue mode.
RI wishes to thank Annie Robin for allowing us privileged access
to the Besan{\c c}on model via UNIX scripts which greatly
facilitated the construction of the foreground model, and also
many thanks to Michele Bellazzini for helpful comments on this work.
AMNF is supported by a Marie Curie Excellence Grant from
the European Commission under contract MCEXT-CT-2005-025869.

\end{document}